\documentclass[twocolumn,runningheads]{svjour3}

\usepackage[latin1]{inputenc}
\usepackage{times}

\usepackage{psfrag,amsmath, amssymb, amscd, amsfonts, latexsym, epsf, graphicx,amscd}
\usepackage{natbib}
\usepackage{breqn}
\usepackage{hyperref}
\usepackage{subcaption}
\usepackage[font=small]{caption}

\usepackage[switch]{lineno}
\def\bbbr{{\mathbb R}} 
 
\def\diag{\operatorname{diag}}

\newcommand{\orth}{\bot}

\def\norm{\scriptsize\mbox{norm}}

\def\vel{\scriptsize\mbox{vel}}

\journalname{arXiv preprint}

\begin{document}

\titlerunning{Direction and speed selectivity properties for the
  generalized Gaussian derivative model for visual receptive fields}
\title{\bf Direction and speed selectivity properties for spatio-temporal
  receptive fields according to the generalized Gaussian derivative
  model for visual receptive fields%
\thanks{The support from the Swedish Research Council 
              (contract 2022-02969) is gratefully acknowledged. }}
\author{Tony Lindeberg}

\institute{Computational Brain Science Lab,
        Department of Computational Science and Technology,
        KTH Royal Institute of Technology,
        SE-100 44 Stockholm, Sweden.
        \email{tony@kth.se}.
      ORCID: 0000-0002-9081-2170.}

\date{}

\maketitle

\begin{abstract}
  \noindent 
  Simple cells and complex cells in the primary visual cortex are
  usually characterized either in terms of explicitly reconstructed
  spatio-temporal receptive fields or their direction and speed selectivity properties.
  Since reconstruction of receptive field profiles can typically be a rather complex
  problem, while accumulation of direction selectivity and speed selectivity curves
  is comparably more straightforward,
  there is a need for bridging the gap between these two types of
  characterizations of motion-sensitive neurons.
  
  This paper bridges that gap by providing an in-depth theoretical analysis
  of both the separate and joint direction and speed selectivity properties
  of idealized models of the spatio-temporal receptive fields of simple
  cells and complex cells, based on the generalized Gaussian derivative model for visual
  receptive fields. According to this theory, the receptive fields are
  modelled as velocity-adapted affine Gaussian derivatives for
  different image velocities, different degrees of elongation
  and different spatial and temporal scales.
  
  By probing such idealized receptive field models of visual neurons
  to moving sine waves with different angular frequencies and image
  velocities, we characterize the responses of the computational models to a
  structurally similar probing method as is used for characterizing
  the direction and speed selective properties of biological neurons.
  We derive a set of idealized low-parameter models of such 
  direction and speed selectivity curves, which quantify how
  much the direction selective properties become sharper
  with increasing order of spatial differentiation and increasing
  degree of elongation in the spatial components of the visual
  receptive fields. Corresponding results for speed selectivity
  properties quantify how much sharper the speed selectivity
  become for different orders of spatial differentiation, while they
  are for the inclination angle $\theta = 0$ independent of
  the degree of elongation.

  By comparisons to neurophysiological measurements of
  direction and speed selectivity for biological neurons in the
  primary visual cortex, we show that our theoretical results are
  consistent with the properties that (i)~velocity-tuned visual neurons 
  are sensitive to particular motion directions and speeds, and (ii)~different
  visual neurons differ in the sharpness of both their direction
  and speed selectivity properties.
  
  Specifically, by modelling direction and speed selectivity curves obtained from previous
  neurophysiological measurements of motion-sensitive neurons in the
  primary visual cortex, we demonstrate that the predictions about direction
  and speed selectivity curves generated from the presented theory do
  very well match the qualitative shapes of corresponding biologically
  recorded direction and speed selectivity curves, based on different
  members of the derived catalogue of idealized models for simple cells
  and complex cells for different orders of spatial differentiation.
  
  In this way, the presented theory can explain the results we get from
  neurophysiological measurements of visual neurons,
  by anchoring those to direct consequences of a normative theory
  for visual receptive fields, that derives the shapes of the
  spatio-temporal receptive fields by necessity, and with the
  only parameters in the idealized models, beyond the order of spatial
  differentiation, being the shape parameters of the receptive fields
  in terms of (i)~the inclination angle between the stimulus and
  the preferred direction of the receptive field,
  (ii)~the ratio between the speed parameters of the stimulus and the receptive field, and
  (iii)~the degree of elongation of the spatial component of the receptive field.
    
  %Our theoretical results, in combination with results from neurophysiological
  %characterizations of motion-sensitive visual neurons, are furthermore
  %consistent with a previously formulated hypothesis that the simple
  %cells in the primary visual cortex ought to be covariant under local
  %Galilean transformations, so as to enable processing of visual
  %stimuli with different motion directions and speeds.
  
  \keywords{Receptive field \and Direction selectivity \and
    Speed selectivity \and Galilean covariance \and
    Gaussian derivative \and Simple cell \and
    Complex cell \and Vision \and
    Theoretical neuroscience}

\end{abstract}

\section{Introduction}
\label{sec-intro}

When aiming to understand the computational function of the
motion-sensitive neurons in the primary visual cortex,
the spatio-temporal receptive fields of these neurons can aptly
be regarded as the main computational units.
Reconstructing a spatio-temporal receptive from neurophysiological
measurements is, however, a non-trivial task, since it constitutes
an inverse problem that requires both a model for the receptive
field and multiple measurements of the same neuron for different
types of stimuli to resolve the inverse problem.

Measuring the direction selectivity and speed selectivity properties
of a neuron does from this perspective represent a substantially
simpler problem, since it can be accomplished merely by subjecting
the neuron to moving stimuli with different directions and speeds.
For this reason, there are significantly more available results
concerning direction selectivity properties of visual neurons in the
primary visual cortex (V1) and the middle temporal visual area (MT)
(Hubel \citeyear{Hub59-JPhys},
Hubel and Wiesel \citeyear{HubWie62-Phys},
Emerson and Gerstein \citeyear{EmeGer77-JNeurPhys},
Albright \citeyear{Alb84-JNeuroPhys},
Albright {\em et al.\/}\ \citeyear{AlbDesGro84-JNeuroPhys},
Ganz and Felder \citeyear{GanFel84-JNeuroPhys},
Mikami {\em et al.\/}\ \citeyear{MikNewWur86a-JNeuroPhys,MikNewWur86b-JNeuroPhys},
Orban {\em et al.\/}\ \citeyear{OrbKenNul86-JNeurPhys},
Lagae {\em et al.\/}\ \citeyear{LagRaiOrb93-JNeurPhys},
Orban \citeyear{Orb97-ExtrStriCortPrim},
Livingstone \citeyear{Liv98-Neuron},
Perrone and Thiele (\citeyear{PerThi01-NatNaeurSci}),
Cavanaugh  {\em et al.\/}\ (\citeyear{CavBaiMov02-JNeuroPhys}),
Livingstone and Conway \citeyear{LivCon03-JNeuroPhys},
Priebe {\em et al.\/} (\citeyear{PriCasLis03-JNeuroSci}),
Born and Bradley \citeyear{BorBra05-AnnRevNeurSci},
Churchland {\em et al.\/}\ \citeyear{ChuPriLis05-JNeuroPhys},
Gur {\em et al.\/}\ \citeyear{GurKagSno05-CerebrCortex},
Moore {\em et al.\/}\ \citeyear{MooAliUsr05-JNeuroPhys},
Priebe and Ferster \citeyear{PriFer05-Neuron},
Priebe  {\em et al.\/}\ (\citeyear{PriLIsMov06-JNeuroSci,PriLamFer10-JNeuroPhys}),
Gur and Snodderly \citeyear{GurSno07-JPhys},
Elstrott and Feller \citeyear{ElsFel09-CurrOpNeuroBiol},
Priebe  {\em et al.\/}\ \citeyear{PriLIsMov06-JNeuroSci,PriLamFer10-JNeuroPhys},
Wang and Yao \citeyear{WanYao11-CerebrCortex},
Hansel and van~Vreeswijk \citeyear{HanVre12-JNeuroSci},
Dai {\em et al.\/}\
\citeyear{DaiWanLiYaZhaWuZhoyuLiWanWanXin25-PLOSBio},
Khamiss {\em et al.\/}\ \citeyear{KhaLemNanNie25-bioRxiv})
than there are publicly available
reconstructed spatio-temporal receptive fields
(DeAngelis {\em et al.\/}\
\citeyear{DeAngOhzFre95-TINS,deAngAnz04-VisNeuroSci},
de~Valois {\em et al.\/}\ \citeyear{ValCotMahElfWil00-VisRes}).

From a theoretical perspective, this imposes a need for bridging the conceptual
gap between these two ways of characterizing motion-sensitive neurons.
In this paper, we derive such explicit relationships for how different shape
parameters in idealized models of the spatio-temporal receptive fields
for simple cells and complex cells lead to different types and
shapes of direction and speed selectivity curves.

We do this for a special type of model of visual receptive fields referred to as
the generalized Gaussian derivative model for visual receptive fields
(Lindeberg \citeyear{Lin21-Heliyon}).
This model has been derived in an axiomatic normative way, by starting
from symmetry properties of the environment combined with internal
consistency requirements to guarantee internally consistent processing
of image representations over multiple spatial and temporal scales.
Specifically, this theoretical model for visual receptive fields has been formulated to
consistently handle the influence of geometric image transformations
on the visual input, to make it possible to, up to first order of approximation,
match the receptive field responses computed from image data that have
been acquired under different viewing conditions in terms of
varying the distance, the viewing direction and the relative motion between
objects in the world and the observer. This is achieved by requiring the family
of receptive fields to be covariant%
\footnote{Covariance in this context means that the result of applying 
  a spatio-temporal receptive field ${\cal R}$ to a geometric image transformation
  ${\cal T}$ of a spatio-temporal video pattern $f$ should be essentially
  the same as applying the same geometric transformation ${\cal T}$
  to an essentially similar type of receptive field ${\cal R}'$ to the same
  spatio-temporal image pattern $f$, in such a way that the receptive field
  essentially commutes with the geometric image transformation,
  such that ${\cal R} \, {\cal T} \, f = {\cal T} \, {\cal R}' \, f$.
  By the second receptive field ${\cal R}'$ being required to be essentially similar to
  the first receptive field ${\cal R}$ means that the two receptive fields
  should either be members of the same family of receptive fields,
  or being related in such a way up to a very simple relationship,
  such as a uniform scaling of the magnitude.
  The notion of covariant receptive fields thereby implies that the
  family of receptive fields is theoretically well-behaved under
  the influence of geometric image transformations.}
under the basic classes of locally linearized geometric image transformations
in terms of spatial scaling transformations, spatial affine transformations,
Galilean transformations and temporal scaling transformations
(Lindeberg \citeyear{Lin23-FrontCompNeuroSci}).

According to the generalized Gaussian derivative model for visual
receptive fields, there ought to specifically be variabilities in the
shapes of the receptive fields according to the degrees of freedom
of the geometric image transformations, to make it possible for
corresponding receptive field responses to be matched under
the influence of geometric image transformations
(Lindeberg \citeyear{Lin26-JCompNeuroSci}).

With respect to the spatio-temporal receptive fields of simple cells
and complex cells, there therefore specifically ought to be variabilities in
the shapes of the receptive fields with respect to Galilean transformations,
to be able to implement Galilean covariance, in order to be able to
consistently process visual stimuli with different relative motions
between objects in the environment and the observer,
Additionally, with regard to the degree of elongation of the spatial
components of the receptive fields, to make it possible to match
the receptive field responses computed from the same surface
patch seen from two different slant angles in relation to the
viewing direction, the spatio-temporal receptive fields also ought to be
affine covariant, which then implies that the receptive fields
ought to have a variability over the degree of elongation
(Lindeberg \citeyear{Lin25-JCompNeurSci-spanelong,Lin25-BICY}).

Due to the very sparse availability of explicitly reconstructed spatio-temporal
receptive fields for simple cells in the primary visual cortex
(DeAngelis {\em et al.\/}\
\citeyear{DeAngOhzFre95-TINS,deAngAnz04-VisNeuroSci},
de~Valois {\em et al.\/}\ \citeyear{ValCotMahElfWil00-VisRes}),
the direct neurophysiological support for the spatio-temporal
receptive fields derived from the normative theory is, however, correspondingly
sparse, although the qualitative shapes of those reconstructed 
receptive fields can be well modelled by idealized receptive fields models
derived from the normative theory, see Figure~18 in 
Lindeberg (\citeyear{Lin21-Heliyon}) regarding space-time separable
and non-separable spatio-temporal receptive fields,
Figures~13--17 in the same paper regarding spatial receptive
fields in the primary visual cortex, and Figure~12 regarding LGN neurons.

This implies a need for further evaluating the normative theory in
relation to neurophysiological measurements.
By comparing idealized models of direction and speed selectivity
curves for simple cells and complex cells derived as direct
consequences of the normative theory for visual receptive fields,
we will in this paper show that the resulting predictions agree very well with
biological speed selectivity and direction selectivity measurements
recorded by Orban {\em et al.\/}\ (\citeyear{OrbKenNul86-JNeurPhys})
and Livingstone and Conway (\citeyear{LivCon03-JNeuroPhys}).
In this way, we demonstrate that predictions from the axiomatically
determined normative theory are consistent with the results of
neurophysiological measurements.

Specifically, with regard to explicit variabilities of the shapes of the
receptive fields over the degrees of freedom of geometric image
transformations, we will demonstrate that the resulting fitted shape parameters of the
receptive fields in the idealized direction and speed selectivity
curves are consistent with variabilities over both the direction,
the speed of the receptive fields over Galilean motions, and
over the degree of elongation of the spatial components of
the receptive fields.
Additionally, with regard to the hypothesis about Galilean-covariant
visual processing for higher mammals, we summarize results of previous findings
regarding the visual areas V1 and MT that are consistent with
an expansion of the receptive field shapes over the degrees of
freedom of local Galilean transformations.

As an immediate consequence of this theoretical analysis, we obtain
very parameter-compact idealized models of direction and speed
selectivity curves, where the parameters of these idealized curves are
solely in terms of shape parameters of the underlying
spatio-temporal receptive fields, specifically
(i)~the inclination angle between the stimulus and the preferred
direction of the receptive field, (ii)~the ratio between the speed parameters of
the stimulus and the receptive field, and (iii) the degree of
elongation of the spatial component of the receptive field.

In this way, we argue that essential aspects of the biologically
recorded direction and speed selectivity curves can be explained
by a very small set of shape parameters of the underlying spatio-temporal
receptive fields.
Complementing with the fact that these shape parameters of the
receptive fields do, in turn, directly connect to the parameters
of the geometric image transformations in the image formation
process according to the unified theory for this
in Lindeberg (\citeyear{Lin25-JMIV}), we can from this perspective
regard the overall shapes of the neurophysiologically recorded
speed and direction selectivity curves as largely determined
by external properties of the environment in relation to the visual observer.
Specifically, the variabilities in the shapes of the experimentally recorded
direction and speed selectivity curves that are observed in biological
experiments can be explained by variabilities in the internal shape
parameters of the spatio-temporal receptive fields,
and with the variability in the shape parameters of the receptive fields,
in turn, explained by the requirement of covariance under geometric
image transformations, to, in turn, be able to match the receptive field responses
computed from the same local surface patch seen under variations
in the relative motion between the object and the observer
and variations in the slant angle in relation to the viewing direction.

In these ways, we do hence address the overreaching goal of bridging the gap
between theoretical models and neurophysiological characterizations of
spatio-temporal receptive fields in the primary visual cortex.

% An underlying motivation for the proposed new type of mathematical
% analysis of properties of computational models of motion-sensitive
% visual neurons is that a deeper understanding of their properties,
% including how these properties relate to neurophysiologically
% measurable entities of biological neurons, could allow for better
% computational models of motion-sensitive neurons, and thereby a better
% understanding of the functional properties of the early vision system.

% To promote opportunities for additional more detailed quantitative modelling of
% the spatio-temporal receptive fields in the primary visual cortex, we will
% also outline directions for further neurophysiological experiments.

\subsection{Contributions and novelty}

This paper comprises the following main theoretical and experimental contributions:
\begin{itemize}
\item
  Derivation of direction and speed selectivity curves
  for 4~submodels of {\em simple cells\/}, based on
  velocity-adapted affine Gaussian derivatives of orders
  between 1 and~4
  (Section~\ref{sec-dir-sel-simpl-cells} and Appendix~\ref{app-anal-simpl-cells}).
\item
  Derivation of direction and speed selectivity properties for a
  set of different models of {\em complex cells\/}, in terms of regionally
  integrated quasi quadrature measures of the output from idealized
  models of simple cells of orders $\{ 1, 2 \}$ or $\{ 3, 4 \}$
  (Section~\ref{sec-dir-sel-compl-cells} and Appendix~\ref{app-anal-compl-cells}).
\item
  Comparisons of the derived idealized models of speed selectivity curves
  to {\em speed selectivity measurements\/} by
  Orban {\em et al.\/}\ (\citeyear{OrbKenNul86-JNeurPhys}),
  demonstrating that the qualitative shapes of these
  neurophysiological recorded speed selectivity curves can be well captured by
  those from corresponding idealized models for simple cells or
  complex cells for different orders of differentiation
  (Section~\ref{sec-rels-neurophys-spec-Orban}).
\item
  Comparisons of the derived idealized models of direction selectivity curves
  to {\em direction selectivity measurements\/} by
  Livingstone and Conway (\citeyear{LivCon03-JNeuroPhys}),
  demonstrating that the qualitative shapes of these
  neurophysiologically recorded direction selectivity curves
  can, for most of the neurons, be well captured by
  the corresponding idealized models from simple cells or
  complex cells for different orders of differentiation
  (Section~\ref{sec-model-livingstone}).
\item
  A theoretically based prediction about instead characterizing direction and speed
  selectivity {\em jointly\/}, instead of as in previous work of using separate
  direction and speed selectivity properties
  (Section~\ref{sec-joint-selectivity-maps}).
\item
  An evaluation of the {\em hypothesis about Galilean covariant visual processing\/}
  with its associated expansion of the receptive field shapes over the degrees of
  freedom of Galilean transformations
  in relation to previous neurophysiological results regarding areas V1 and MT
  (Section~\ref{sec-biol-rel-gal-cov}).
\item
  A set of {\em proposals for further directions of neurophysiological experiments\/}
  to investigate predictions raised from the presented theoretical framework
  (Section~\ref{sec-sugg-neuro-exps}).
\end{itemize}
These theoretical contributions and the corresponding experimental
evaluations thereof do specifically extend the normative theory for
visual receptive fields in
Lindeberg (\citeyear{Lin21-Heliyon,Lin23-FrontCompNeuroSci,Lin25-BICY,Lin25-JMIV,Lin26-JMIV,Lin26-JCompNeuroSci})
to direction and speed selectivity properties of idealized models of
simple cells and complex cells in the primary visual cortex.

To make the paper easier for less technical readers, the mathematically
technical material in terms of the actual derivations
of the main theoretical results has been put in
Appendices~\ref{app-anal-simpl-cells} and~\ref{app-anal-compl-cells},
with the main results then summarized on more compact forms in
Sections~\ref{sec-dir-sel-simpl-cells} and~\ref{sec-dir-sel-compl-cells},
including the interpretations of these theoretical results with regard to theoretical
modelling of spatio-temporal receptive fields and their
direction and speed selectivity properties.
The biological interpretations of these results, including
extended theoretical predictions, are then described
in Sections~\ref{sec-rels-biol-vision} and~\ref{sec-sugg-neuro-exps}.

\section{Relations to previous work}
\label{sec-rel-work}

The direction and speed selectivity properties of neurons in the primary
visual cortex (V1), the secondary visual cortex (V2)
and the middle temporal visual area (MT) have
been mapped in several neurophysiological studies by
Hubel (\citeyear{Hub59-JPhys}),
Hubel and Wiesel (\citeyear{HubWie62-Phys}),
Emerson and Gerstein (\citeyear{EmeGer77-JNeurPhys}),
Albright (\citeyear{Alb84-JNeuroPhys}),
Albright {\em et al.\/}\ (\citeyear{AlbDesGro84-JNeuroPhys}),
Ganz and Felder (\citeyear{GanFel84-JNeuroPhys}),
Foster {\em et al.\/} (\citeyear{FosGasNagPol85-JPhys}),
Mikami {\em et al.\/}\
(\citeyear{MikNewWur86a-JNeuroPhys,MikNewWur86b-JNeuroPhys}),
Orban {\em et al.\/}\ (\citeyear{OrbKenNul86-JNeurPhys}),
Lagae {\em et al.\/}\ (\citeyear{LagRaiOrb93-JNeurPhys}),
Livingstone (\citeyear{Liv98-Neuron}),
Perrone and Thiele (\citeyear{PerThi01-NatNaeurSci}),
Cavanaugh  {\em et al.\/}\ (\citeyear{CavBaiMov02-JNeuroPhys}),
Livingstone and Conway (\citeyear{LivCon03-JNeuroPhys}),
Priebe {\em et al.\/} (\citeyear{PriCasLis03-JNeuroSci}),
Churchland {\em et al.\/} (\citeyear{ChuPriLis05-JNeuroPhys}),
Gur {\em et al.\/}\ (\citeyear{GurKagSno05-CerebrCortex}),
Moore {\em et al.\/}\ (\citeyear{MooAliUsr05-JNeuroPhys}),
Priebe and Ferster (\citeyear{PriFer05-Neuron}),
Priebe  {\em et al.\/}\ (\citeyear{PriLIsMov06-JNeuroSci,PriLamFer10-JNeuroPhys}),
Gur and Snodderly (\citeyear{GurSno07-JPhys}),
Elstrott and Feller (\citeyear{ElsFel09-CurrOpNeuroBiol}),
Priebe  {\em et al.\/}\ (\citeyear{PriLamFer10-JNeuroPhys}),
Wang and Yao (\citeyear{WanYao11-CerebrCortex}),
Hansel and van~Vreeswijk (\citeyear{HanVre12-JNeuroSci}),
Dai {\em et al.\/}\
(\citeyear{DaiWanLiYaZhaWuZhoyuLiWanWanXin25-PLOSBio}) and
Khamiss {\em et al.\/}\ (\citeyear{KhaLemNanNie25-bioRxiv})
with excellent reviews by
Orban (\citeyear{Orb97-ExtrStriCortPrim}) and
Born and Bradley (\citeyear{BorBra05-AnnRevNeurSci}).

Specifically, Priebe  {\em et al.\/}\
(\citeyear{PriLIsMov06-JNeuroSci})
have performed an extensive mapping of
how direction selectivity properties are related to both
spatial and temporal frequency characteristics of biological neurons.
Since the magnitude maps of the
spatial and temporal frequency characteristics are, in general,
however, not sufficient to fully characterize the spatio-temporal
response properties to any given input data,
unless the local phase information is also mapped,
we in this paper instead characterize how intrinsic parameters
of the explicit spatio-temporal receptive field, in
terms of the degree of elongation and the preferred
speed of the neuron in relation to the speed of the stimulus,
determine the direction and speed selectivity properties.
We derive explicit expressions for this on compact closed form,
in this way more addressing theoretical properties between
spatio-temporal receptive fields and their
direction and speed selectivity properties.

Computational models of motion analysis with close relations
neurophysiological results regarding motion-sensitive neurons
in the primary visual cortex and the middle temporal visual area
have been formulated by
Adelson and Bergen (\citeyear{AdeBer85-JOSA}),
Heeger (\citeyear{Hee87-JOSA}),
Wilson {\em et al.\/}\ (\citeyear{WilFerYo93-VisNeuroSci}),
Nowlan and Sejnowski (\citeyear{NowSej97-JNeuroSci}),
Simoncelli and Heeger (\citeyear{SimHee98-VR}) and
Lisberger and Movshon (\citeyear{LisMov99-JNeuroSci}).
Srinivasan (\citeyear{Sri90-BICY}) developed an early model that
uses spatiotemporal processing for detecting the direction of
motion and the image velocity.
A recent overview of experimentally driven computational models of
area MT is given by
Zarei~Eskikand {\em et al.\/}\ (\citeyear{ZarGraKamBurIbb24-RevNeuroSci}).
The goal of the theoretical analysis in this paper is, however,
not to address the motion estimation problem,
but instead to understand how the direction and speed selectivity properties
of visual neurons can be related to properties of the underlying
spatio-temporal receptive fields, based on highly idealized models thereof.

Our knowledge about the functional properties of the receptive fields
of simple cells in the primary visual cortex originates from the
pioneering work by Hubel and Wiesel
(\citeyear{HubWie59-Phys,HubWie62-Phys,HubWie68-JPhys,HubWie05-book})
followed by more detailed characterizations by
DeAngelis {\em et al.\/}\
(\citeyear{DeAngOhzFre95-TINS,deAngAnz04-VisNeuroSci}),
de~Valois {\em et al.\/}\ (\citeyear{ValCotMahElfWil00-VisRes}),
Ringach (\citeyear{Rin01-JNeuroPhys,Rin04-JPhys}),
Conway and Livingstone (\citeyear{ConLiv06-JNeurSci}),
Johnson {\em et al.\/}\ (\citeyear{JohHawSha08-JNeuroSci}),
Walker {\em et al.\/}
(\citeyear{WalSinCobMuhFroFahEckReiPitTol19-NatNeurSci}) and
De and Horwitz (\citeyear{DeHor21-JNPhys}).

Computational models of simple cells have specifically
been expressed in terms of Gabor filters 
by Marcelja (\citeyear{Mar80-JOSA}),
Jones and Palmer (\citeyear{JonPal87a,JonPal87b}),
Porat and Zeevi (\citeyear{PorZee88-PAMI}),
Ringach (\citeyear{Rin01-JNeuroPhys,Rin04-JPhys}),
Serre {\em et al.\/} (\citeyear{SerWolBilRiePog07-PAMI}),
Baspinar {\em et al.\/} (\citeyear{BasCitSar18-JMIV,BasSarCit20-MathNeuroSci})
and
De and Horwitz (\citeyear{DeHor21-JNPhys}),
and in terms of Gaussian derivatives by
Koenderink and van Doorn (\citeyear{Koe84,KoeDoo87-BC,KoeDoo92-PAMI}),
Young (\citeyear{You87-SV}),
Young {\em et al.\/}\ (\citeyear{YouLesMey01-SV,YouLes01-SV}) and
Lindeberg (\citeyear{Lin13-BICY,Lin21-Heliyon}).
Theoretical models of early visual processes have also been
formulated based on Gaussian derivatives by
Lowe (\citeyear{Low00-BIO}),
May and Georgeson (\citeyear{MayGeo05-VisRes}),
Hesse and Georgeson (\citeyear{HesGeo05-VisRes}),
Georgeson  {\em et al.\/}\ (\citeyear{GeoMayFreHes07-JVis}),
Hansen and Neumann (\citeyear{HanNeu09-JVis}),
Wallis and Georgeson (\citeyear{WalGeo09-VisRes}),  
Wang and Spratling (\citeyear{WanSpra16-CognComp}),
Pei {\em et al.\/}\ (\citeyear{PeiGaoHaoQiaAi16-NeurRegen}),
Ghodrati {\em et al.\/}\ (\citeyear{GhoKhaLeh17-ProNeurobiol}),
Kristensen and Sandberg (\citeyear{KriSan21-SciRep}),
Abballe and Asari (\citeyear{AbbAsa22-PONE}),
Ruslim {\em et al.\/}\ (\citeyear{RusBurLia23-bioRxiv}) and
Wendt and Faul (\citeyear{WenFay24-JVis}).
Modelling of spatio-temporal receptive fields using Gabor
functions has, in turn, been proposed by
Cocci {\em et al.\/} (\citeyear{CocBarSar11-JOSA}) and
Barbieri {\em et al.\/} (\citeyear{BarCitCocSar14-JMIV}).

The neurophysiological properties of complex cells have been studied by
Movshon {\em et al.\/}\ (\citeyear{MovThoTol78-JPhys}), 
Emerson {\em et al.\/}\ (\citeyear{EmeCitVauKle87-JNeuroPhys}),
Martinez and Alonso (\citeyear{MarAlo01-Neuron}),
Touryan {\em et al.\/}\ (\citeyear{TouLauDan02-JNeuroSci,TouFelDan05-Neuron}),
Rust {\em et al.\/}\ (\citeyear{RusSchMovSim05-Neuron}),
van~Kleef {\em et al.\/}\ (\citeyear{KleCloIbb10-JPhys}),  
Goris {\em et al.\/}\ (\citeyear{GorSimMov15-Neuron}),
Li {\em et al.\/}\ (\citeyear{LiLiuChoZhaTao15-JNeuroSci}) and
Almasi {\em et al.\/}\ (\citeyear{AlmMefCloWonYunIbb20-CerCort}),
as well as been modelled computationally by
Adelson and Bergen (\citeyear{AdeBer85-JOSA}),
Heeger (\citeyear{Hee92-VisNeuroSci}),
Serre and Riesenhuber (\citeyear{SerRie04-AIMemo}),
Einh{\"a}user {\em et  al.\/} (\citeyear{EinKayKoeKoe02-EurJNeurSci}),
Kording {\em et al.\/}\ (\citeyear{KorKayWinKon04-JNeuroPhys}),
Merolla and Boahen (\citeyear{MerBoa04-NIPS}),
Berkes and Wiscott (\citeyear{BerWis05-JVis}),
Carandini (\citeyear{Car06-JPhys}),
Hansard and Horaud (\citeyear{HanHor11-NeurComp}),
Franciosini {\em et al.\/}\ (\citeyear{FraBouPer19-AnnCompNeurSciMeet}),
Lindeberg (\citeyear{Lin20-JMIV}),
Lian {\em et  al.\/} (\citeyear{LiaAlmGraKamBurMef21-PLOSCompBiol}),
Oleskiw {\em et al.\/}\ (\citeyear{OleLieSimMov23-bioRxiv}),
Yedjour and Yedjour (\citeyear{YedYed24-CognNeurDyn})
and Nguyen {\em et al.\/}\ (\citeyear{NguSooHuaBak24-PLOSCompBio}).

Notably, in relation to the generalized quadratic models of complex
cells in V1 to be considered in this paper,
Rowekamp and Sharpee (\citeyear{RowSha25-PLOSCompBiol})
have found that quadratic computations strongly increase both the predictive power
of their models of visual neurons in V1, V2 and V4 as well
as their neural selectivity to natural stimuli.

More explicit neural models of spatio-temporal receptive fields in the primary
visual cortex have also been formulated by
Heitmann and Ermentrout (\citeyear{HeiErm16-BIONETICS}),
Chizhov and Merkulyeva (\citeyear{ChiMer20-PLOSCompBio}),
Chariker {\em et al.\/}
(\citeyear{ChaShaHawYou21-PNAS,ChaShaHawYou22-JNeuroSci}),
Freeman (\citeyear{Fre21-JNeuroSci})  
and
Larisch and Hamker (\citeyear{LarHam25-NeurNetw}),
including studies of direction selectivity properties.

More detailed specific modelling of direction selectivity properties
has been performed by Adelson and Bergen (\citeyear{AdeBer85-JOSA}),
Emerson {\em et al.\/} (\citeyear{EmeBerAde92-VR}),
McLean {\em et al.\/} (\citeyear{McLRaaPal94-VisNeuroSci}),
Hamada {\em et al.\/} (\citeyear{HamYamKat97-BICY}),
Wimbauer {\em et al.\/} (\citeyear{WimWenMilHem97-BICY}) and
Chariker {\em et al.\/} (\citeyear{ChaShaHawYou21-PNAS})
with more explicit cortical models by
Peterson {\em et al.\/} (\citeyear{PetLiFre04-JNeuroSci}),
Wenisch {\em et al.\/} (\citeyear{WenNolHem05-BICY}),
Chizhov and Merkulyeva (\citeyear{ChiMer20-PLOSCompBiol}),
Freeman (\citeyear{Fre21-JNeuroSci});
see the introduction in
(Larisch and Hamker \citeyear{LarHam25-NeurNetw})
for a very good more general survey.

Compared to such explicit neural models of
spatio-temporal receptive fields,
a conceptual difference with our approach is that we instead
analyze
compact functional models of the spatio-temporal receptive fields,
which have been determined from an axiomatically determined
normative theory of visual receptive fields.
In this way, we are aiming at more compact descriptions of the
functional phenomena, as parameterized by a small set of shape
parameters of the spatio-temporal receptive fields, in contrast to
instead having to determine a much larger number of filter weights and
other dynamics parameters in the more explicit neural models
and computing the results by numerical simulations.

The orientation selectivity properties of idealized models of simple
and complex cells have been analyzed in detail in
Lindeberg (\citeyear{Lin25-JCompNeurSci-orisel,Lin25-PONE}).
The notions of direction selectivity and speed selectivity,
that we address in this work, are
related to the notion of orientation selectivity, but are structurally
different properties. The notion of orientation selectivity concerns
behaviour over a spatial domain, whereas the notions of direction
and speed selectivity concern behaviour over the joint spatio-temporal domain.

In this work, we theoretically analyze the results of how
the output from idealized models of simple cells and complex cells are
affected by varying both the direction and the speed of motion stimuli.
We also relate these results to direction and speed selectivity
properties of biological neurons, and to a hypothesis that the
receptive fields in the primary visual cortex ought to be covariant
under local Galilean transformations
(Lindeberg \citeyear{Lin21-Heliyon,Lin23-FrontCompNeuroSci,Lin26-JCompNeuroSci}).

\begin{figure*}[hbtp]
  \begin{center}
    \begin{tabular}{cccccc}
      & $v = -1$ & $v = -1/2$ & $v = 0$ & $v = 1/2$ & $v = 1$
      \\
      $\scriptsize{\sigma_x = 2, \sigma_t = 2}$
      & \includegraphics[width=0.12\textwidth]{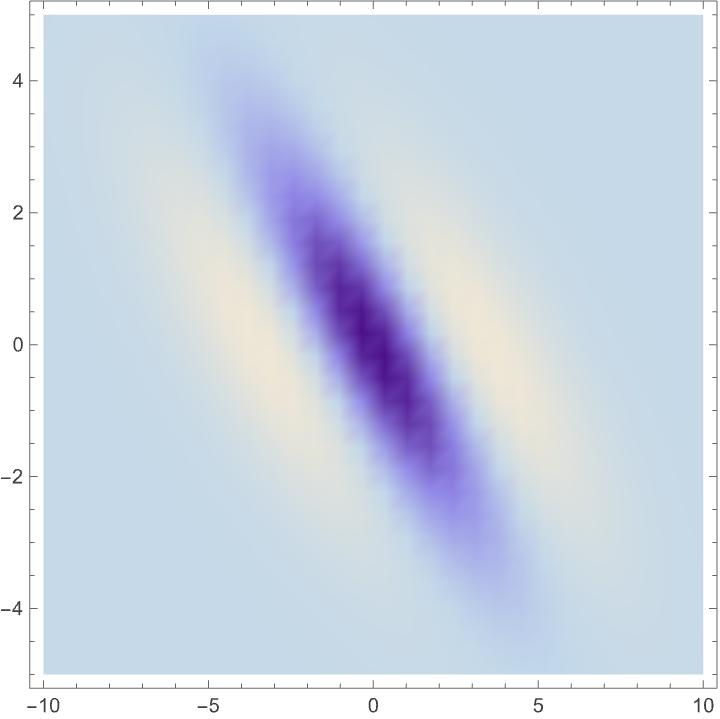}
      & \includegraphics[width=0.12\textwidth]{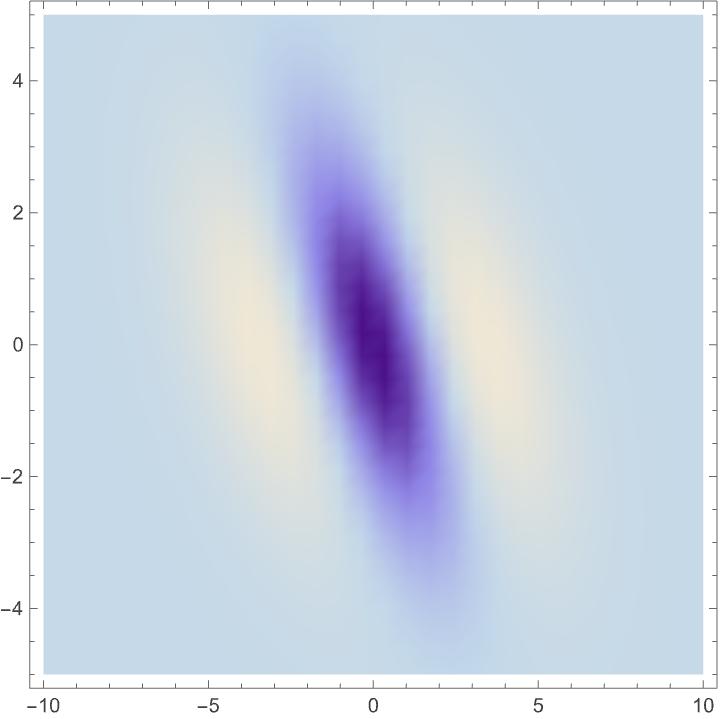}
      & \includegraphics[width=0.12\textwidth]{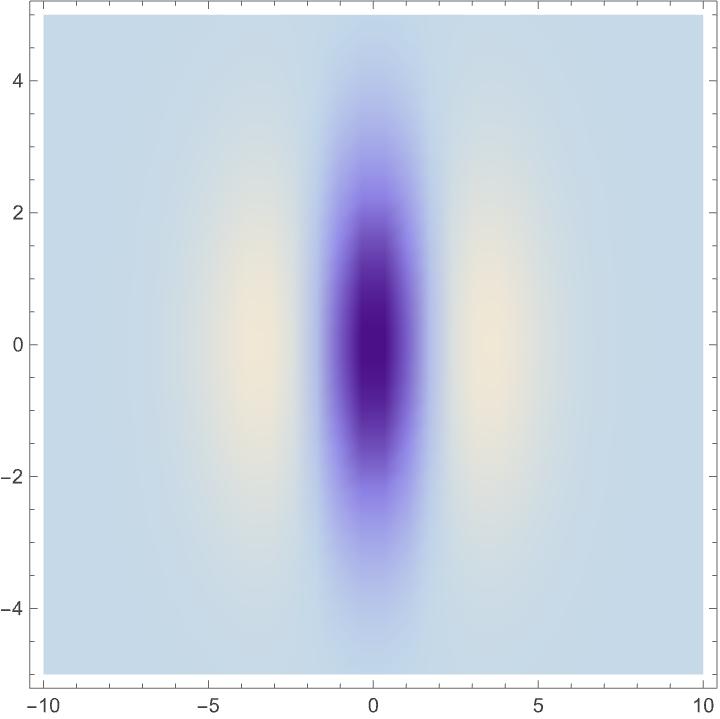}
      & \includegraphics[width=0.12\textwidth]{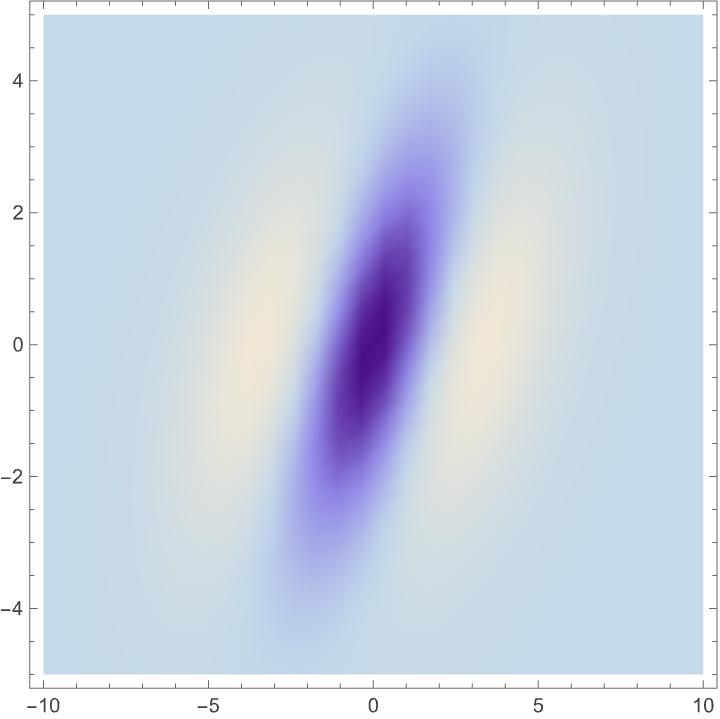}
      & \includegraphics[width=0.12\textwidth]{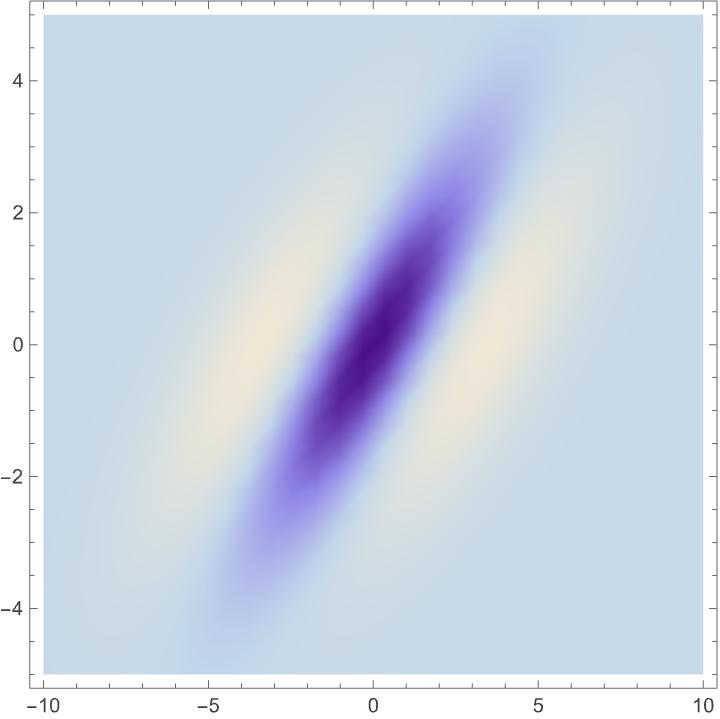}                                          
      \\
      $\scriptsize{\sigma_x = 1, \sigma_t = 2}$
      & \includegraphics[width=0.12\textwidth]{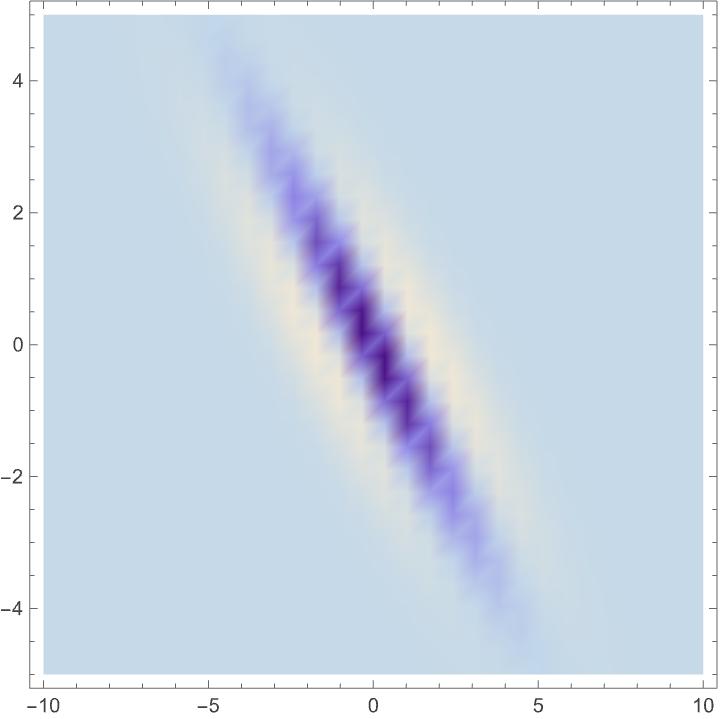}
      & \includegraphics[width=0.12\textwidth]{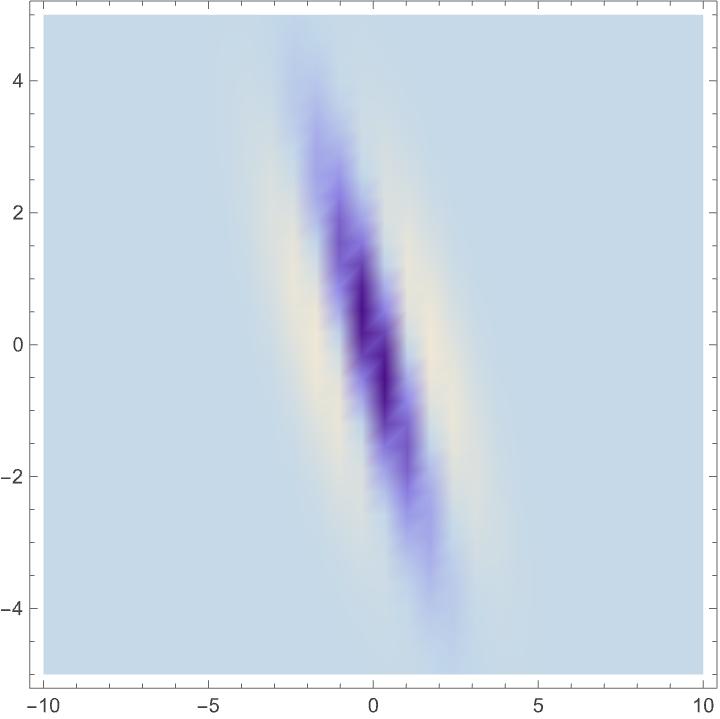}
      & \includegraphics[width=0.12\textwidth]{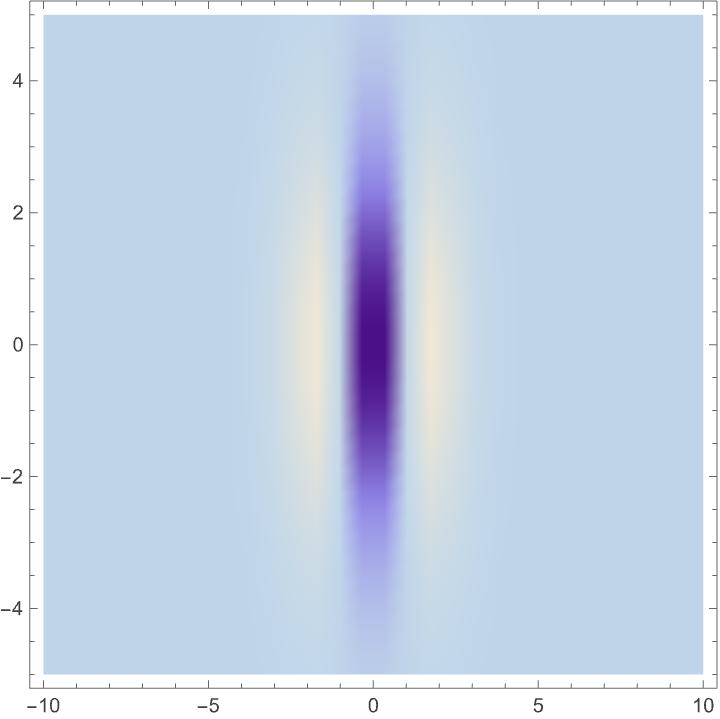}
      & \includegraphics[width=0.12\textwidth]{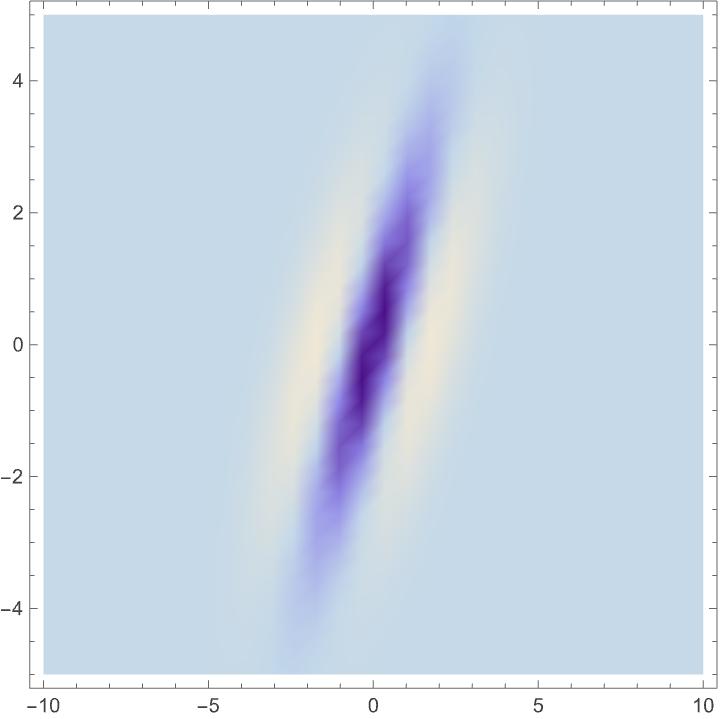}
      & \includegraphics[width=0.12\textwidth]{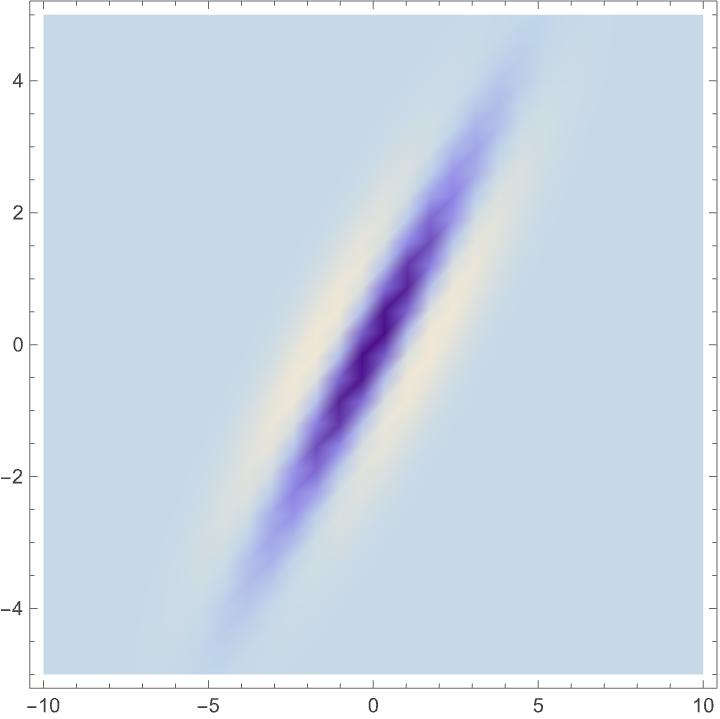}                               
      \\
      $\scriptsize{\sigma_x = 2, \sigma_t = 1}$
      & \includegraphics[width=0.12\textwidth]{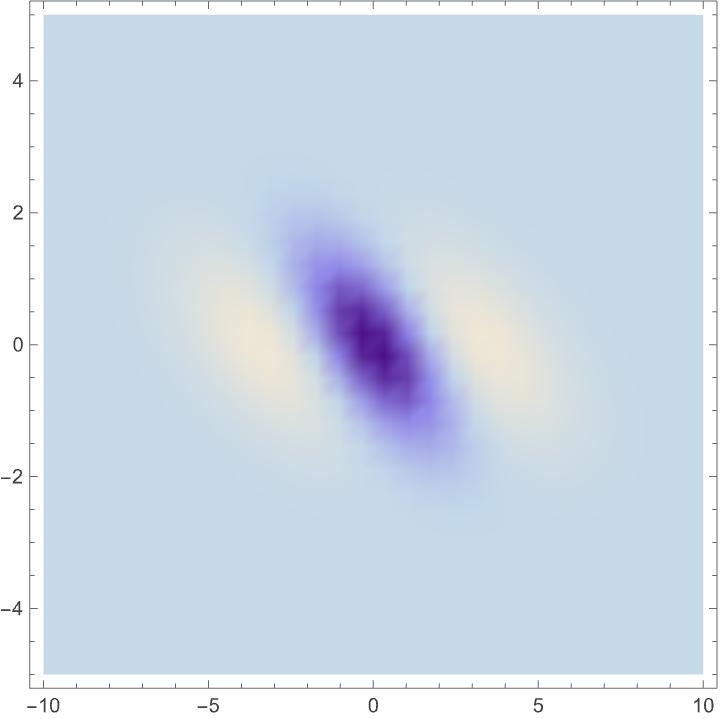}
      & \includegraphics[width=0.12\textwidth]{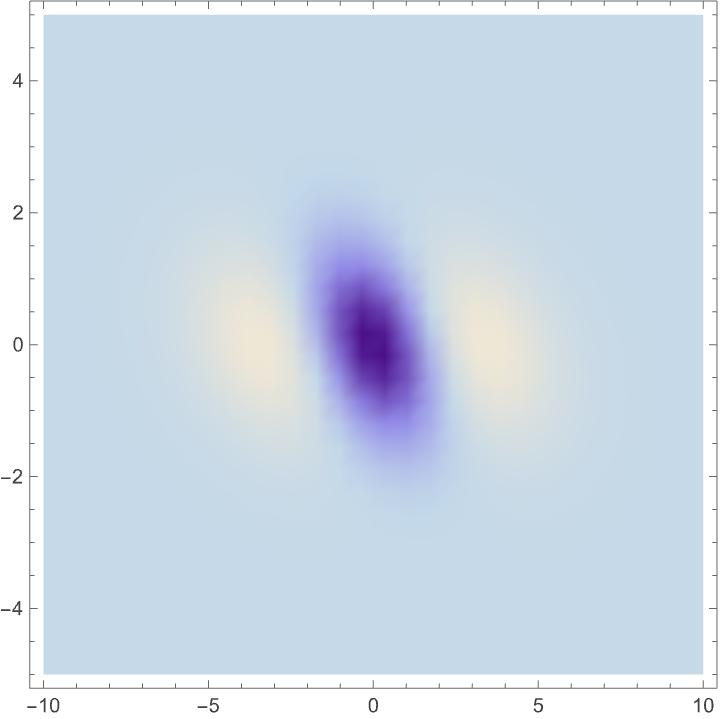}
      & \includegraphics[width=0.12\textwidth]{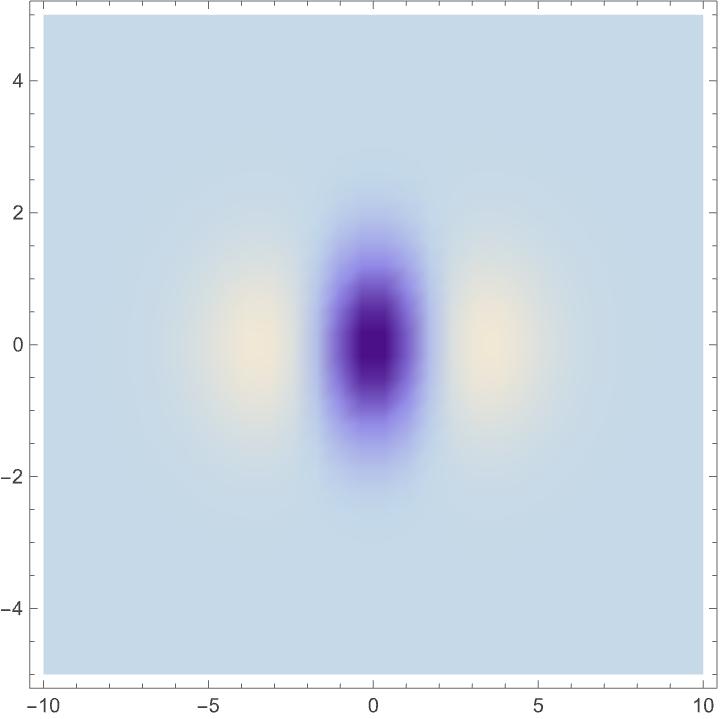}
      & \includegraphics[width=0.12\textwidth]{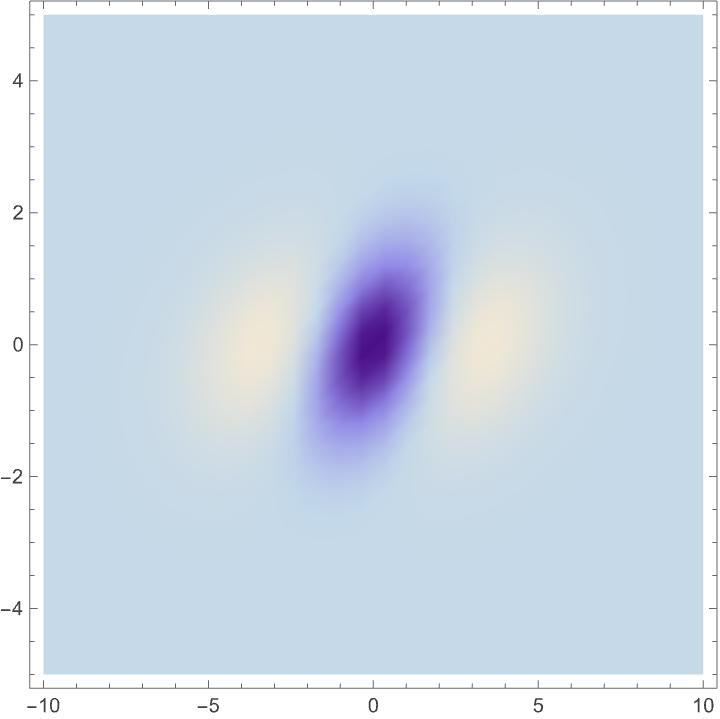}
      & \includegraphics[width=0.12\textwidth]{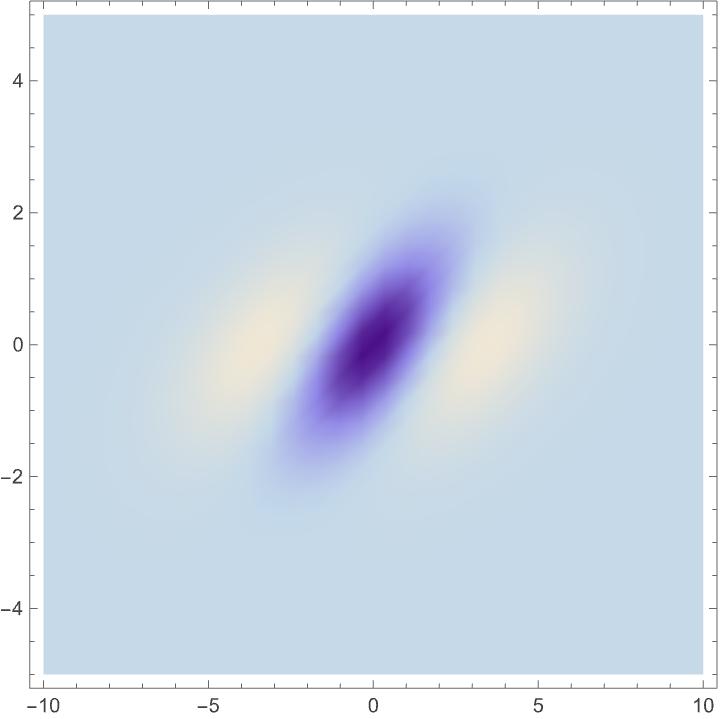}                                          
      \\
      $\scriptsize{\sigma_x = 1, \sigma_t = 1}$
      & \includegraphics[width=0.12\textwidth]{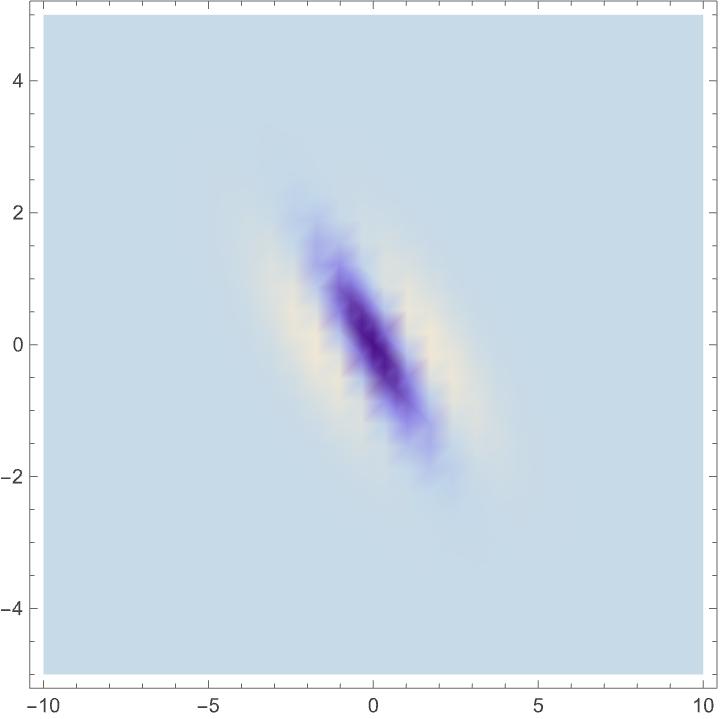}
      & \includegraphics[width=0.12\textwidth]{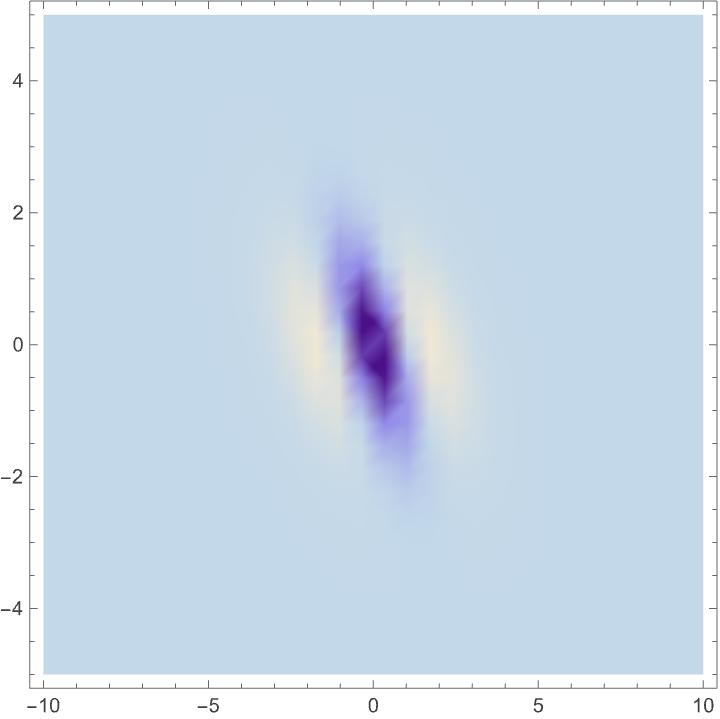}
      & \includegraphics[width=0.12\textwidth]{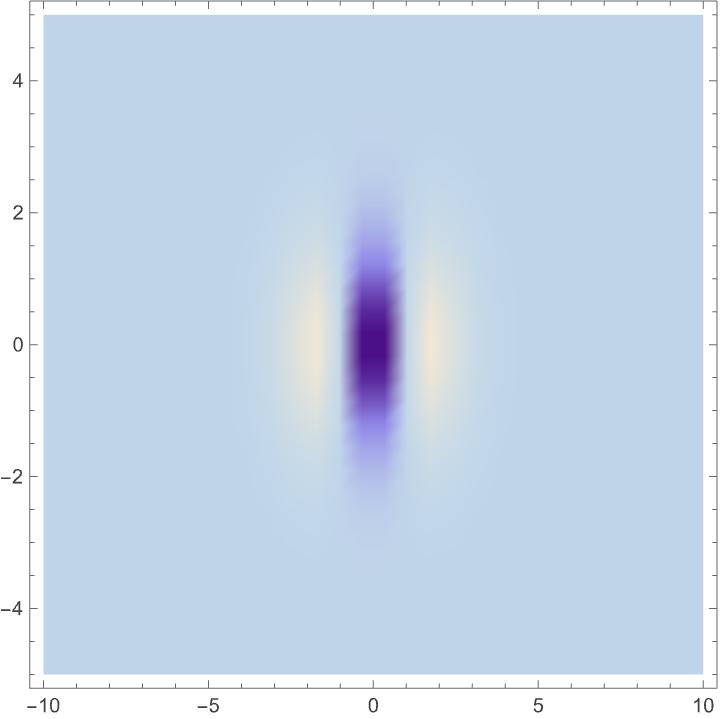}
      & \includegraphics[width=0.12\textwidth]{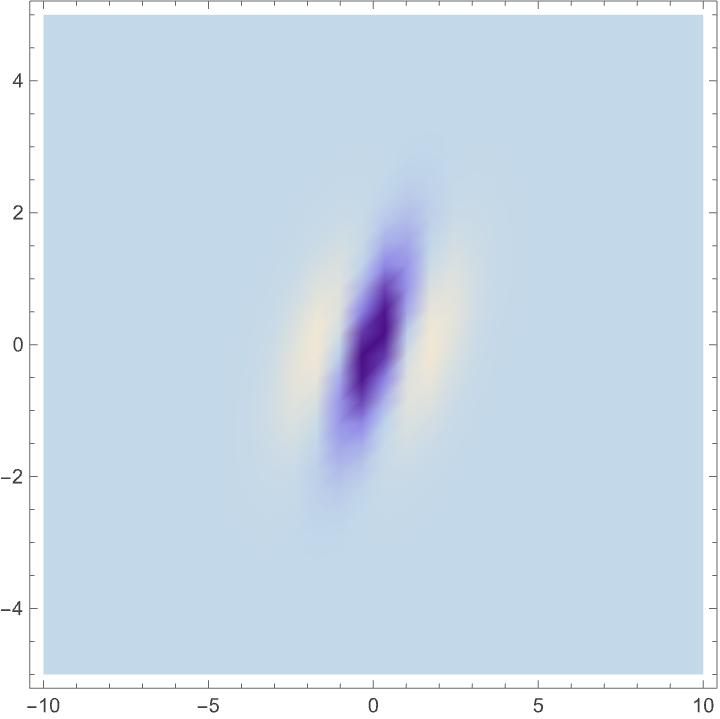}
      & \includegraphics[width=0.12\textwidth]{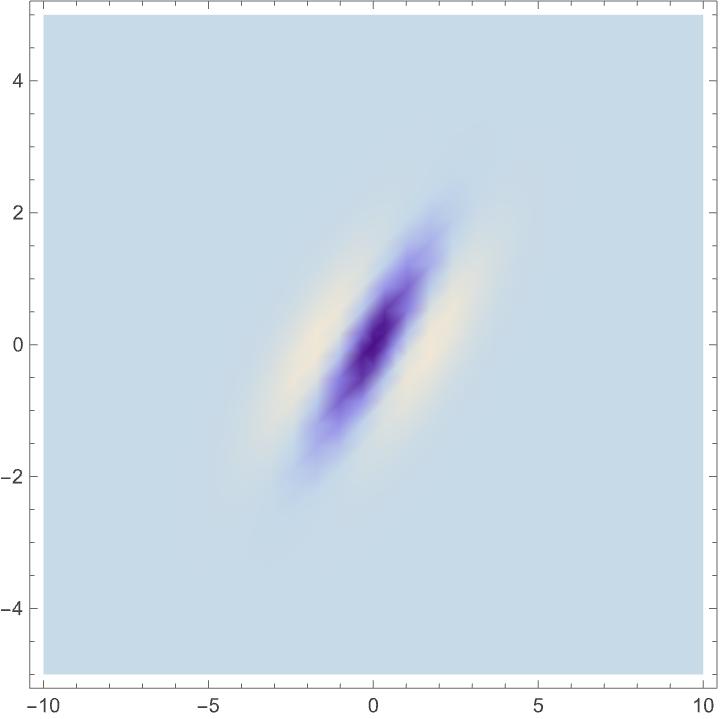}                                          
      \\
    \end{tabular}
  \end{center}
  \caption{Non-causal joint spatio-temporal receptive fields over
    a 1+1D spatio-temporal domain in terms of the second-order 
    spatial derivative of the form $T_{xx,\text{norm}}(x, t;\; s, \tau, v)$
    according to (\ref{eq-spat-temp-RF-model-der-norm-caus}) of the product of a
    velocity-adapted 1-D Gaussian kernel
    over the spatial domain and the non-causal temporal kernel over the
    temporal domain according to (\ref{eq-non-caus-temp-gauss}).
    The spatio-temporal receptive fields are shown for different
    values of the spatial scale parameter $\sigma_x = \sqrt{s}$ and
    the temporal scale parameter $\sigma_t = \sqrt{\tau}$ in
    dimensions of $[\mbox{length}]$ and $[\mbox{time}]$.
    (Horizontal axes: Spatial image coordinate $x \in [-10, 10]$.
  Vertical axes: Temporal variable $t \in [-4, 4]$.)}
  \label{fig-noncaus-strfs}

  \bigskip
  
  \begin{center}
    \begin{tabular}{cccccc}
      & $v = -1$ & $v = -1/2$ & $v = 0$ & $v = 1/2$ & $v = 1$
      \\
      $\scriptsize{\sigma_x = 2, \sigma_t = 2}$
      & \includegraphics[width=0.12\textwidth]{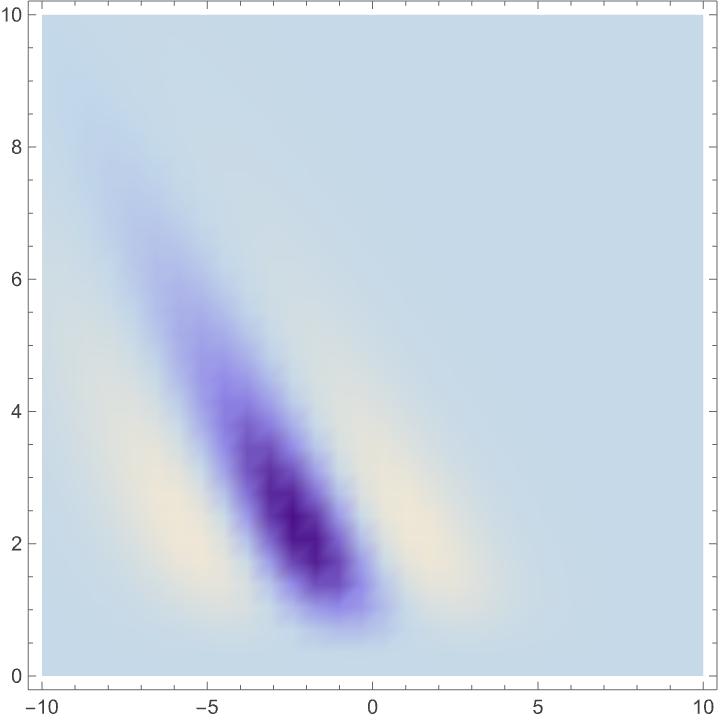}
      & \includegraphics[width=0.12\textwidth]{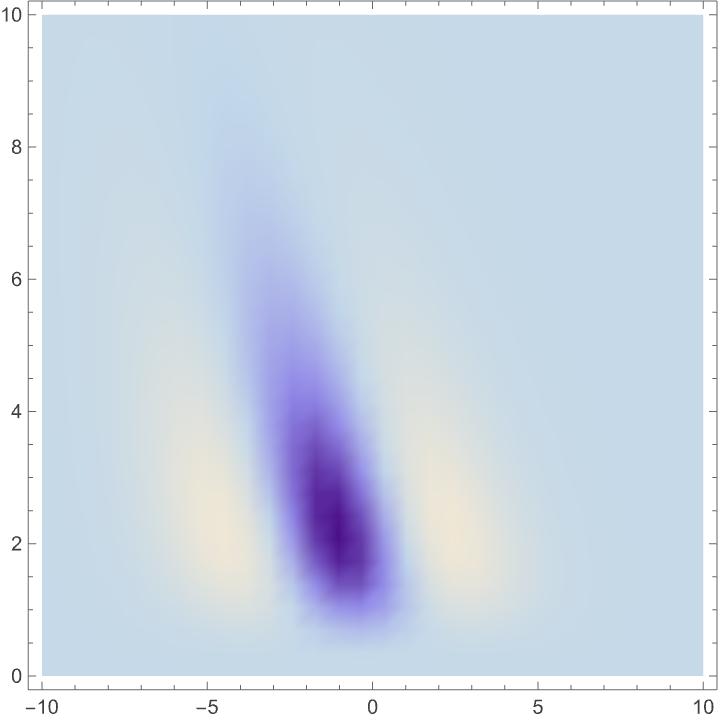}
      & \includegraphics[width=0.12\textwidth]{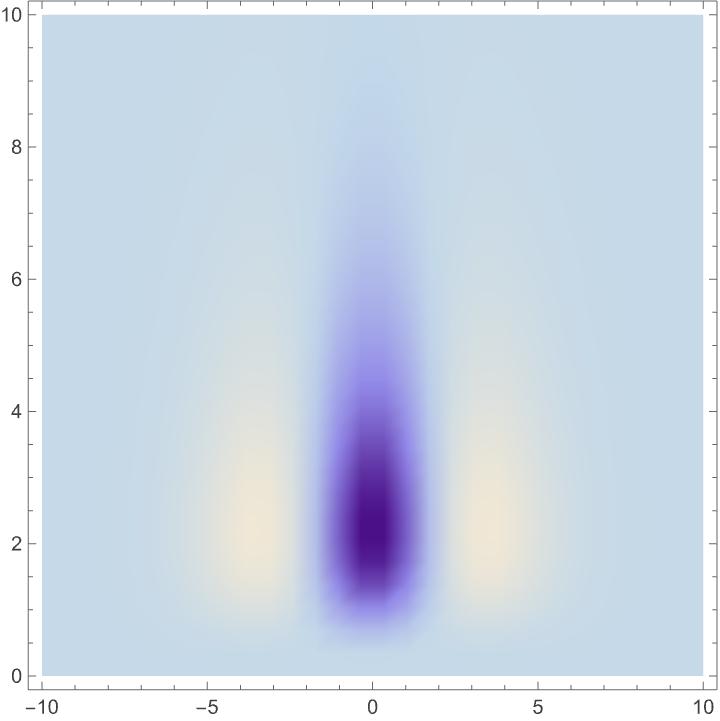}
      & \includegraphics[width=0.12\textwidth]{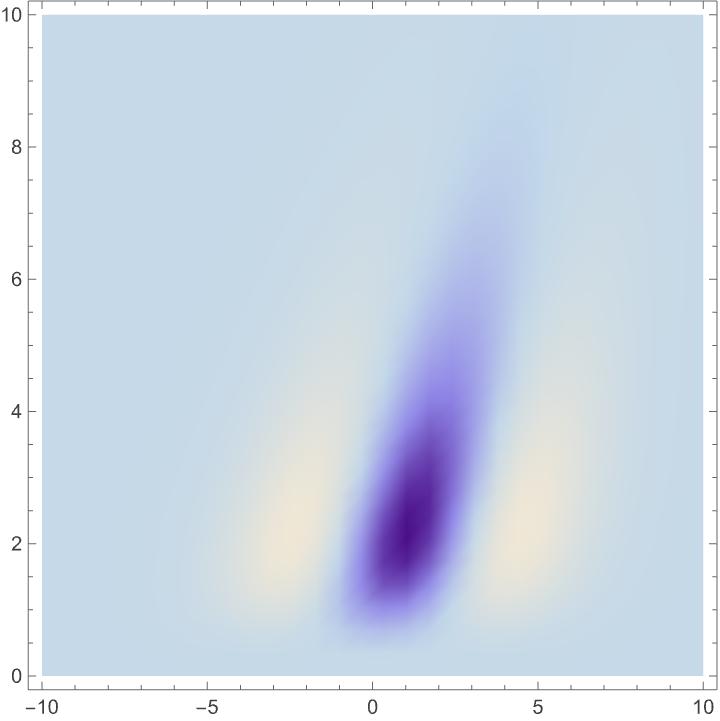}
      & \includegraphics[width=0.12\textwidth]{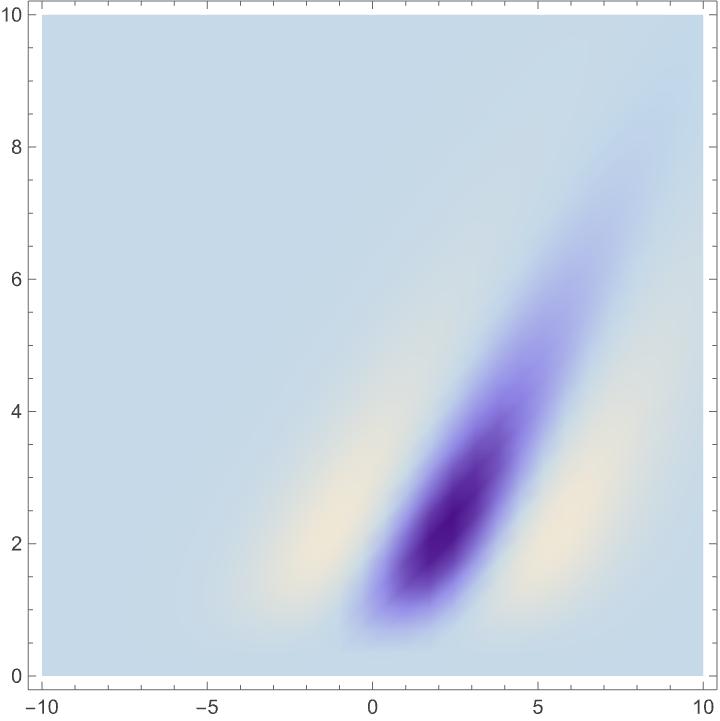}                                          
      \\
      $\scriptsize{\sigma_x = 1, \sigma_t = 2}$
      & \includegraphics[width=0.12\textwidth]{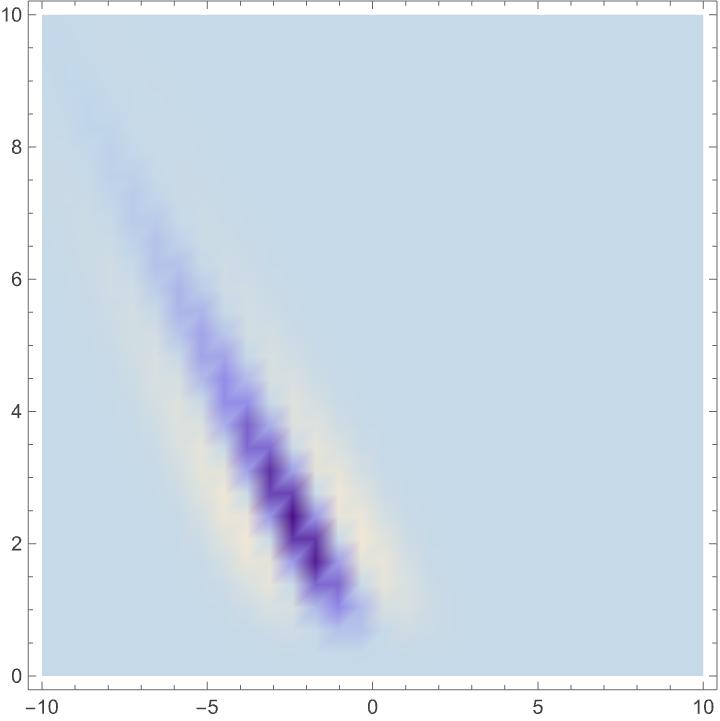}
      & \includegraphics[width=0.12\textwidth]{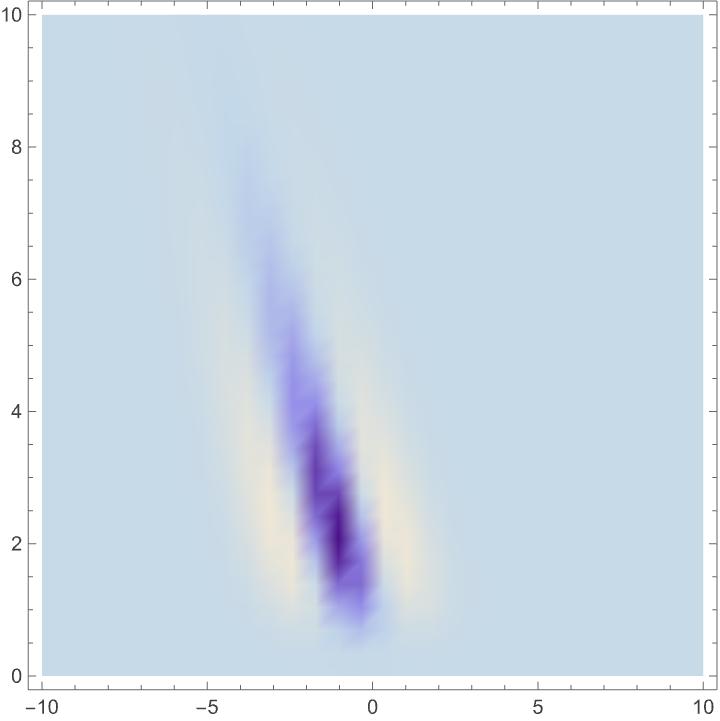}
      & \includegraphics[width=0.12\textwidth]{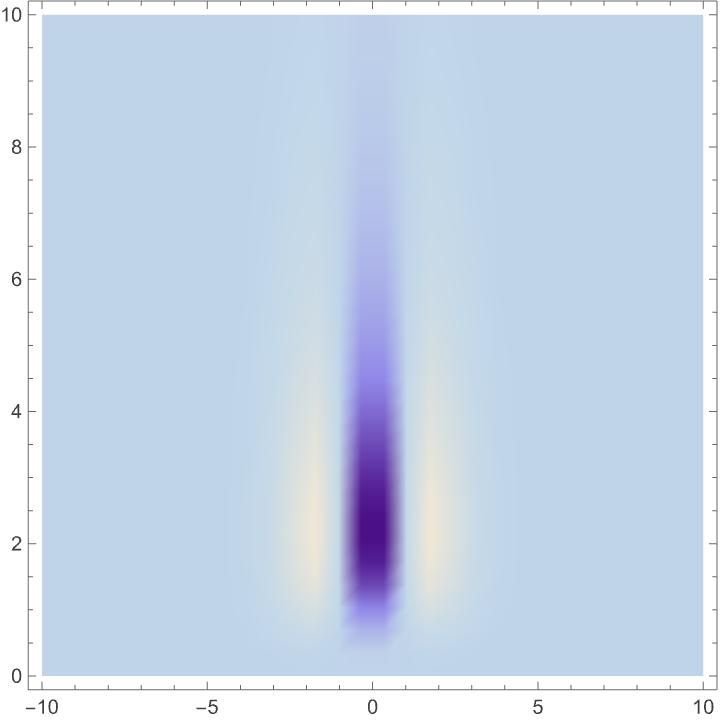}
      & \includegraphics[width=0.12\textwidth]{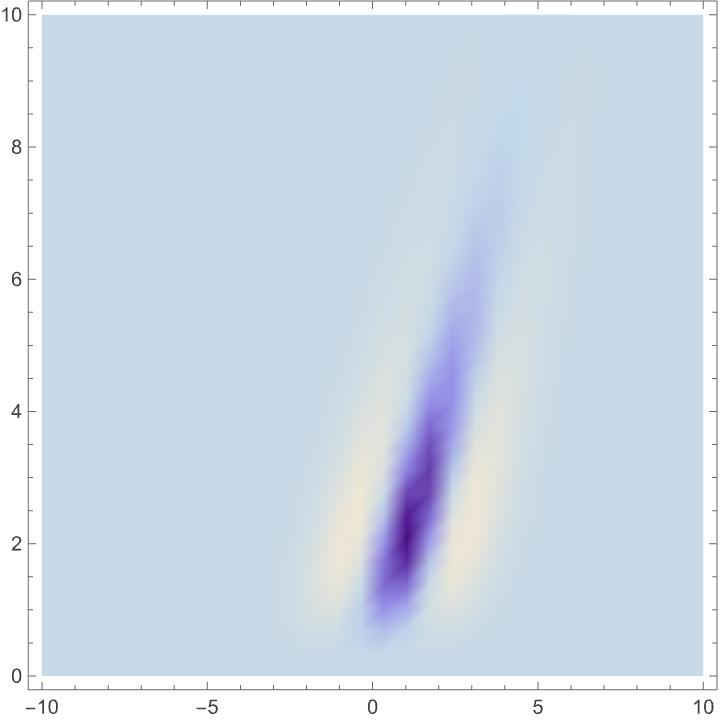}
      & \includegraphics[width=0.12\textwidth]{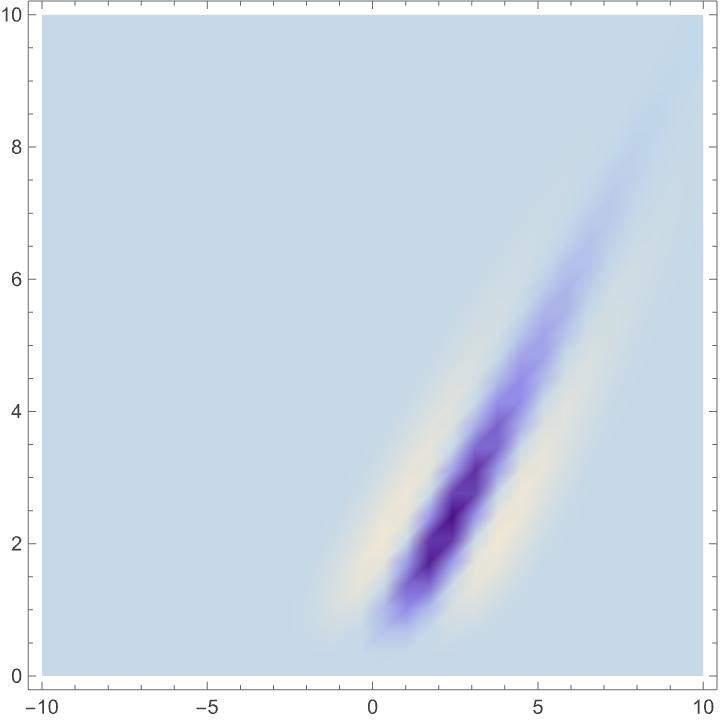}                               
      \\
      $\scriptsize{\sigma_x = 2, \sigma_t = 1}$
      & \includegraphics[width=0.12\textwidth]{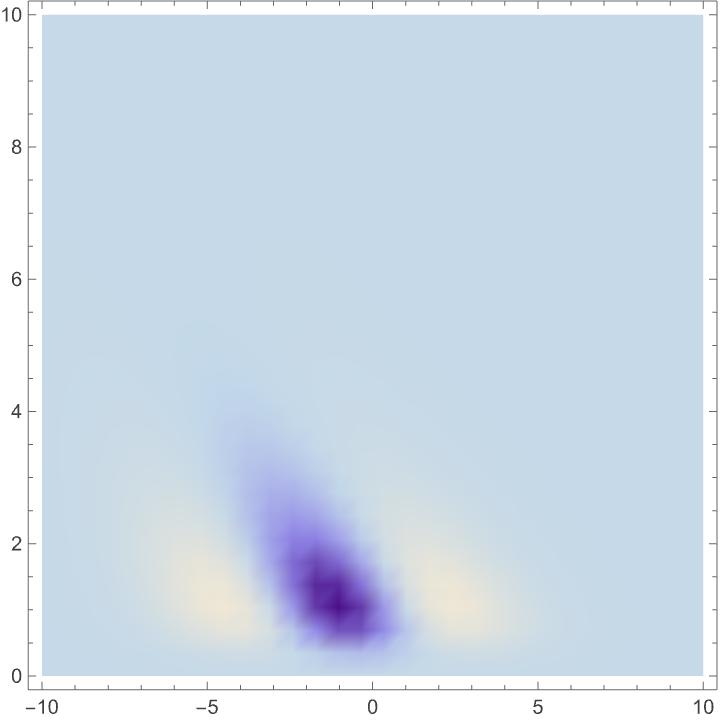}
      & \includegraphics[width=0.12\textwidth]{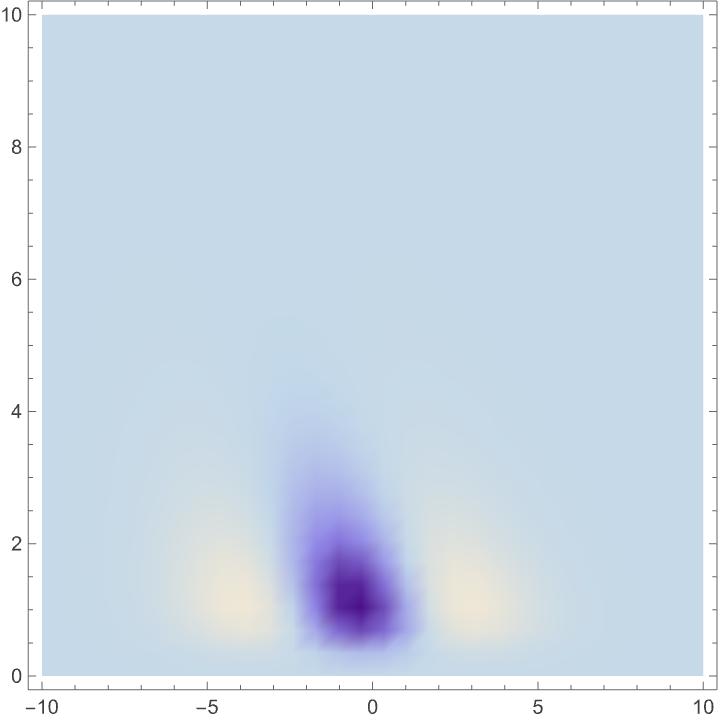}
      & \includegraphics[width=0.12\textwidth]{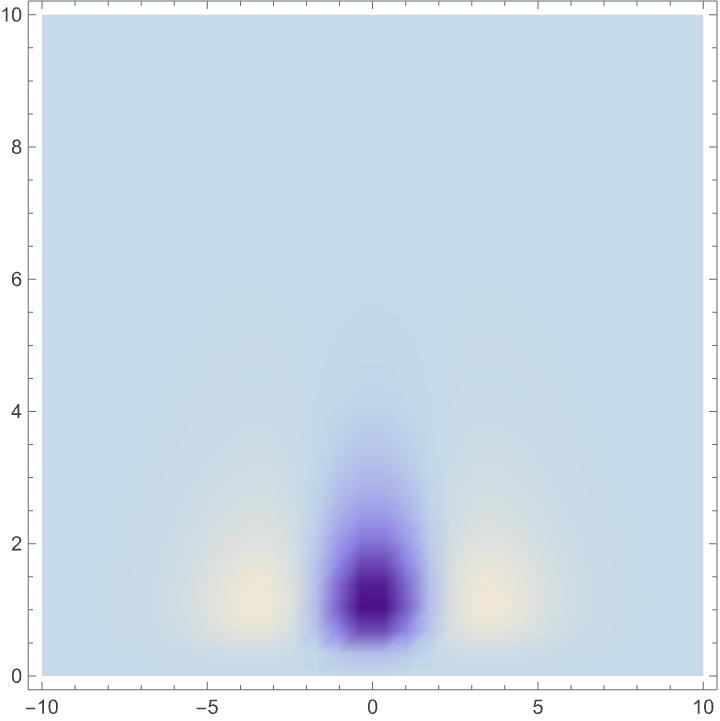}
      & \includegraphics[width=0.12\textwidth]{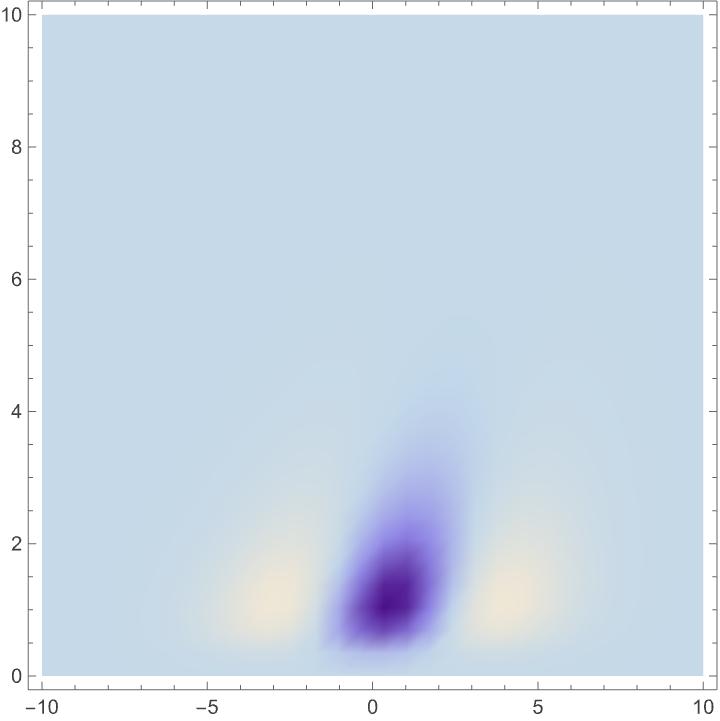}
      & \includegraphics[width=0.12\textwidth]{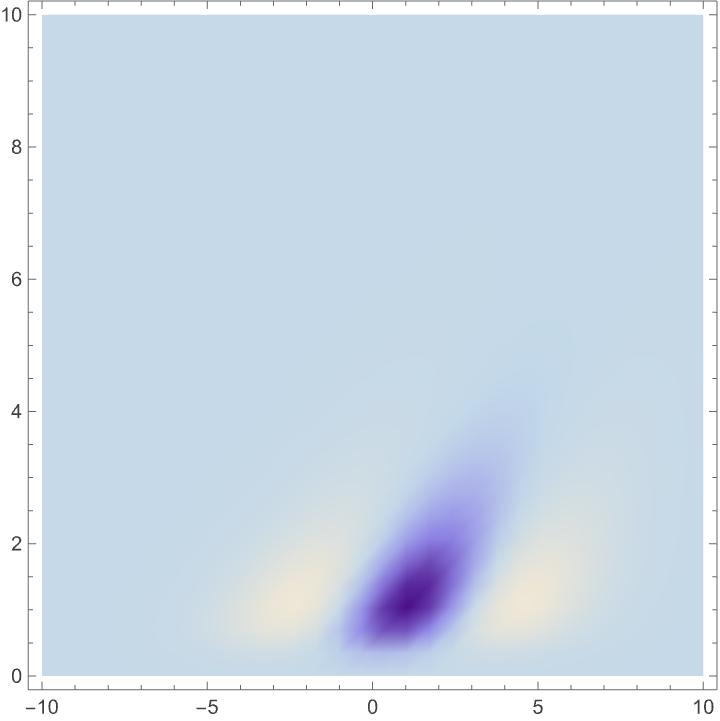}                                          
      \\
      $\scriptsize{\sigma_x = 1, \sigma_t = 1}$
      & \includegraphics[width=0.12\textwidth]{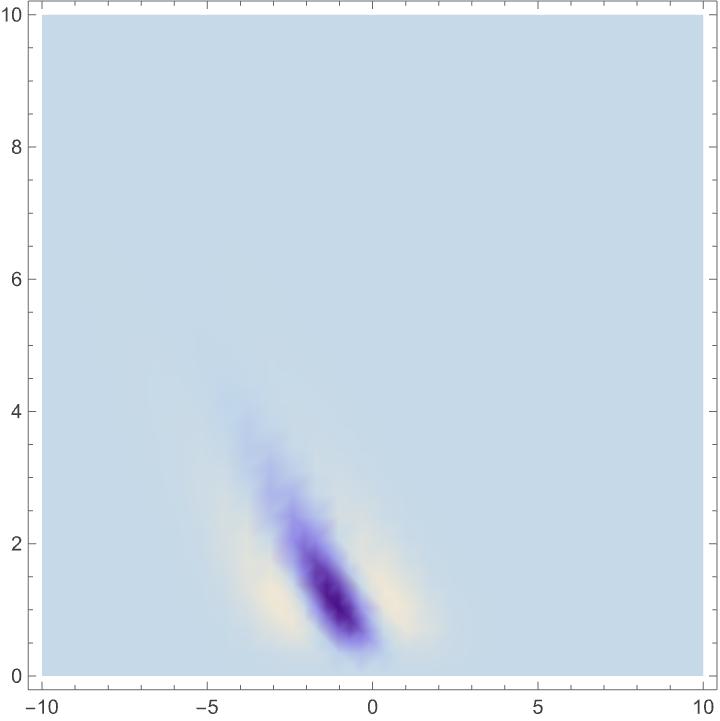}
      & \includegraphics[width=0.12\textwidth]{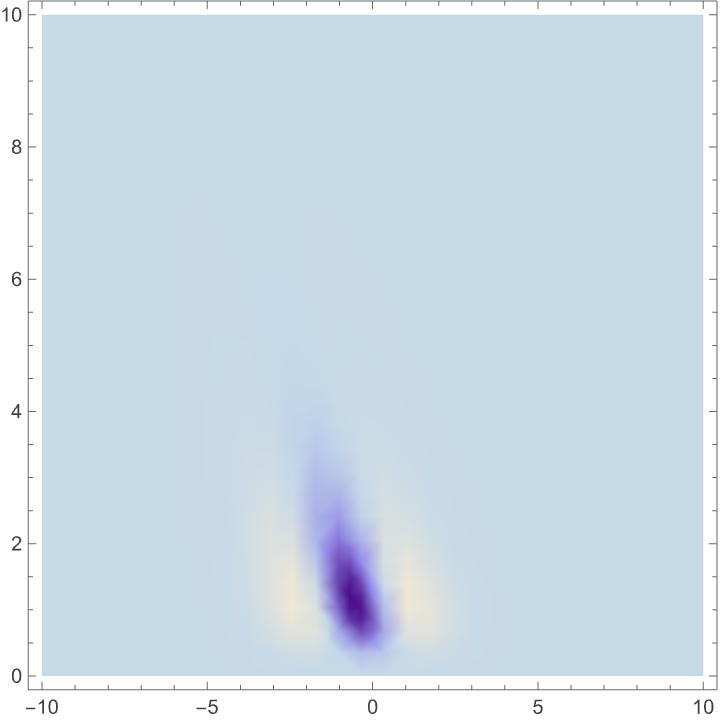}
      & \includegraphics[width=0.12\textwidth]{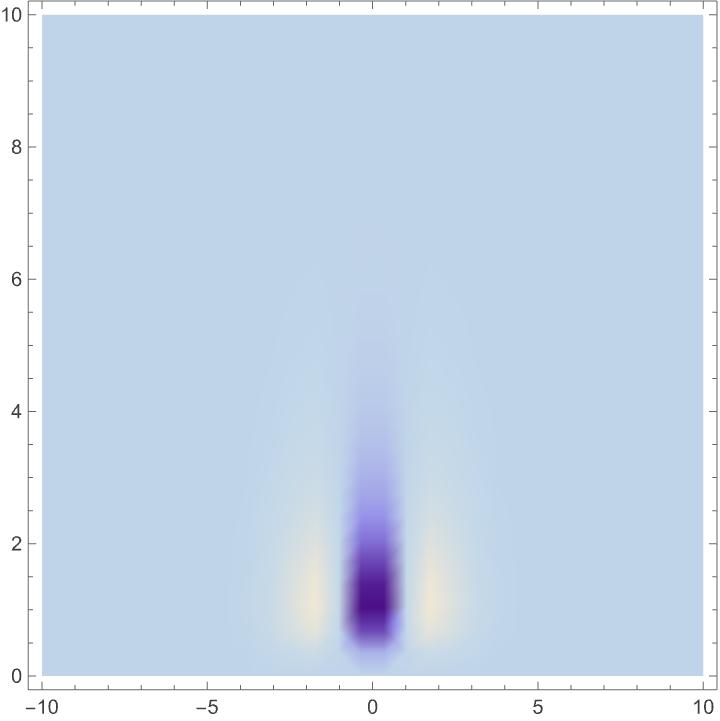}
      & \includegraphics[width=0.12\textwidth]{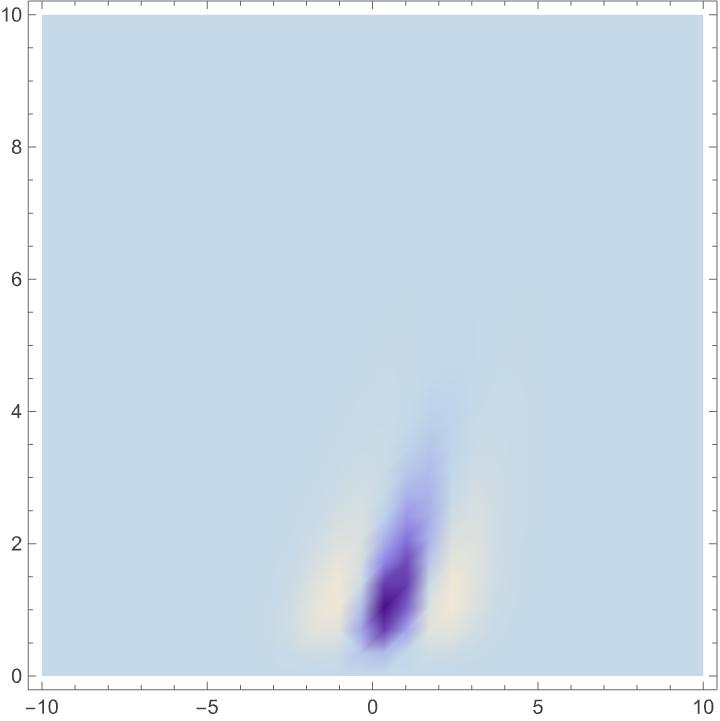}
      & \includegraphics[width=0.12\textwidth]{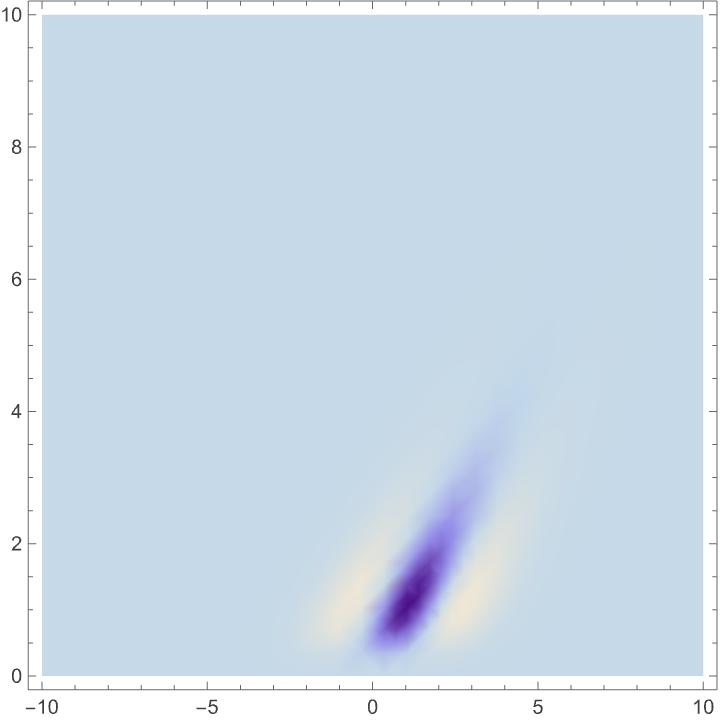}                                          
      \\
    \end{tabular}
  \end{center}
  \caption{Time-causal joint spatio-temporal receptive fields over
    a 1+1D spatio-temporal domain in terms of the second-order
    spatial derivative of the form $T_{xx,\text{norm}}(x, t;\; s, \tau, v)$
    according to (\ref{eq-spat-temp-RF-model-der-norm-caus}) of the product of a
    velocity-adapted 1-D Gaussian kernel
    over the spatial domain and the time-causal limit kernel over the
    temporal domain according to (\ref{eq-time-caus-lim-kern}).
    The spatio-temporal receptive fields are shown for different
    values of the spatial scale parameter $\sigma_x = \sqrt{s}$ and
    the temporal scale parameter $\sigma_t = \sqrt{\tau}$ in
    dimensions of $[\mbox{length}]$ and $[\mbox{time}]$.
    (Horizontal axes: Spatial image coordinate $x \in [-10, 10]$.
  Vertical axes: Temporal variable $t \in [0, 8]$.)}
  \label{fig-timecaus-strfs}
\end{figure*}

The latter will then imply a
variability of the shapes of the spatio-temporal receptive fields
under the degrees of freedom of local Galilean transformations.
In practice, this would
imply that the there would be multiple copies of similar types of
spatio-temporal receptive fields for different values of the velocity
parameter $v = (v_1, v_2)^T$ in the idealized model
(\ref{eq-spat-temp-RF-model-der-norm-caus}) of simple cells, as illustrated in
Figures~\ref{fig-noncaus-strfs} and~\ref{fig-timecaus-strfs}.

As will be described further in Section~\ref{sec-rels-biol-vision},
where we relate properties of our theoretically derived models to
results from previous experiments regarding biological vision,
the neurophysiological measurements that we relate to are consistent
with such an expansion of the receptive field shapes over the
parameters of local Galilean motion, thus consistent with treating the
primary visual cortex of higher mammals as providing a Galilean
covariant representation of spatio-temporal receptive field responses.

\section{The generalized Gaussian derivative model for visual
  receptive fields}
\label{sec-gen-gauss-der-strfs}

In this section, we will describe the idealized models of simple cells
and complex cells, whose direction and speed selectivity properties
will then be analyzed in
Sections~\ref{sec-dir-sel-simpl-cells} and~\ref{sec-dir-sel-compl-cells}.

\subsection{Idealized model for simple cells}
\label{sec-model-vel-tuned-simple-cells}

According to the generalized Gaussian derivative model for visual
receptive fields, linear spatio-temporal receptive fields
corresponding to the simple cells in the primary visual cortex are
modelled by spatio-temporal derivatives of velocity-adapted%
\footnote{Concerning the terminology ``velocity-adapted'' used for
  affine Gaussian kernels, temporal derivatives and spatio-temporal receptive fields
  in this paper as well as in earlier work,
  please do not confuse this notion with the terminology ``motion adaptation''
  in vision science, in the sense of the loss of sensitivity that arises from
  extended viewing of an image moving in a fixed direction.
  Here, instead the notion of ``velocity-adapted'' means that a
  receptive field is adapted to a particular Galilean motion direction
  in joint space-time or that a temporal derivative is computed along
  a particular velocity direction in joint space-time.}
affine Gaussian kernels in the following idealized way
(Lindeberg \citeyear{Lin10-JMIV,Lin13-BICY,Lin16-JMIV,Lin21-Heliyon}):
\begin{align}
  \begin{split}
    \label{eq-spat-temp-RF-model-der-norm-caus}
    T_{\text{simple}}(x_1, x_2, t;\; \sigma_{\varphi}, \sigma_t, \varphi, v, \Sigma_{\varphi}, m, n) 
  \end{split}\nonumber\\
  \begin{split}
   & = T_{{\varphi}^m, {\bar t}^n,\text{norm}}(x_1, x_2, t;\; \sigma_{\varphi}, \sigma_t, v, \Sigma_{\varphi})
  \end{split}\nonumber\\
  \begin{split}
   &  = \sigma_{\varphi}^{m} \, 
          \sigma_t^{n} \, 
  %\end{split}\nonumber\\
  %\begin{split}
  %    \phantom{=} \,\,\,\,
         \partial_{\varphi}^{m} \,\partial_{\bar t}^n 
          \left( g(x_1 - v_1 t, x_2 - v_2 t;\; \Sigma_{\varphi}) \,
           h(t;\; \sigma_t) \right),
  \end{split}
\end{align}
where
\begin{itemize}
\item
  $\varphi \in [-\pi, \pi]$ is the preferred direction
  of the receptive field, which we here assume is also aligned
  with the orientation of the spatial component of the receptive field,
\item
  $\sigma_{\varphi} \in \bbbr_+$ is the amount of spatial smoothing
  (in units of the spatial standard deviation),
\item
  $\partial_{\varphi}^m =
  (\cos \varphi \, \partial_{x_1} + \sin  \varphi \, \partial_{x_2})^m$
  is an $m$th-order directional derivative operator
   in the direction $\varphi$,
 \item
   $\Sigma_{\varphi}$ is a $2 \times 2$ symmetric positive definite covariance matrix, with
   one of its eigenvectors in the direction of $\varphi$, 
 \item
   $g(x;\; \Sigma_{\varphi})$ is a 2-D affine Gaussian kernel, with its shape
   determined by the spatial covariance matrix $\Sigma_{\varphi}$
   \begin{equation}
     \label{eq-2D-aff-gauss}
     g(x;\; \Sigma_{\varphi})
     = \frac{1}{2 \pi \sqrt{\det \Sigma_{\varphi}}}
         e^{-x^T \Sigma_{\varphi}^{-1} x/2}
    \end{equation}
    for $x = (x_1, x_2)^T \in \bbbr^2$.
\item
  $\sigma_t$ represents the amount of temporal smoothing (in units of
  the temporal standard deviation),
\item
  $v = (v_1, v_2)^T$ represents a local motion vector in the
  direction $\varphi$ of the spatial direction of the receptive field,
\item
  $\partial_{\bar t}^n = (\partial_t + v_1 \, \partial_{x_1} + v_2 \, \partial_{x_2})^n$
  represents an $n$th-order velocity-adapted temporal derivative
  operator, and
\item
  $h(t;\; \sigma_t)$ represents a temporal smoothing kernel with temporal
  standard deviation $\sigma_t$.
\end{itemize}
In the case when the temporal domain is regarded as non-causal
(implying that the future relative to any temporal moment can be
accessed, as it can be for pre-recorded video data),
the temporal kernel can be chosen as the 1-D Gaussian kernel
\begin{equation}
  \label{eq-non-caus-temp-gauss}
  h(t;\; \sigma_t) = \frac{1}{\sqrt{2 \pi} \sigma_t} \, e^{-t^2/2\sigma_t^2},
\end{equation}
whereas in the case when the temporal is truly time-causal
(corresponding to the more realistic real-time scenario, where the
future cannot be accessed),
the temporal kernel can determined as the time-causal limit kernel
(Lindeberg \citeyear{Lin16-JMIV} Section~5;
Lindeberg \citeyear{Lin23-BICY} Section~3)
\begin{equation}
  \label{eq-time-caus-lim-kern}
  h(t;\; \sigma_t) = \psi(t;\; \sigma_t, c),
\end{equation}
characterized by having a Fourier transform of the form
\begin{equation}
  \label{eq-FT-comp-kern-log-distr-limit}
     \hat{\Psi}(\omega;\; \sigma_t, c) 
     = \prod_{k=1}^{\infty} \frac{1}{1 + i \, c^{-k} \sqrt{c^2-1} \, \sigma_t \, \omega}.
\end{equation}
Figure~\ref{fig-noncaus-strfs} shows examples of such receptive fields
over a non-causal 1+1-D spatio-temporal domain, for the case when the
spatial differentiation order is $m = 2$ and the temporal
differentiation order is $n = 0$.
Figure~\ref{fig-timecaus-strfs} shows corresponding examples of such receptive fields
over a time-causal 1+1-D spatio-temporal domain, for the same orders
of spatial and temporal differentiation.

Notably, these idealized models for spatio-temporal receptive fields
are scale covariant, implying that according to the most idealized
version of the theory there should be multiple rescaled copies of
the spatio-temporal receptive fields over variabilities in their
spatial and scale parameters. As described in more detail in
Lindeberg (\citeyear{Lin26-JMIV}), due to cascade smoothing
properties over the spatial and scale parameters, the receptive
field representations at coarser spatial and temporal scales
can, however, also be computed by applying spatial and
temporal smoothing operations in receptive field representations
at finer spatial and temporal scales. Thereby, a vision system
could also choose to only explicitly compute the receptive field
representations in at the finest spatial and temporal scales in
the sensorium, corresponding to the primary visual cortex,
and then compute visual representations at coarser spatial
and temporal scales in an implicit manner at higher levels
in the visual hierarchy.

\begin{figure*}[hbtp]
  \begin{center}
    \begin{subfigure}{0.99\textwidth}
      \begin{tabular}{cccc}
        $T_{\varphi}(x_1, x_2;\; \sigma_{\varphi}, \sigma_{\orth})$
        & $T_{\varphi\varphi}(x_1, x_2;\; \sigma_{\varphi}, \sigma_{\orth})$
        & $T_{\varphi\varphi\varphi}(x_1, x_2;\; \sigma_{\varphi}, \sigma_{\orth})$
        & $T_{\varphi\varphi\varphi\varphi}(x_1, x_2;\;  \sigma_{\varphi}, \sigma_{\orth})$ \\
        \includegraphics[width=0.22\textwidth]{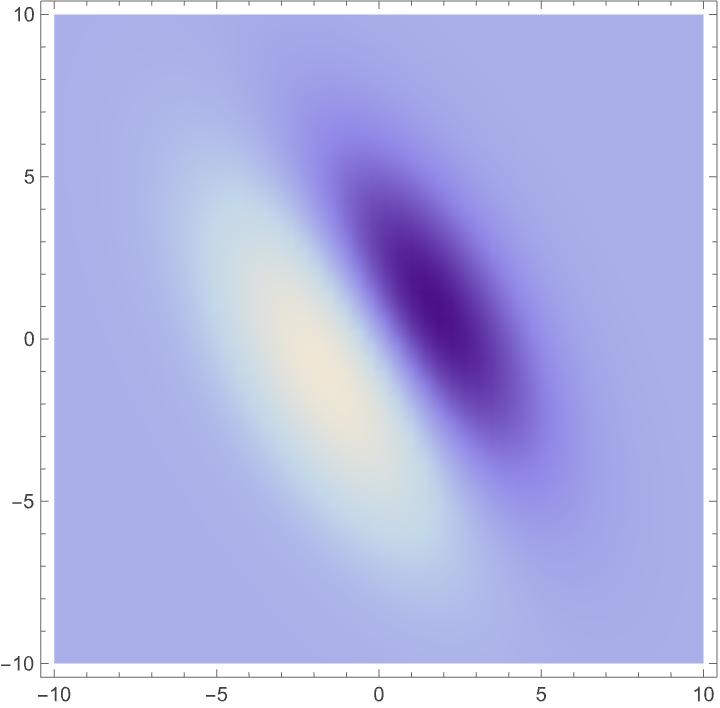}
        & \includegraphics[width=0.22\textwidth]{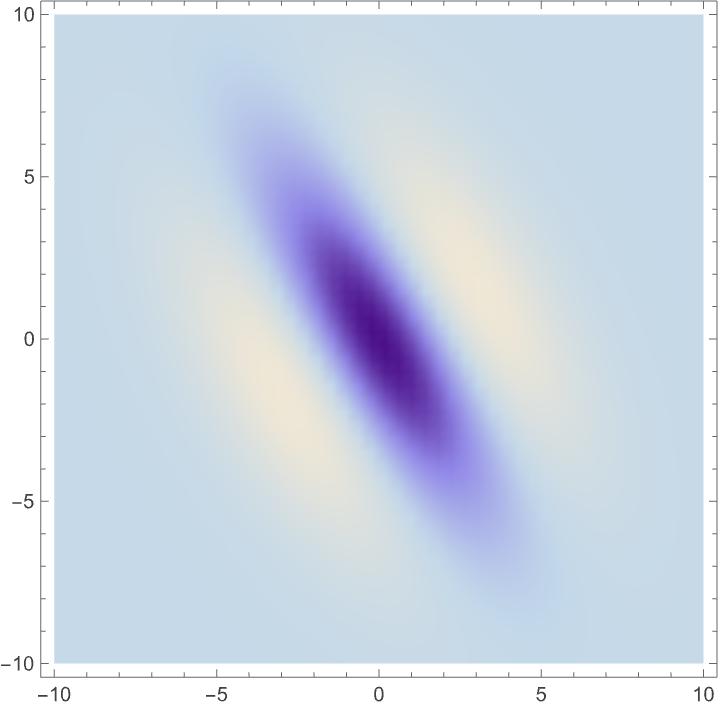}
        & \includegraphics[width=0.22\textwidth]{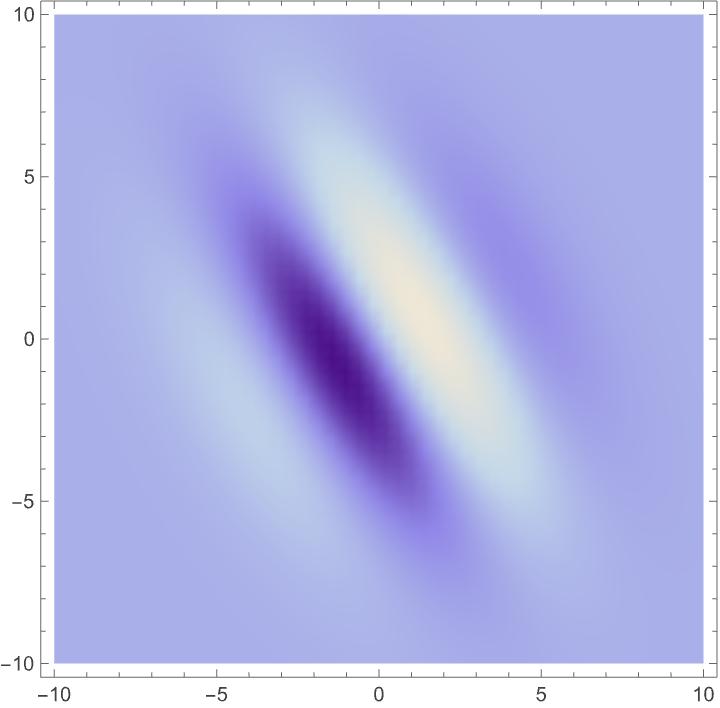}
        & \includegraphics[width=0.22\textwidth]{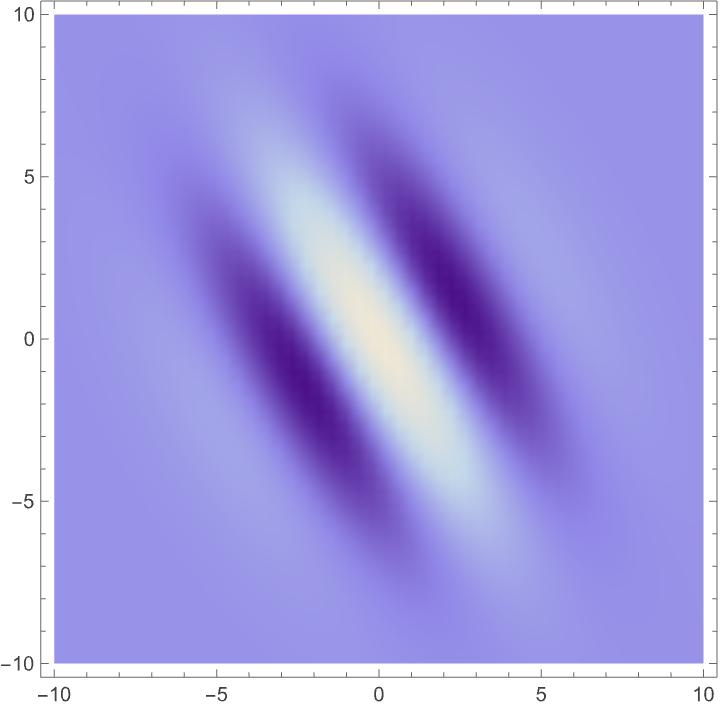}
      \end{tabular}
      \caption{2-D spatial cross-sections of the of the form
        $T_{\varphi^m}(x_1, x_2;\; \sigma_{\varphi}, \sigma_{\orth})$
        of the 2+1-D spatio-temporal receptive field model
        (\ref{eq-spat-temp-RF-model-der-norm-caus}),
        with $\sigma_{\varphi}$ and $\sigma_{\orth}$ corresponding to
        the amount of smoothing in the spatial covariance matrix
        $\Sigma$ along the preferred and the orthogonal directions of
        the receptive field, here for $\varphi = \pi/6$,
        $\sigma_{\varphi}=2$  and $\sigma_{\orth}=4$.
        (Horizontal axes: spatial coordinate $x_1 \in [-10, 10]$.
        Vertical axes: spatial coordinate $x_2 \in [-10, 10]$.)}
      
    \end{subfigure}

    \medskip
    
    \begin{subfigure}{0.99\textwidth}
      \begin{tabular}{cccc}
        {\footnotesize $T_{\varphi}(x_{\varphi}, t;\; \sigma_{\varphi}, \sigma_{t}, v)$}
        & {\footnotesize $T_{\varphi\varphi}(x_{\varphi}, t;\; \sigma_{\varphi}, \sigma_{t}, v)$}
        & {\footnotesize $T_{\varphi\varphi\varphi}(x_{\varphi}, t;\; \sigma_{\varphi}, \sigma_{t}, v)$}
        & {\footnotesize $T_{\varphi\varphi\varphi\varphi}(x_{\varphi}, t;\; \sigma_{\varphi}, \sigma_{t}, v)$} \\
        \includegraphics[width=0.22\textwidth]{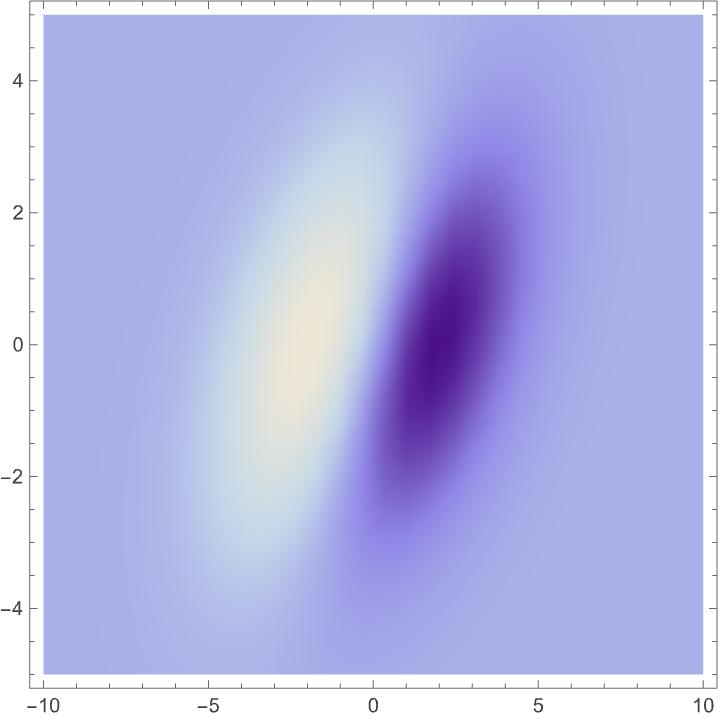}
        & \includegraphics[width=0.22\textwidth]{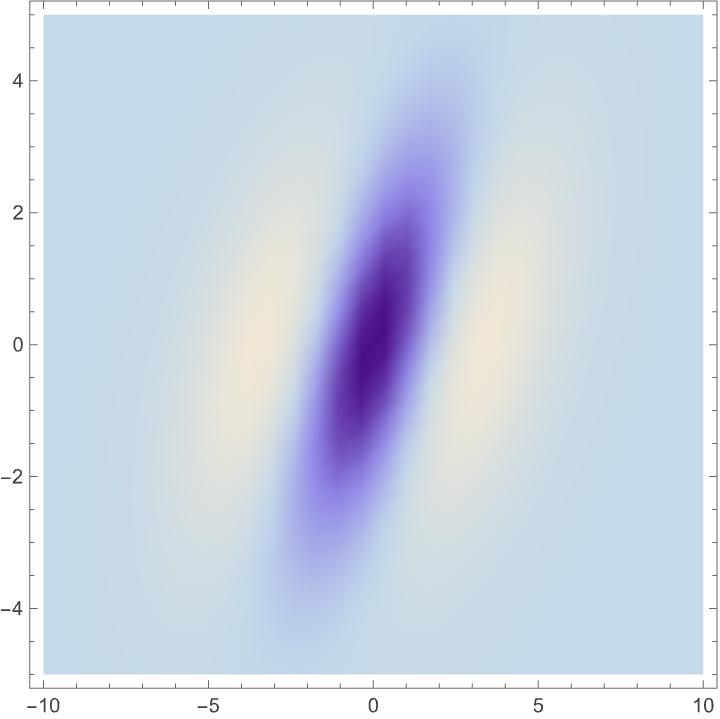}
        & \includegraphics[width=0.22\textwidth]{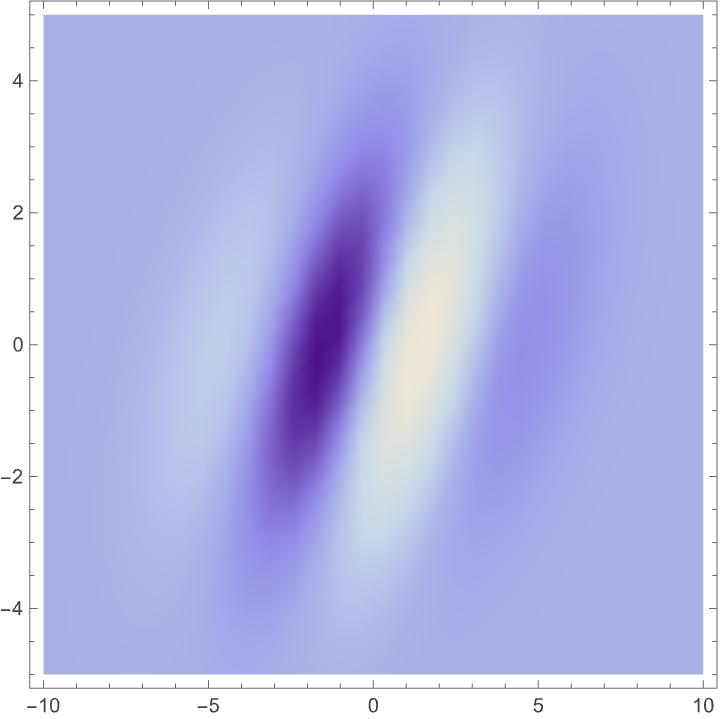}
        & \includegraphics[width=0.22\textwidth]{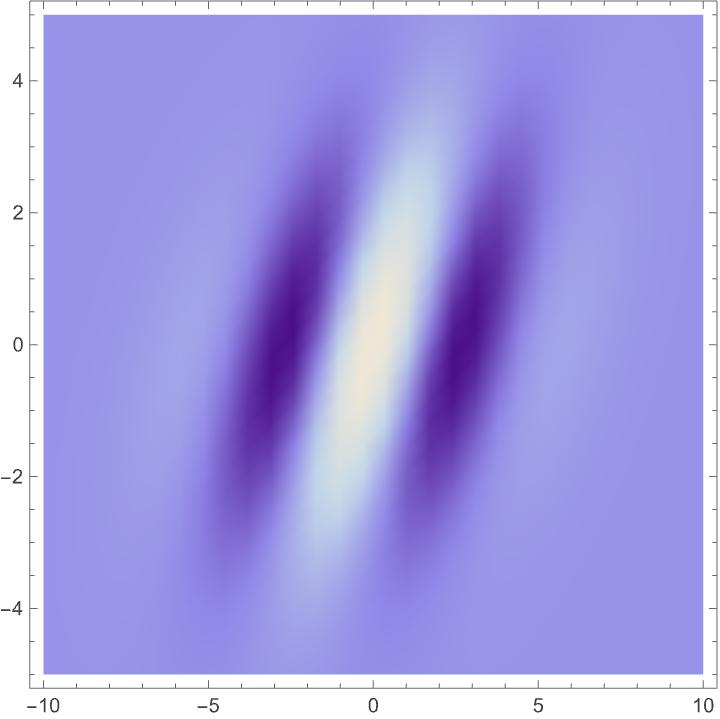}
      \end{tabular}
      \caption{1+1-D spatio-temporal cross-sections of the form
        $T_{\varphi^m}(x_1, t;\; \sigma_{\varphi}, \sigma_{t}, v)$
        of the 2+1-D spatio-temporal receptive field model
        (\ref{eq-spat-temp-RF-model-der-norm-caus})
        along the preferred direction over the spatial domain,
        here for $\varphi = \pi/6$, $\sigma_{\varphi}=2$,
        $\sigma_t=2$ and $v=1/2$.
        (Horizontal axes: spatial coordinate along the preferred
        direction of the receptive field $x_{\varphi} \in [-10, 10]$.
        Vertical axes: time $t \in [-5, 5]$.)}
    \end{subfigure}
  \end{center}
  \caption{Visualizations of the velocity-adapted affine Gaussian
    derivative filters used both as (i)~idealized models of simple cells
    of the form (\ref{eq-spat-temp-RF-model-der-norm-caus})
    and (ii)~for computing the directional quasi
    quadrature measures according to
    (\ref{eq-quasi-quad-dir-vel-adapt-spat-temp-12}) and
   (\ref{eq-quasi-quad-dir-vel-adapt-spat-temp-34}) in the idealized
   models of complex cells.}
   \label{fig-filters-compl-cell-models}
\end{figure*}

\subsection{Idealized models for velocity-tuned complex cells based on
  the output from velocity-tuned simple cells}
\label{sec-model-vel-tuned-complex-cells}

When defining idealized models of complex cells from the output from
simple cells, one basic fact to consider is that if one applies
velocity-adapted spatio-temporal receptive fields for non-zero orders
$n$ of temporal differentiation, then the resulting temporal
derivatives will be zero if the velocity parameter $v = (v_1, v_2)^T$ of the
spatio-temporal receptive field is equal to the image velocity
$u = (u_1, u_2)^T$ of
the image stimulus. For this reason, if we want to have a model of
complex cells, that is to respond to image data with large values for matching
image velocities without introducing a magnitude inversion step, it
is therefore natural to formulate an idealized model of complex cells based on
the output from simple cells for zero order $n = 0$ of temporal
differentiation.

Based on Lindeberg (\citeyear{Lin25-JCompNeurSci-orisel}) Equation~(23),
with the complementary scale normalization parameter $\Gamma$
in that more general expression 
set to $\Gamma = 0$, for simplicity, we will consider the following
quasi quadrature combination of first- and second-order spatial derivatives
\begin{equation}
  \label{eq-quasi-quad-dir-vel-adapt-spat-temp-12}
  ({\cal Q}_{\varphi,12,\vel,\norm} L)^2
  = L_{\varphi,\norm}^2 + \, C_{\varphi} \, L_{\varphi\varphi,\norm}^2
\end{equation}
and the following new quasi quadrature combination of third- and
fourth-order spatial derivatives
\begin{equation}
  \label{eq-quasi-quad-dir-vel-adapt-spat-temp-34}
  ({\cal Q}_{\varphi,34,\vel,\norm} L)^2
  = L_{\varphi\varphi\varphi,\norm}^2 + \, C_{\varphi} \, L_{\varphi\varphi\varphi\varphi,\norm}^2,
\end{equation}
where the individual components in this expression are defined
from velocity-adapted spatio-temporal receptive fields according to
\begin{multline}
  \label{eq-spat-temp-quasi-vel-adapt-comp1}
    L_{\varphi,\norm}(\cdot, \cdot, \cdot;\;  \sigma_{\varphi}, \sigma_t, v, \Sigma_{\varphi}) =\\
    = T_{\varphi,\norm}(\cdot, \cdot, \cdot;\; \sigma_{\varphi}, \sigma_t, v, \Sigma_{\varphi}) *
        f(\cdot, \cdot, \cdot),
 \end{multline}
\begin{multline}
    \label{eq-spat-temp-quasi-vel-adapt-comp2}
    L_{\varphi\varphi,\norm}(\cdot, \cdot, \cdot;\;  \sigma_{\varphi}, \sigma_t, v, \Sigma_{\varphi}) =\\
    = T_{\varphi\varphi,\norm}(\cdot, \cdot, \cdot;\; \sigma_{\varphi}, \sigma_t, v, \Sigma_{\varphi}) *
        f(\cdot, \cdot, \cdot),
\end{multline}
and
\begin{multline}
  \label{eq-spat-temp-quasi-vel-adapt-comp3}
    L_{\varphi\varphi\varphi,\norm}(\cdot, \cdot, \cdot;\;  \sigma_{\varphi}, \sigma_t, v, \Sigma_{\varphi}) =\\
    = T_{\varphi\varphi\varphi,\norm}(\cdot, \cdot, \cdot;\; \sigma_{\varphi}, \sigma_t, v, \Sigma_{\varphi}) *
        f(\cdot, \cdot, \cdot),
 \end{multline}
\begin{multline}
    \label{eq-spat-temp-quasi-vel-adapt-comp4}
    L_{\varphi\varphi\varphi\varphi,\norm}(\cdot, \cdot, \cdot;\;  \sigma_{\varphi}, \sigma_t, v, \Sigma_{\varphi}) =\\
    = T_{\varphi\varphi\varphi\varphi,\norm}(\cdot, \cdot, \cdot;\; \sigma_{\varphi}, \sigma_t, v, \Sigma_{\varphi}) *
        f(\cdot, \cdot, \cdot),
\end{multline}
with the underlying velocity-adapted spatio-temporal receptive fields
$T_{\varphi^m, t^n,\norm}(x_1, x_2, t;\; \sigma_{\varphi}, \sigma_t, v, \Sigma_{\varphi}) $
according to (\ref{eq-spat-temp-RF-model-der-norm-caus}) for $n = 0$.
Figure~\ref{fig-filters-compl-cell-models} gives illustrates the types
of 2+1-D filters used for computing these directional quasi quadrature
measures, in terms of either purely 2-D spatial cross-sections or joint
1+1-D spatio-temporal cross-sections.
Figure~\ref{fig-ft-filters-compl-cell-models} in
Appendix~\ref{app-fourier-transf} does additionally show
the Fourier transforms of these filters.

This model is structurally related to the energy model for complex
cells developed by Adelson and Bergen (\citeyear{AdeBer85-JOSA})
and Heeger (\citeyear{Hee87-JOSA}) and to the proposal by
Koenderink and van Doorn (\citeyear{KoeDoo90-BC}) to sum
up the squares of first- and second-order derivative responses,
in a similar way as for sine and cosine waves, to get more
phase-independent responses.
This model is also inspired by the finding that odd- and even-shaped
receptive fields have been observed to occur in pairs
(De~Valois {\em et al.\/}\ \citeyear{ValCotMahElfWil00-VR}).
The quasi quadrature measures serve as approximations of
quadrature pairs, as can be defined from a Hilbert transform
(Bracewell \citeyear{Bra99}, pp.~267--272).

Here, we conceptually extend these notions to instead
basing the quadrature measures on the responses from
velocity-adapted affine Gaussian derivatives,
with explicit variabilities with respect to both the
degree of elongation and local Galilean transformations,
that have not been addressed in the previous work
by Adelson and Bergen (\citeyear{AdeBer85-JOSA})
and Heeger (\citeyear{Hee87-JOSA}) or in previous
other work regarding Gabor models.
Hence, in relation to previous work, our work
is based on a more general theoretical model
for the variabilities in the shapes of the
spatio-temporal receptive fields.

According to the general theory for quasi quadrature measures in
Lindeberg (\citeyear{Lin18-SIIMS}), we then use the parameter
choice $C_{\varphi} = 1/\sqrt{2}$, as derived to the result in Equation~(27)
in that paper for the specific choice of the
parameter $\Gamma$ in that paper being $\Gamma = 0$.

While the quasi quadrature measures ${\cal Q}_{\varphi,12,\vel,\norm}L$
and ${\cal Q}_{\varphi,34,\vel,\norm} L$
have been designed to reduce the spatial variations in the underlying
pairs of first-order and second-order
or third-order and fourth-order spatial derivative responses, as 
approximations to the notion of quadrature pairs, we will here instead
consider corresponding extensions to spatially integrated quasi
quadrature measures according to
\begin{multline}
  \label{eq-quasi-quad-dir-vel-adapt-spat-temp-12-int}
  ((\overline{{\cal Q}}_{\varphi,12,\vel,\norm} L)
  (\cdot, \cdot, t;\;  \sigma_{\varphi}, \sigma_t, v,
  \Sigma_{\varphi}))^2 = \\
  = g(\cdot, \cdot;\; \gamma^2 \Sigma_{\varphi})
  * (({\cal Q}_{\varphi,12,\vel,\norm} L) (\cdot, \cdot, t;\;  \sigma_{\varphi}, \sigma_t, v, \Sigma_{\varphi}))^2
\end{multline}
and
\begin{multline}
  \label{eq-quasi-quad-dir-vel-adapt-spat-temp-34-int}
  ((\overline{{\cal Q}}_{\varphi,34,\vel,\norm} L)
  (\cdot, \cdot, t;\;  \sigma_{\varphi}, \sigma_t, v,
  \Sigma_{\varphi}))^2 = \\
  = g(\cdot, \cdot;\; \gamma^2 \Sigma_{\varphi})
  * (({\cal Q}_{\varphi,34,\vel,\norm} L)(\cdot, \cdot, t;\;  \sigma_{\varphi}, \sigma_t, v, \Sigma_{\varphi}))^2,
\end{multline}
where $\gamma > 0$ is a relative integration scale, that we will
henceforth choose%
\footnote{The motivations for using the value $\gamma = \sqrt{2}$
  for the relative integration scale parameter $\gamma$ used
  for defining the second-stage integration scale as a multiple of the
  local scale for differentiation are that: (i)~a multiplicative relationship
  between the local scale parameter and the integration scale parameter
  preserves the scale covariance properties, and (ii)~the specific
  value $\gamma = \sqrt{2}$ has in previous work on shape-from-texture
  (Lindeberg and G{\aa}rding \citeyear{LG93-ICCV},
  G{\aa}rding and Lindeberg \citeyear{GL94-IJCV})
  been found to constitute a reasonable minimum value of $\gamma$
  that still suppresses most of the local perturbations in integrated quadratic-type
  image descriptors based on local Gaussian derivatives.}
as $\gamma = \sqrt{2}$.
In this way, the results will have a much lower variability with
respect to the spatial positions in the image domain.

The underlying motivation for these constructions is that the
resulting spatio-temporal quasi quadrature measures\\
$\overline{{\cal Q}}_{\varphi,12,\vel,\norm} L$
and
$\overline{{\cal Q}}_{\varphi,34,\vel,\norm} L$
should then assume their maximum values when the velocity parameter
$v = (v_1, v_c)^T$ of the receptive field is equal to the image
velocity $u = (u_1, u_2)^T$ of the input stimulus.

\section{Direction and speed selective properties for idealized models
  of simple cells}
\label{sec-dir-sel-simpl-cells}

\begin{figure}[hbtp]
  \begin{center}
    \includegraphics[width=0.40\textwidth]{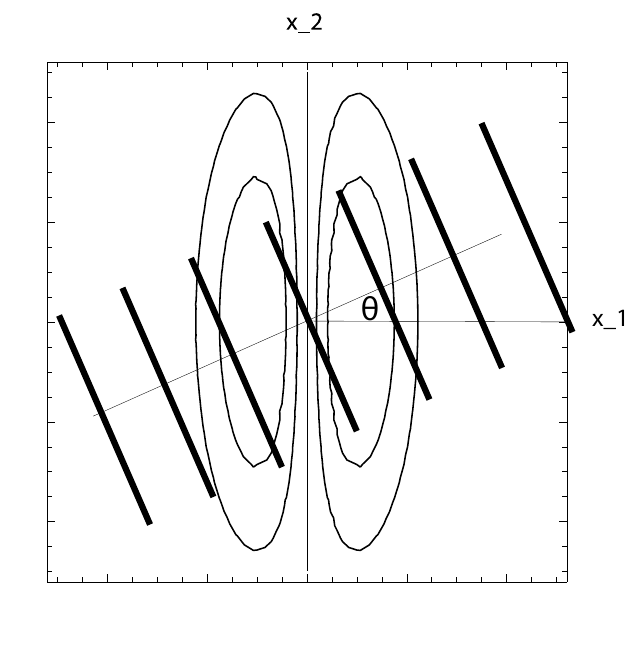}
  \end{center}
  \caption{Schematic illustration of the modelling situation studied
    in the theoretical analysis, where
    the coordinate system is aligned to the preferred direction
    $\varphi = 0$ of the receptive field, and the receptive field is
    then exposed to a moving sine wave
    pattern with inclination angle $\theta$. In this figure, the sine
    wave pattern is schematically illustrated by a set of level lines,
  overlayed onto a few level curves of a first-order affine Gaussian
  derivative kernel. (Horizontal axis: spatial coordinate $x_1$.
  Vertical axis: spatial coordinate $x_2$.)}
  \label{fig-schem-ill-model-rf-sine}
\end{figure}

For probing the direction selectivity properties of the different
forms of spatio-temporal receptive fields described in
Section~\ref{sec-gen-gauss-der-strfs}, we will in this paper
throughout make use of stimuli in terms of
moving sine waves of the form
\begin{multline}
   \label{eq-sine-wave-model-vel-adapt-anal}
  f(x_1, x_2, t) = \\ 
  = \sin
     \left(
       \omega \cos (\theta) \, x_1  + \omega \sin (\theta) \, x_2 -  |u| \, t + \beta
     \right),
\end{multline}
where $\omega$ is the magnitude of the angular frequency,
$\theta$ its inclination angle, $|u|$ is the image speed, and $\beta$ is a phase angle,
see Figure~\ref{fig-schem-ill-model-rf-sine} for an illustration.

\begin{figure*}[hbtp]
  \begin{equation}
    \label{eq-dir-sel-simple-1-general}
    A_{\varphi}(\theta, |u|;\; \sigma_{\varphi}, \kappa, \sigma_t, |v|) =
    \frac{\cos (\theta ) \sqrt{\sigma_{\varphi}^2+\sigma_t^2 (|u|-|v|)^2}}{\sqrt{\kappa ^2
        \sigma_{\varphi}^2 \sin ^2(\theta )+\cos ^2(\theta )
        \left(\sigma_{\varphi}^2+\sigma_t^2 |v|^2\right)-2 \sigma_t^2 |u| |v| \cos (\theta
        )+\sigma_t^2 |u|^2}}
  \end{equation}
  
  \begin{equation}
    \label{eq-dir-sel-simple-2-general}    
    A_{\varphi\varphi}(\theta, |u|;\; \sigma_{\varphi}, \kappa, \sigma_t, |v|) =
    \frac{\cos ^2(\theta ) \left(\sigma_{\varphi}^2+\sigma_t^2 (|u|-|v|)^2\right)}{\kappa ^2
      \sigma_{\varphi}^2 \sin ^2(\theta )+\cos ^2(\theta )
      \left(\sigma_{\varphi}^2+\sigma_t^2 |v|^2\right)-2 \sigma_t^2 |u| |v| \cos (\theta
      )+\sigma_t^2 |u|^2}
  \end{equation}

  \begin{equation}
    \label{eq-dir-sel-simple-3-general}    
    A_{\varphi\varphi\varphi}(\theta, |u|;\; \sigma_{\varphi}, \kappa, \sigma_t, |v|) =
    \frac{\cos ^3(\theta ) \left(\sigma_{\varphi}^2+\sigma_t^2
        (|u|-|v|)^2\right)^{3/2}}{\left(\kappa ^2 \sigma_{\varphi}^2 \sin ^2(\theta )+\cos ^2(\theta
        ) \left(\sigma_{\varphi}^2+\sigma_t^2 |v|^2\right)-2 \sigma_t^2 |u| |v| \cos (\theta
        )+\sigma_t^2 |u|^2\right)^{3/2}}
  \end{equation}

  \begin{equation}
    \label{eq-dir-sel-simple-4-general}    
    A_{\varphi\varphi\varphi\varphi}(\theta, |u|;\; \sigma_{\varphi}, \kappa, \sigma_t, |v|) =
    \frac{\cos ^4(\theta ) \left(\sigma_{\varphi}^2+\sigma_t^2
        (|u|-|v|)^2\right)^2}{\left(\kappa ^2 \sigma_{\varphi}^2 \sin ^2(\theta )+\cos ^2(\theta )
        \left(\sigma_{\varphi}^2+\sigma_t^2 |v|^2\right)-2 \sigma_t^2 |u| |v| \cos (\theta
        )+\sigma_t^2 |u|^2\right)^2}
  \end{equation}

  \caption{Explicit expressions for the direction selectivity curves
    for idealized models of simple cells for different orders
    $m \in \{1, 2, 3, 4 \}$ of spatial differentiation,
    without any coupling between the spatial scale parameter $\sigma_{\varphi}$
    and the temporal scale parameter $\sigma_t$.}
  \label{eq-dir-sel-expr-no-rel-sigma-xt}
\end{figure*}

While it may be more common in some neurophysiological experiments to
instead probe the direction selectivity properties of visual neurons
with moving bars, a main theoretical reason for using moving sine
waves is that the responses of the spatio-temporal receptive fields
can then be computed in closed form, for the spatio-temporal receptive
fields based on the generalized Gaussian derivative model.

Furthermore, to make the results independent of the relationship between the
spatial and temporal scale levels used in the models of the
spatio-temporal receptive fields, we will 
\begin{itemize}
\item
  for each receptive field choose the angular frequency of the probing
  sine wave in such a way that it maximizes the response of the
  spatio-temporal receptive field over all the possible choices of the
  angular frequency, and
\item
  for the spatio-temporal receptive fields with velocity adaptation,
  show the direction and speed selectivity graphs only for the special case when
  the spatial and the temporal scale levels are coupled in
  relation to the magnitude $|v|$ of the velocity parameter
  $v = (v_1, v_2)^T$ of the  receptive field according to
  \begin{equation}
    \label{eq-rel-sigma-xt-v}
    \frac{\sigma_x}{\sigma_t} = |v|,
  \end{equation}
  which is natural from the viewpoint of regarding the velocity
  estimation problem for a visual neuron as depending on the extent of
  the underlying receptive field over the spatial and the temporal domains.
\end{itemize}
In these ways, the resulting measures of the direction selectivity
properties of the idealized models of the spatio-temporal receptive
fields can be regarded as reflecting inherent properties of their
computational functions.

To handle the possible reservation, in case the coupling of the
spatial and temporal scale parameters $\sigma_{\varphi}$ and $\sigma_t$
according to
(\ref{eq-rel-sigma-xt-v}) would constitute a too strong
simplification, we do, however, also provide explicit mathematical models
for the combined direction and speed selectivity properties for the idealized models of
simple cells in Section~\ref{sec-prel-summary-simpl}.

Concerning the types of idealized models of motion-sensitive neurons,
we do in Appendix~\ref{app-anal-simpl-cells} perform separate
analyses of simple cells for different orders $m \in \{1, 2, 3, 4\}$.
Then, in Section~\ref{sec-dir-sel-compl-cells}, we will perform 
corresponding analyses of direction and speed selectivity properties for a set of models
of complex cells.
The motivation for studying such a large variety of models of visual
neurons is that in a neurophysiological experiment, where the
direction and speed selectivity properties of a neuron in the primary visual
cortex, it may not be {\em a priori\/} clear if that neuron would
correspond to a simple cell or a complex cell, and furthermore
regarding the models of simple cells, it may not be {\em a priori\/}
clear what order of spatial differentiation the neuron would correspond to.
Thereby, it is necessary to analyze a rather large set of
idealized models of simple cells and complex cells with a similar
theoretical probing scheme.

\begin{figure*}[hbtp]
  \begin{center}
    \begin{tabular}{ccccc}
      & {\em\footnotesize First-order simple cell}
      & {\em\footnotesize Second-order simple cell}
      & {\em\footnotesize Third-order simple cell }
      & {\em\footnotesize Fourth-order simple cell} \\
      & {\em\footnotesize $A_{\varphi}(\theta;\; r=1, \kappa)$}
      & {\em\footnotesize $A_{\varphi\varphi}(\theta;\; r=1, \kappa)$}
      & {\em\footnotesize $A_{\varphi\varphi\varphi}(\theta;\; r=1, \kappa)$}
      & {\em\footnotesize $A_{\varphi\varphi\varphi\varphi}(\theta;\; r=1, \kappa)$} \\       
      {\footnotesize $\kappa = 1$}
      & \includegraphics[height=0.14\textheight]{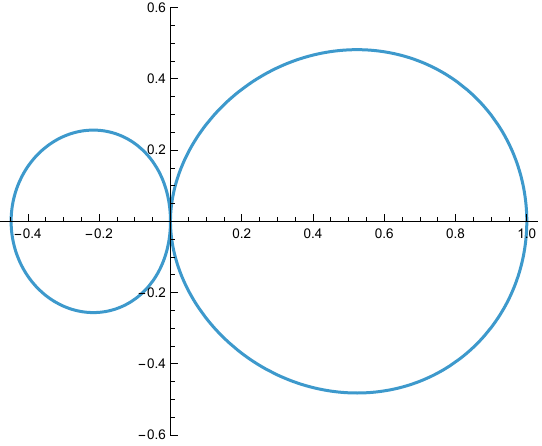}
      & \includegraphics[height=0.14\textheight]{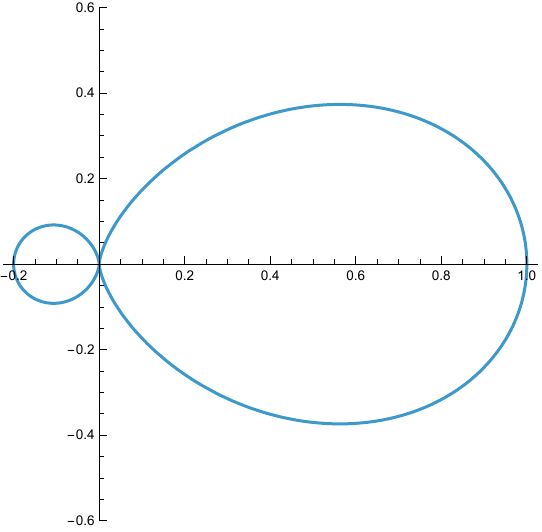}
      & \includegraphics[height=0.14\textheight]{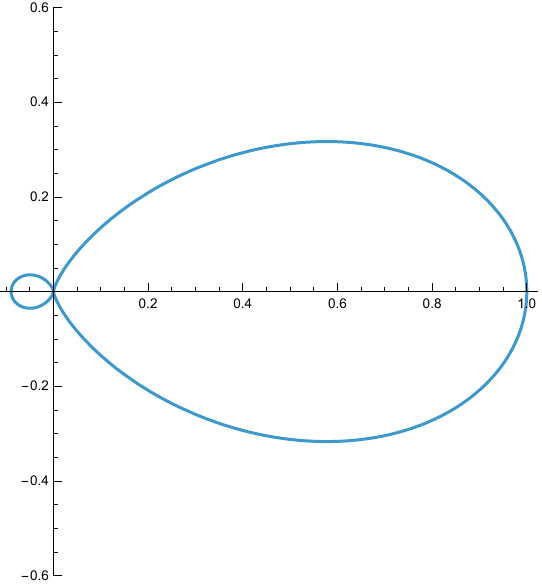}
      & \includegraphics[height=0.14\textheight]{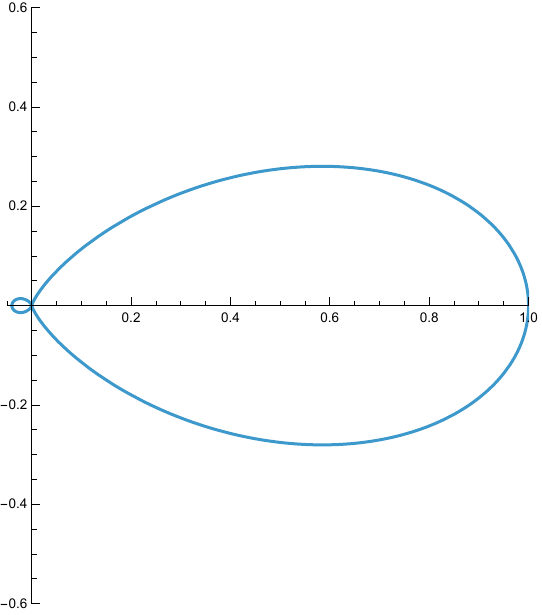} \\
      {\footnotesize $\kappa = 2$}
      & \includegraphics[height=0.14\textheight]{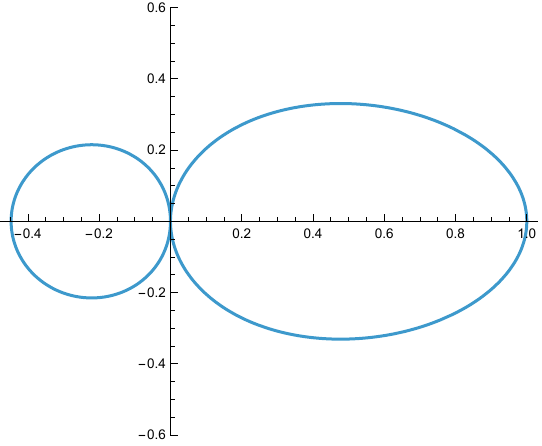}
      & \includegraphics[height=0.14\textheight]{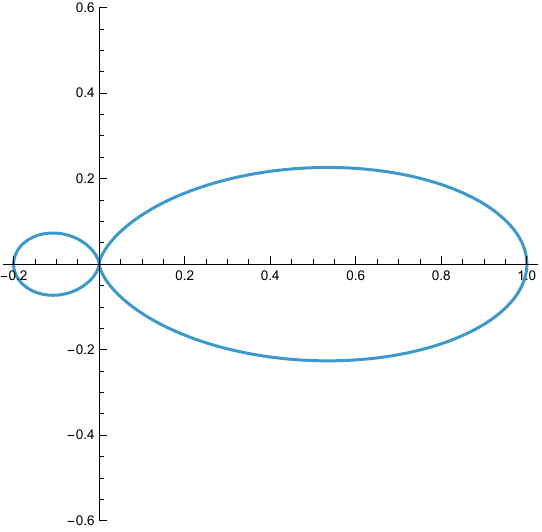}
      & \includegraphics[height=0.14\textheight]{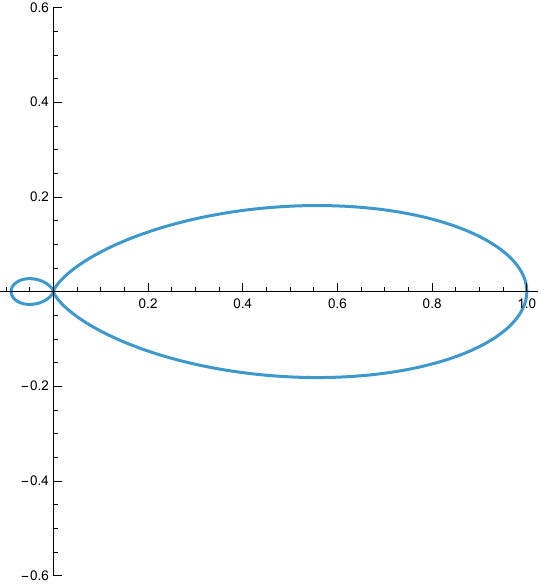}
      & \includegraphics[height=0.14\textheight]{diradaptphiphiphi-dirsel-omegasel-kappa2-eps-converted-to.pdf} \\
      {\footnotesize $\kappa = 4$}
      & \includegraphics[height=0.14\textheight]{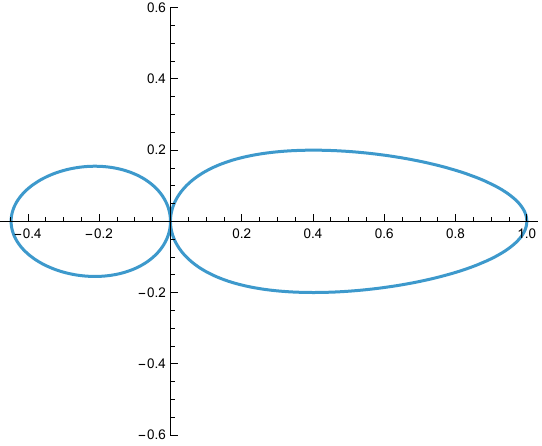}
      & \includegraphics[height=0.14\textheight]{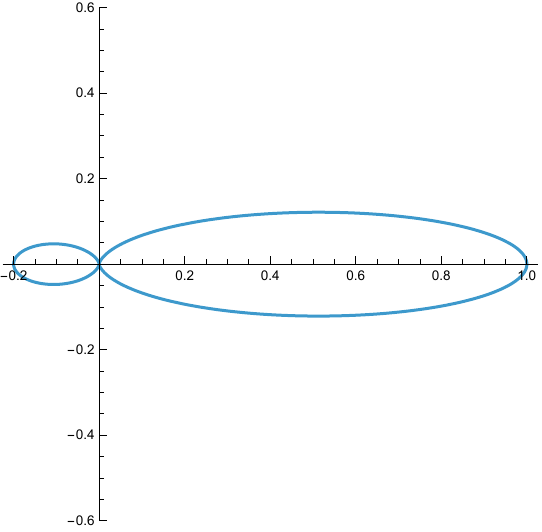}
      & \includegraphics[height=0.14\textheight]{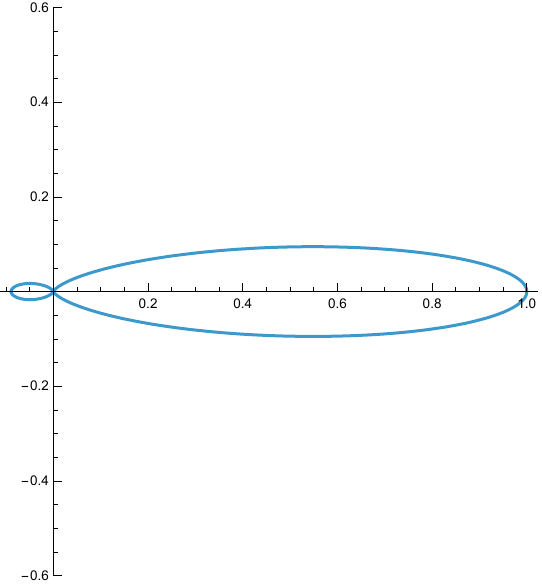}
      & \includegraphics[height=0.14\textheight]{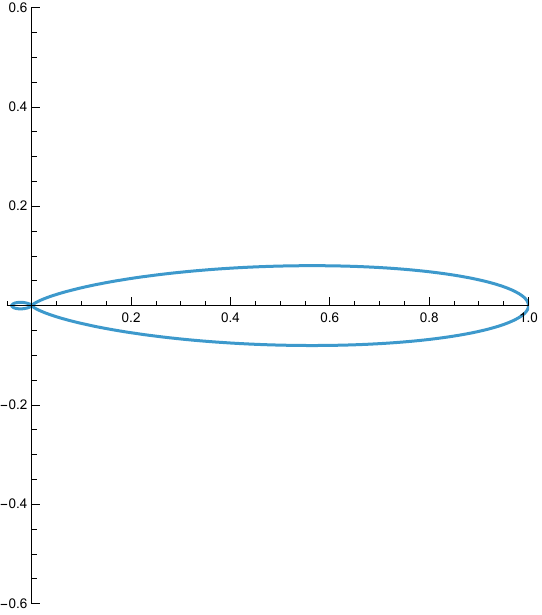} \\
      {\footnotesize $\kappa = 8$}
      & \includegraphics[height=0.14\textheight]{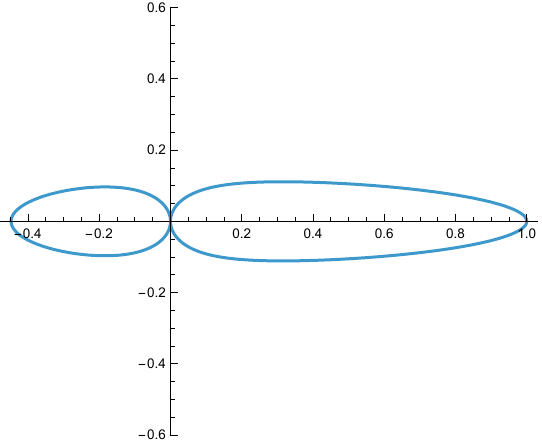}
      & \includegraphics[height=0.14\textheight]{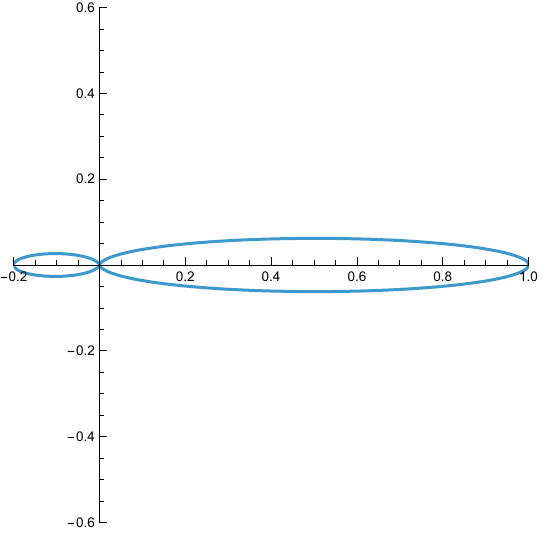}
      & \includegraphics[height=0.14\textheight]{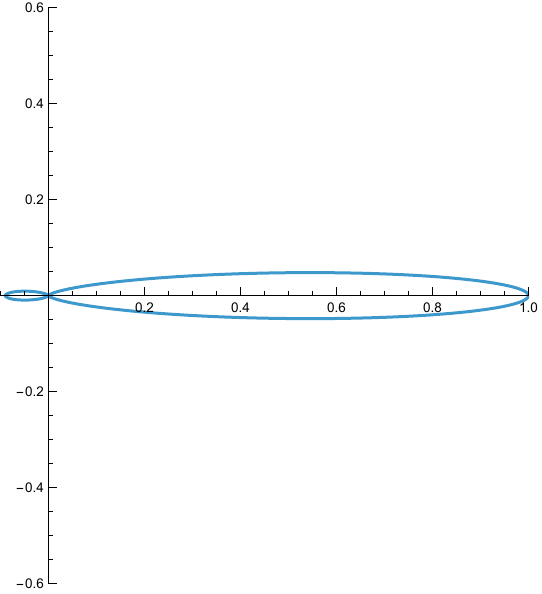}
      & \includegraphics[height=0.14\textheight]{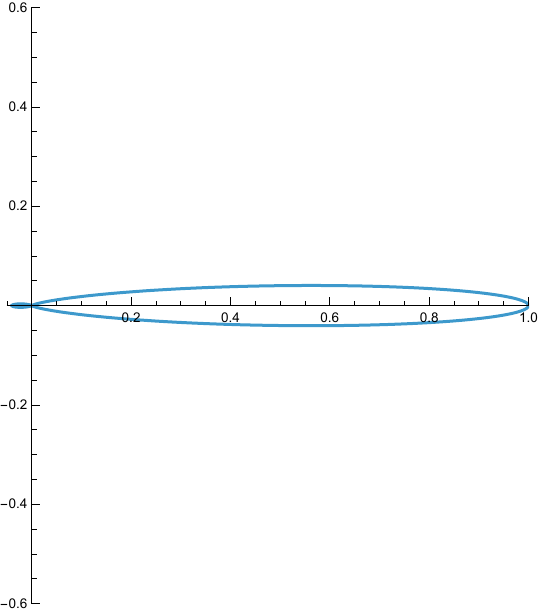} \\
     \end{tabular}
  \end{center}
  \caption{Graphs of the {\em direction selectivity for idealized
      models of velocity-tuned simple cells\/} of the form
    $A_{\varphi^m}(\theta;\; r, \kappa)  = A_{\varphi^m,\max}(\theta, |u|;\; \kappa, |v|)$
    for $r = |u|/|v|$ and with $A_{\varphi^m,\max}(\theta, |u|;\; \kappa, |v|)$
    according to (\ref{eq-cond-summ-dir-sel-prop})
    based on {\bf (left column)} first-order
    directional derivatives of affine Gaussian kernels combined with
    zero-order temporal Gaussian kernels, {\bf (middle left column)} 
    second-order directional derivatives of affine Gaussian
    kernels combined with zero-order
    temporal Gaussian kernels, {\bf (middle right column)} 
    third-order directional derivatives of affine Gaussian
    kernels combined with zero-order
    temporal Gaussian kernels, 
    or {\bf (right column)}
    fourth-order directional derivatives of affine Gaussian
    kernels combined with zero-order
    temporal Gaussian kernels, 
    {\em shown for different values of the degree of elongation
    $\kappa$\/} between the spatial scale parameters in the vertical
    {\em vs.\/}\ the horizontal directions. Observe how the direction
    selectivity varies strongly depending on the eccentricity
    $\epsilon = 1/\kappa$ of the receptive fields.
    {\bf (top row)} Results for $\kappa = 1$.
    {\bf (second row)} Results for $\kappa = 2$.
    {\bf (third row)} Results for $\kappa = 4$.
    {\bf (bottom row)} Results for $\kappa = 8$.}
  \label{fig-dir-sel-simple-veladapt-anal-const-u}
\end{figure*}

\begin{figure*}[hbtp]
  \begin{center}
    \begin{tabular}{ccccc}
      & {\em\footnotesize First-order simple cell\/}
      & {\em\footnotesize Second-order simple cell\/}
      & {\em\footnotesize Third-order simple cell\/}
      & {\em\footnotesize Fourth-order simple cell\/} \\
      & {\em\footnotesize $A_{\varphi}(\theta;\;  r, \kappa=2)$}
      & {\em\footnotesize $A_{\varphi\varphi}(\theta;\; r, \kappa=2)$}
      & {\em\footnotesize $A_{\varphi\varphi\varphi}(\theta;\; r, \kappa=2)$}
      & {\em\footnotesize $A_{\varphi\varphi\varphi\varphi}(\theta;\; r, \kappa=2)$} \\          
      {\footnotesize $r = 1/4$}
      & \includegraphics[height=0.12\textheight]{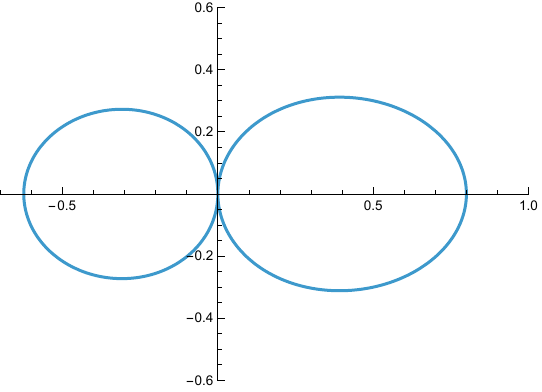}
      & \includegraphics[height=0.12\textheight]{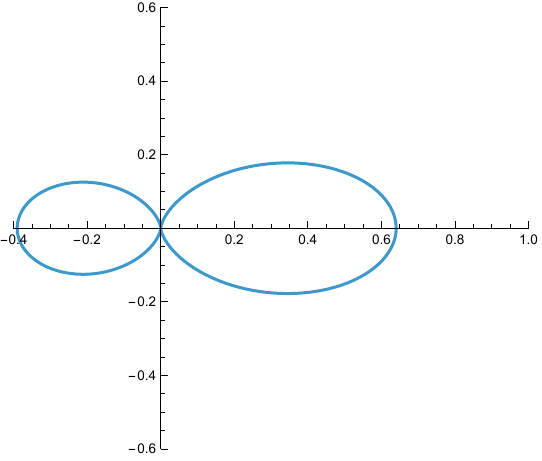}
      & \includegraphics[height=0.12\textheight]{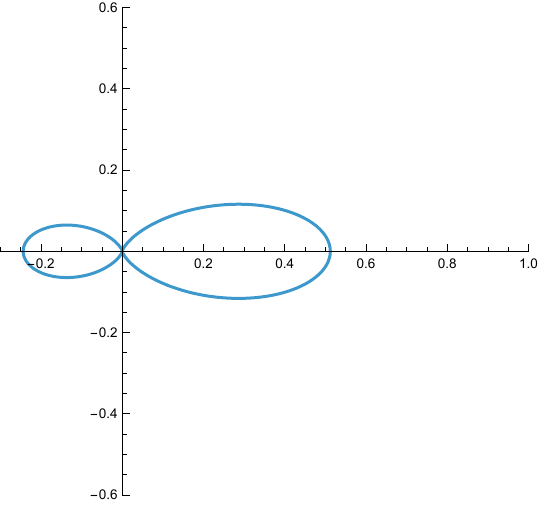}
      & \includegraphics[height=0.12\textheight]{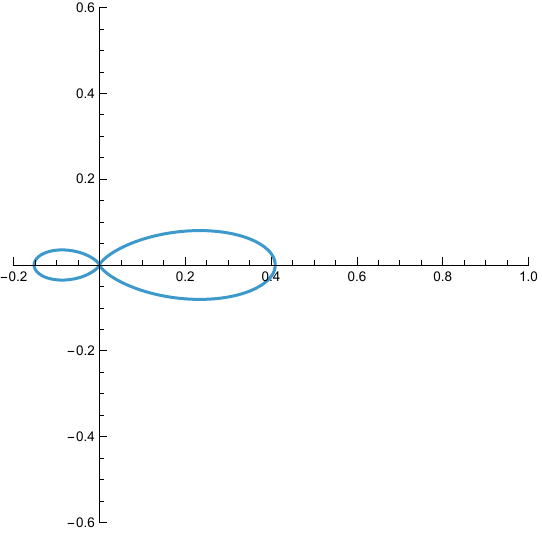} \\
      {\footnotesize $r = 1/2$}
      & \includegraphics[height=0.12\textheight]{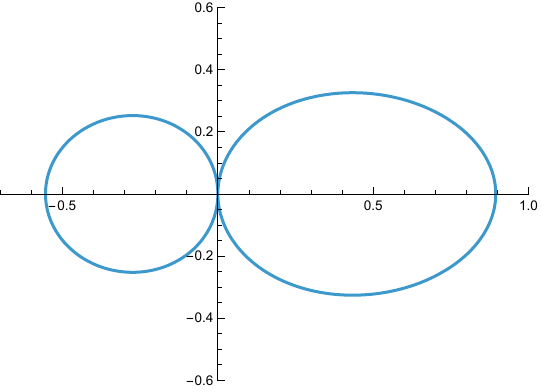}
      & \includegraphics[height=0.12\textheight]{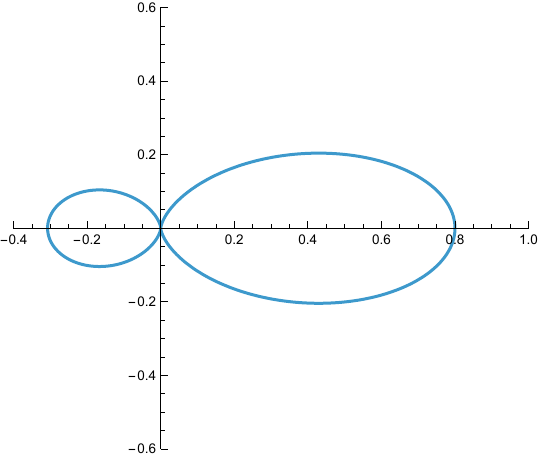}
      & \includegraphics[height=0.12\textheight]{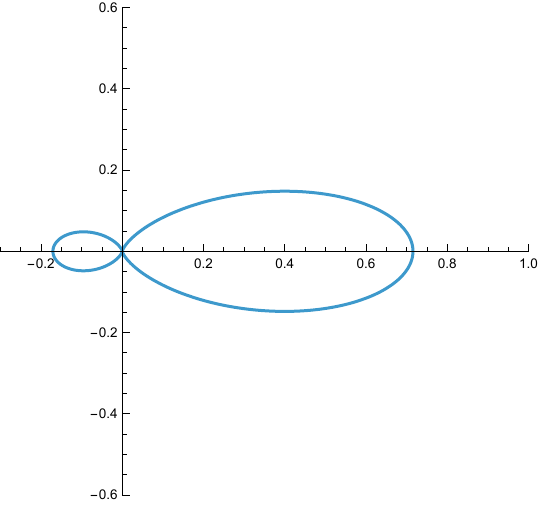}
      & \includegraphics[height=0.12\textheight]{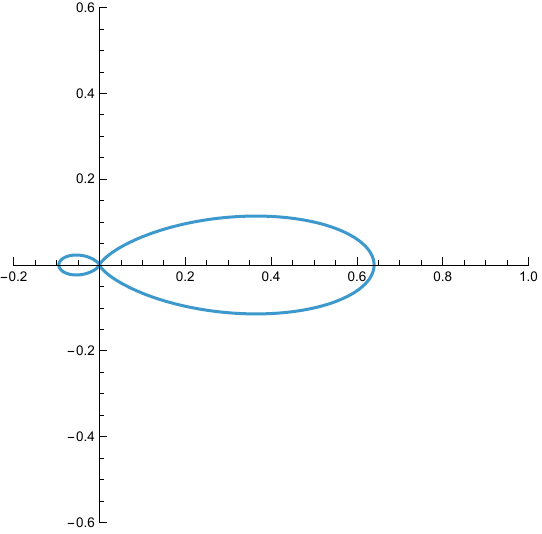} \\
      {\footnotesize $r = 1$}
      & \includegraphics[height=0.12\textheight]{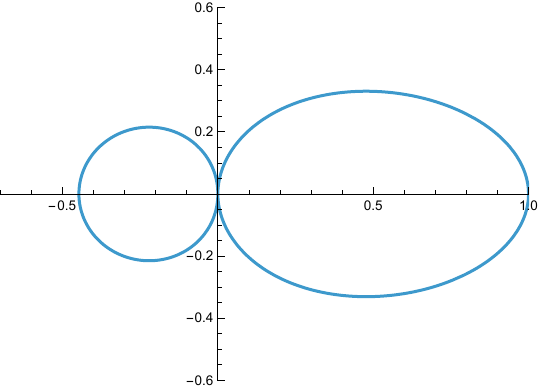}
      & \includegraphics[height=0.12\textheight]{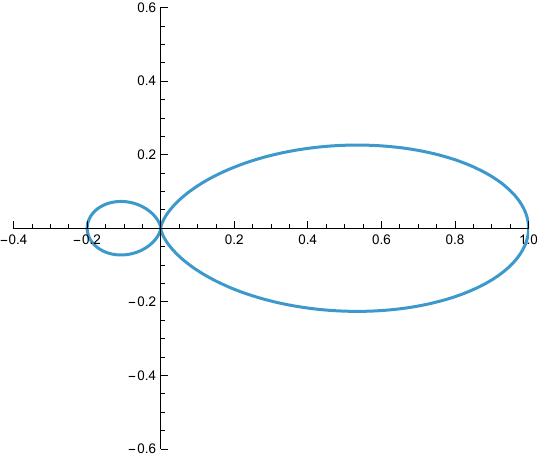}
      & \includegraphics[height=0.12\textheight]{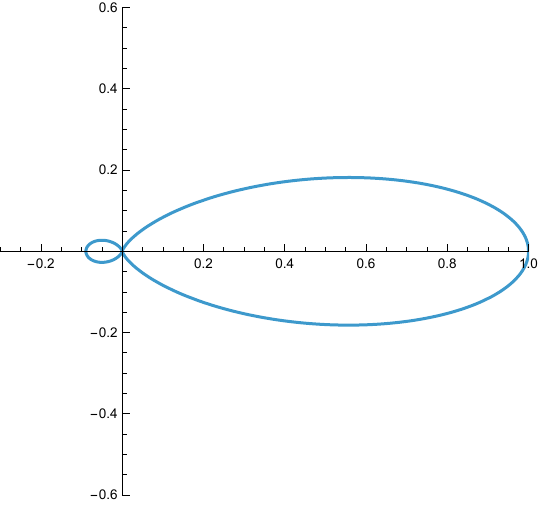}
      & \includegraphics[height=0.12\textheight]{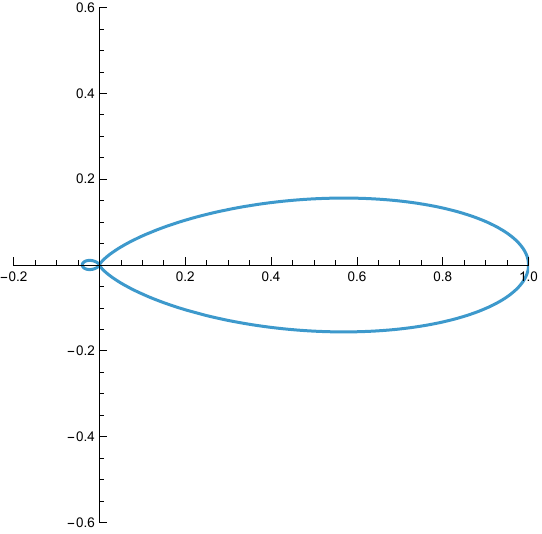} \\
      {\footnotesize $r = 2$}
      & \includegraphics[height=0.12\textheight]{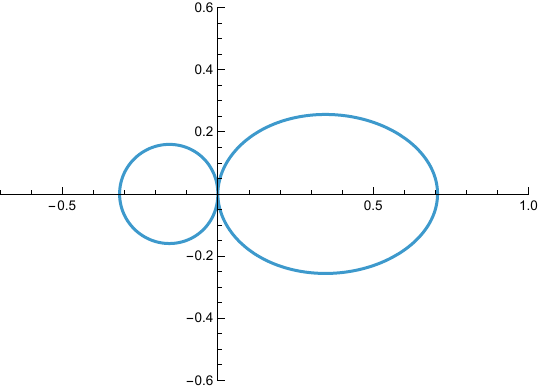}
      & \includegraphics[height=0.12\textheight]{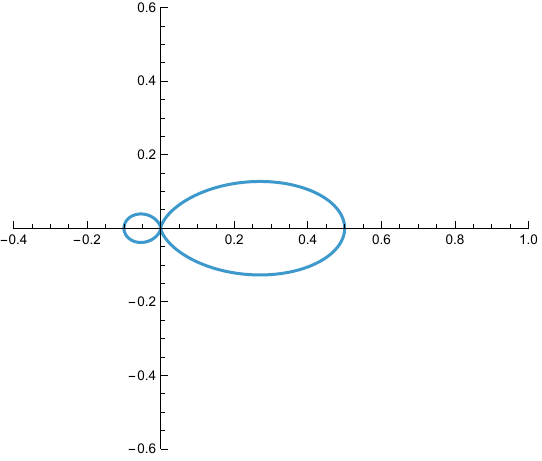}
      & \includegraphics[height=0.12\textheight]{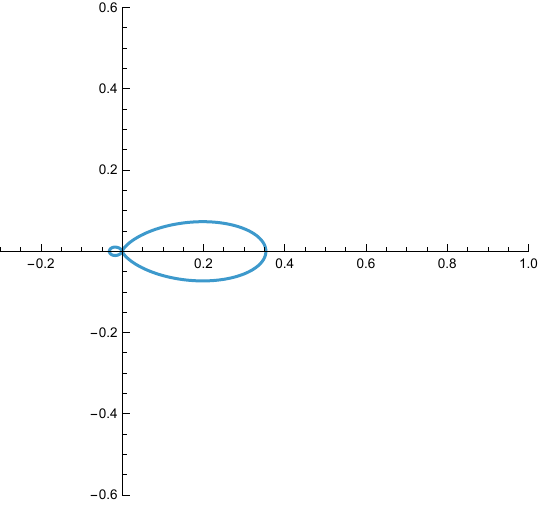}
     & \includegraphics[height=0.12\textheight]{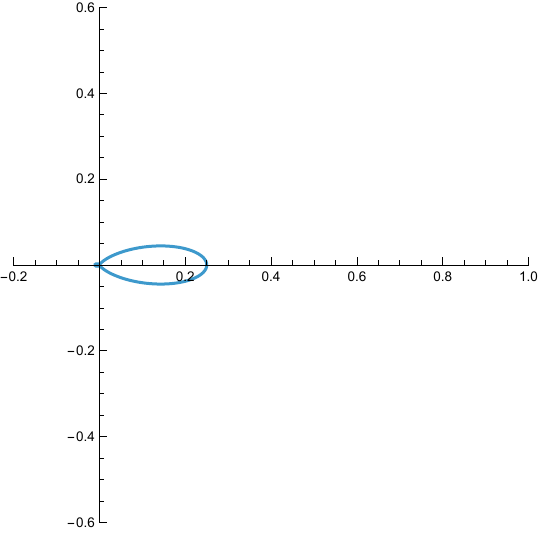} \\
      {\footnotesize $r = 4$}
      & \includegraphics[height=0.12\textheight]{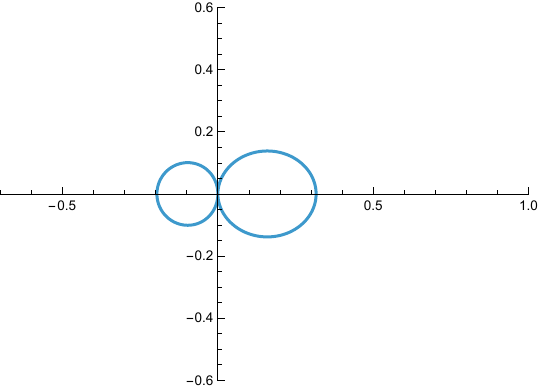}
      & \includegraphics[height=0.12\textheight]{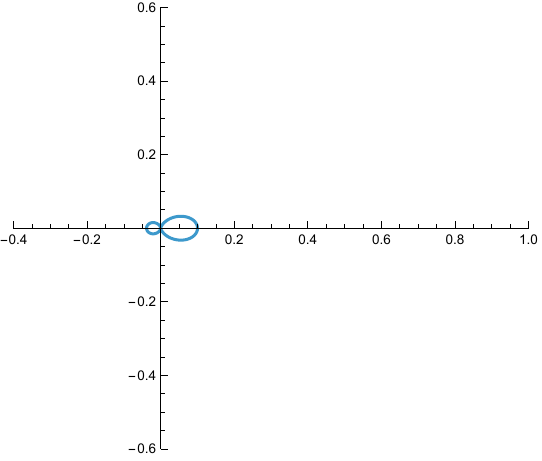}
      & \includegraphics[height=0.12\textheight]{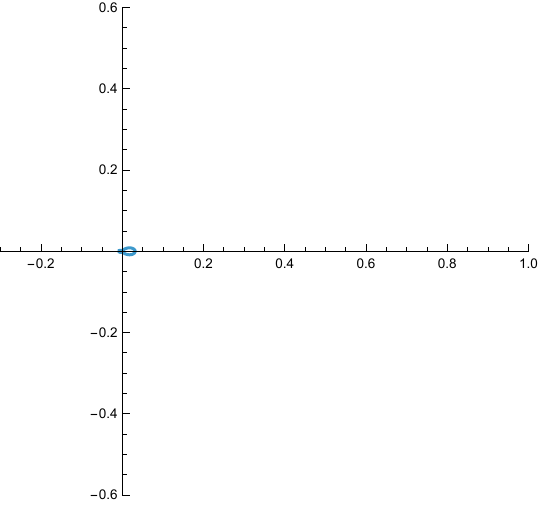}
      & \includegraphics[height=0.12\textheight]{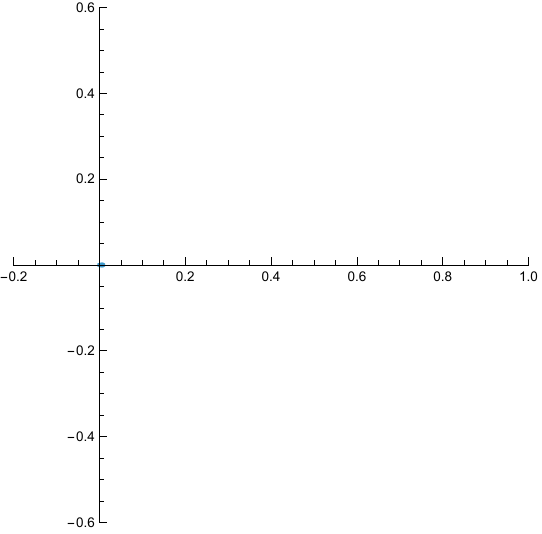} \\
     \end{tabular}
  \end{center}
  \caption{Graphs of the {\em direction selectivity for idealized
      models of velocity-tuned simple cells\/} of the form
    $A_{\varphi^m}(\theta;\; r, \kappa)  = A_{\varphi^m,\max}(\theta, |u|;\; \kappa, |v|)$
    for $r = |u|/|v|$ and with $A_{\varphi^m,\max}(\theta, |u|;\; \kappa, |v|)$
    according to (\ref{eq-cond-summ-dir-sel-prop})
    based on {\bf (left column)} first-order
    directional derivatives of affine Gaussian kernels combined with
    zero-order temporal Gaussian kernels, {\bf (middle left column)} 
    second-order directional derivatives of affine Gaussian
    kernels combined with zero-order
    temporal Gaussian kernels, {\bf (middle right column)}
    third-order directional derivatives of affine Gaussian
    kernels combined with zero-order
    temporal Gaussian kernels, 
    or {\bf (right column)}
    fourth-order directional derivatives of affine Gaussian
    kernels combined with zero-order
    temporal Gaussian kernels, 
    {\em shown for different values of the ratio $r$\/} in the relationship
    $|u| = r \, |v|$ between the speed $|u|$ of the motion stimulus
    and the speed $|v|$ of the receptive field, for a fixed value of
    $\kappa = 2$ between the spatial scale parameters in the vertical
    {\em vs.\/}\ the horizontal directions. Observe how the direction
    selectivity varies strongly depending on the eccentricity
    $\epsilon = 1/\kappa$ of the receptive fields.
    {\bf (top row)} Results for $r = 1/4$.
    {\bf (second row)} Results for $r = 1/2$.
    {\bf (third row)} Results for $r = 1 $.
    {\bf (fourth row)} Results for $r = 2 $.
    {\bf (bottom row)} Results for $r = 4$.}
  \label{fig-dir-sel-simple-veladapt-anal-const-kappa}
\end{figure*}

\begin{figure*}[hbtp]
  \begin{center}
    \begin{tabular}{cccc}
      {\em\footnotesize First-order simple cell $R_{\varphi}(r)$}
      & {\em\footnotesize Second-order simple cell $R_{\varphi\varphi}(r)$}
      & {\em\footnotesize Third-order simple cell $R_{\varphi\varphi\varphi}(r)$}
      & {\em\footnotesize Fourth-order simple cell $R_{\varphi\varphi\varphi\varphi}(r)$} \\       
      \includegraphics[width=0.22\textwidth]{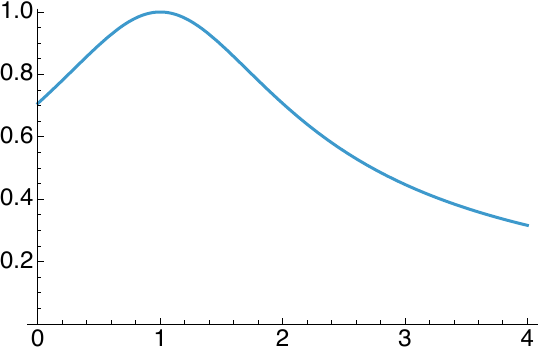}
      & \includegraphics[width=0.22\textwidth]{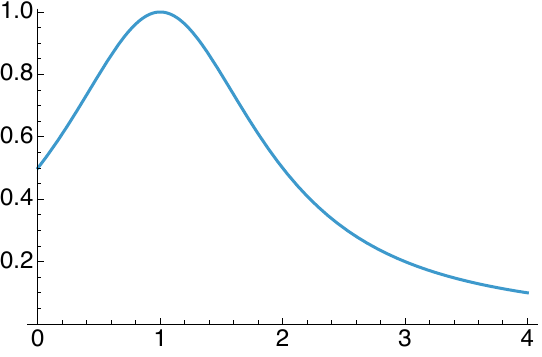}
      & \includegraphics[width=0.22\textwidth]{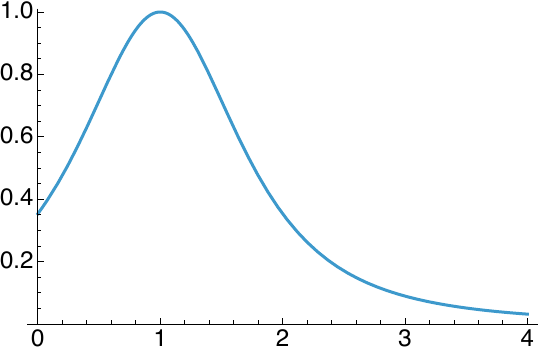}
      & \includegraphics[width=0.22\textwidth]{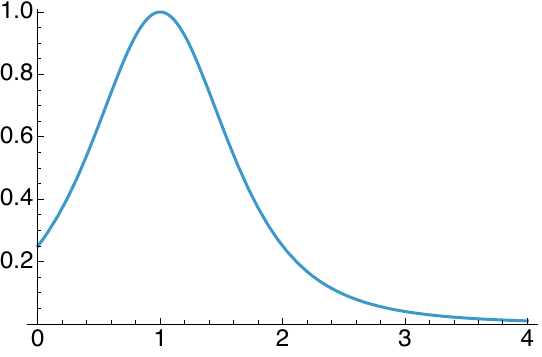} \\
            \includegraphics[width=0.22\textwidth]{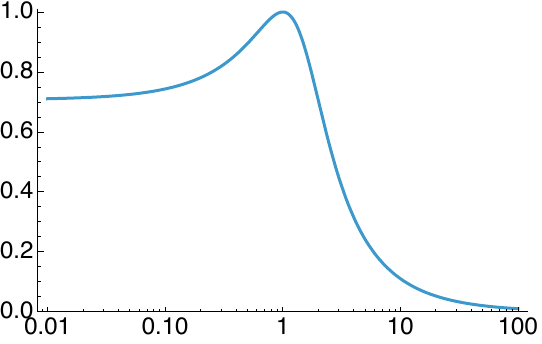}
      & \includegraphics[width=0.22\textwidth]{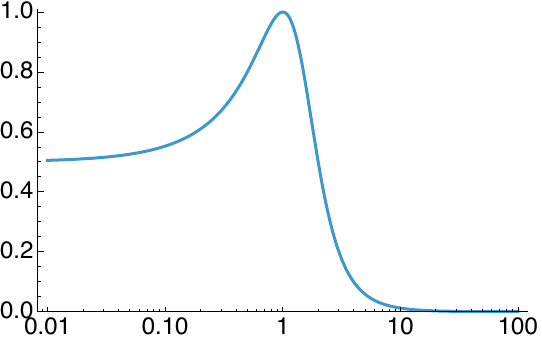}
      & \includegraphics[width=0.22\textwidth]{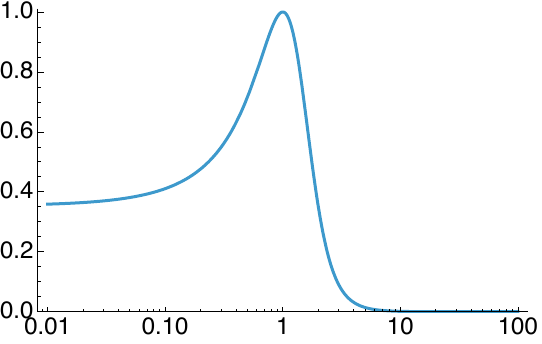}
      & \includegraphics[width=0.22\textwidth]{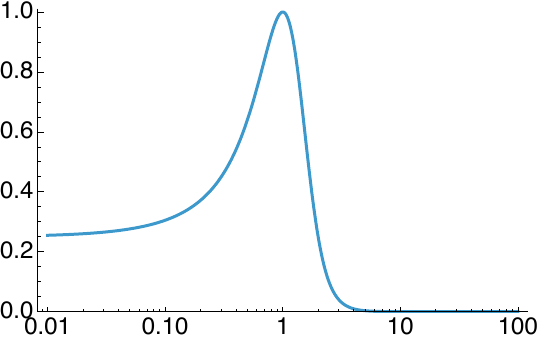} \\
     \end{tabular}
  \end{center}
  \caption{Graphs of the {\em speed selectivity curves\/} $R_{\varphi^m}(r)$
    according to (\ref{eq-rcurve-order-all})
    %(\ref{eq-rcurve-order1}),  (\ref{eq-rcurve-order2}),
    % (\ref{eq-rcurve-order3}) and (\ref{eq-rcurve-order4})
    for the inclination angle $\theta = 0$ for {\em idealized
      models of velocity-tuned simple cells\/} using {\bf (top row)}  a linear speed
    scale and {\bf (bottom row)} a logarithmic speed scale.
    The underlying theoretical models are based on {\bf (left)} first-order
    directional derivatives of affine Gaussian kernels combined with
    zero-order temporal Gaussian kernels, {\bf (middle left)}
    second-order directional derivatives of affine Gaussian
    kernels combined with zero-order
    temporal Gaussian kernels, {\bf (middle right)} 
    third-order directional derivatives of affine Gaussian
    kernels combined with zero-order
    temporal Gaussian kernels, or {\bf (right)}
    fourth-order directional derivatives of affine Gaussian
    kernels combined with zero-order
    temporal Gaussian kernels, 
    {\em shown for different values of the parameter $r$} in the relationship
    $|u| = r \, |v|$ between the speed $|u|$ of the motion stimulus
    and the speed $|v|$ of the receptive field.
    Notably, these measures are independent of the degree of elongation
    $\kappa$ of the spatio-temporal receptive fields.
    (Horizontal axes in the top row: relative speed parameter $r \in [0, 4]$.
    Horizontal axes in the bottom row: relative speed parameter $r \in [0.01, 100]$.
    Vertical axes: speed selectivity function $R_{\varphi^m}(r)$.)}
  \label{fig-dir-sel-simple-veladapt-anal-theta-0}

  \bigskip
  
  \begin{equation}
    \label{eq-speed-sel-curve-compl-12}
    R_{{\cal Q}_{12}}(r) = 
    \frac{e^{\frac{(r-1)^2-2 \gamma ^2}{\sqrt{2} \left(r^2-2 r+2\right)}} \sqrt{\left(r^2-2
          r+3\right) e^{\frac{2 \sqrt{2} \gamma ^2}{r^2-2 r+2}}+(r-1)^2}}{\sqrt{2} \left(r^2-2
        r+2\right)}
  \end{equation}
  \begin{equation}
    \label{eq-speed-sel-curve-compl-34}
    R_{{\cal Q}_{34}}(r) = 
    \sqrt{\frac{e^{\frac{2 \sqrt{3} \left(2 \gamma ^2+1\right) (r-1)^2}{r^2-2 r+2}}
        \left(\left(r^2-2 r+\sqrt{6}+2\right) e^{\frac{4 \sqrt{3} \gamma ^2}{r^2-2
              r+2}}+r^2-2 r-\sqrt{6}+2\right)}{\left(\left(1+\sqrt{6}\right) e^{4 \sqrt{3} \gamma
            ^2}+1-\sqrt{6}\right) \left(r^2-2 r+2\right)^4}}
  \end{equation}
  \caption{Dependencies on the relative speed parameter $r$ for the
    integrated quasi quadrature measures
    $\overline{{\cal Q}}_{\varphi,12,\vel,\norm} L$ according to
    (\ref{eq-quasi-quad-dir-vel-adapt-spat-temp-12-int}) and
    $\overline{{\cal Q}}_{\varphi,34,\vel,\norm} L$ according to
    (\ref{eq-quasi-quad-dir-vel-adapt-spat-temp-34-int}), when
    subjected to a moving sine wave with inclination angle $\theta =
    0$. The parameter $\gamma$ is the relative integration scale
    factor for the spatial integration of the corresponding pointwise quasi
    quadrature measure with an affine Gaussian kernel.}
  \label{fig-alg-expr-R-curves-int-compl-cells}

  \bigskip
  
  \begin{center}
    \begin{tabular}{cc}
      {\em\footnotesize Velocity-tuned complex cell (12) $R_{{\cal Q}_{12}}(r)$}
      & {\em\footnotesize Velocity-tuned complex cell (34) $R_{{\cal Q}_{34}}(r)$}\\
      \includegraphics[width=0.22\textwidth]{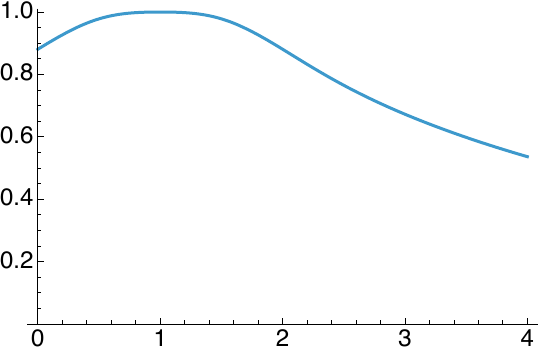}
      & \includegraphics[width=0.22\textwidth]{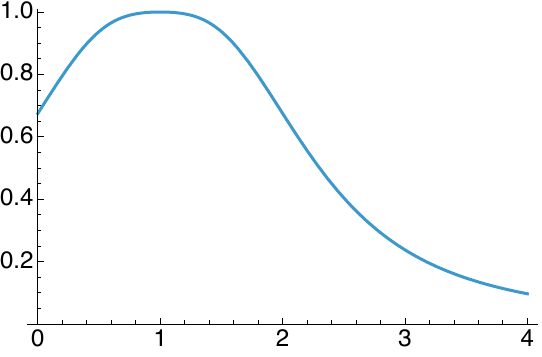} \\
     \includegraphics[width=0.22\textwidth]{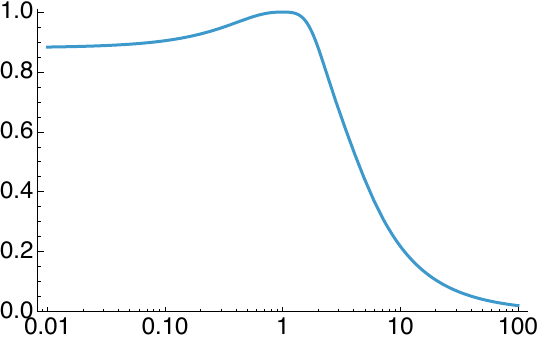}
      & \includegraphics[width=0.22\textwidth]{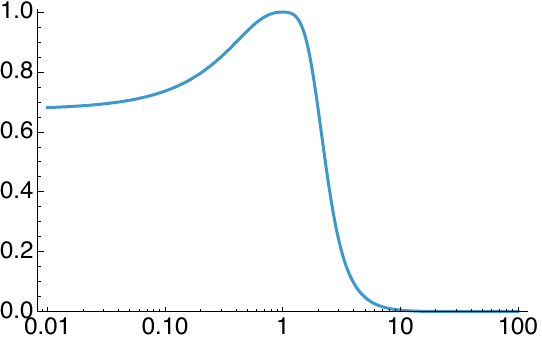} \\
      \end{tabular}
  \end{center}
  \caption{Graphs of the {\em speed selectivity curves\/} $R_{{\cal Q}_{12}}(r)$
    according to (\ref{eq-speed-sel-curve-compl-12}) and
    $R_{{\cal Q}_{34}}(r)$
    according to (\ref{eq-speed-sel-curve-compl-34})
    for the inclination angle $\theta = 0$ for {\em idealized
    models of velocity-tuned complex cells cells\/} using {\bf (top row)} a linear speed
    scale and {\bf (bottom row)} a logarithmic speed scale.
    The underlying theoretical models are based on {\bf (left)} an integrated
  quadratic combination of first- and second-order
    directional derivatives of affine Gaussian kernels combined with
    zero-order temporal Gaussian kernels, {\bf (right)}
    an integrated
    quadratic combination of third- and fourth-order
    directional derivatives of affine Gaussian kernels combined with
    zero-order temporal Gaussian kernels,
    {\em shown for different values of the parameter $r$} in the relationship
    $|u| = r \, |v|$ between the speed $|u|$ of the motion stimulus
    and the speed $|v|$ of the receptive field.
    Notably, these measures are independent of the degree of elongation
    $\kappa$ of the spatio-temporal receptive fields.
    (Horizontal axes in the top row: relative speed parameter $r \in [0, 4]$.
    Horizontal axes in the bottom row: relative speed parameter $r \in [0.01, 100]$.
    Vertical axes: speed selectivity function $R_{\cal Q}(r)$.)}
  \label{fig-dir-sel-complex-veladapt-anal-theta-0}
\end{figure*}

\subsection{Condensed summary of direction and speed selectivity
  properties for the idealized models of simple cells}
\label{sec-prel-summary-simpl}

The main theoretical result in this paper concerns the derivation
of explicit closed-form expressions for the direction and speed selectivity
properties for our idealized models of simple cells and complex
cells, in terms of only a very small set of shape parameters
of the spatio-temporal receptive fields.
Since that topic, by necessity, turns out to be somewhat
technical in terms of the required mathematical analysis,
we have put those derivations in
Appendices~\ref{app-anal-simpl-cells}--\ref{app-anal-compl-cells},
and do instead here only summarize those results
and their interpretation in this section as well as in the
following Section~\ref{sec-dir-sel-compl-cells}, so that
a reader more interested in the actual results than the
machinery of the theoretical analysis should be able to digest
these results in a more easy manner.
This material will then constitute the theoretical foundation
for the biological modelling and the extended biological
interpretations and predictions presented in
Sections~\ref{sec-rels-biol-vision}--\ref{sec-sugg-neuro-exps}.

Summarizing the results for the direction selectivity properties of the different
idealized models of simple cells analyzed in Appendix~\ref{app-anal-simpl-cells},
specifically the results derived in Equations~(\ref{eq-Aphi-max}),
(\ref{eq-Aphiphi-max}), (\ref{eq-ori-sel-simple-3der})
and (\ref{eq-ori-sel-simple-4der}), we find that provided that the
spatial and the temporal scale parameters $\sigma_{\varphi}$ and $\sigma_t$
are coupled to the velocity parameter $v= (v_1, v_2)^T$ of the spatio-temporal
receptive field according to
\begin{equation}
  \label{eq-couple-sigmaphi-sigmat}
  \frac{\sigma_{\varphi}}{\sigma_t} = |v| = \sqrt{v_1^2 + v_2^2},
\end{equation}
then with the degree of elongation of the spatial component
of the receptive field defined as
\begin{equation}
  \kappa = \frac{\sigma_{\orth}}{\sigma_{\varphi}}
\end{equation}
with $\sigma_{\orth}$ denoting the scale parameter corresponding
to the amount of affine Gaussian smoothing in the direction orthogonal to the
preferred direction of the receptive field,
the combined direction
and speed selectivity in the motion direction $\theta$ and image
velocity $u = (u_1, u_2)^T = (|u| \, \cos \theta, |u| \, \sin \theta)$
of the stimulus is of the form
\begin{multline}
  \label{eq-cond-summ-dir-sel-prop}
  A_{\varphi^m,\max}(\theta, |u|;\; \kappa, |v|) = \\
  = \alpha_m \left( \frac{\cos (\theta )}{\sqrt{\kappa ^2 \sin ^2(\theta )-\frac{2 |u| \cos (\theta
   )}{|v|}+2 \cos ^2(\theta )+\frac{|u|^2}{|v|^2}}} \right)^m,
\end{multline}
as depending on the order $m \in \{ 1, 2, 3, 4 \}$ of spatial differentiation in the
spatial component of the idealized receptive field model.
Here, it can be specifically noted that the motion parameters
$|u|$ and $|v|$ of the stimulus and the receptive field, respectively,
throughout occur jointly in terms of the relative speed ratio $r = |u|/|v|$.

Specifically, in the special case when the inclination angle
$\theta$ between the velocity parameters $u = (u_1, u_2)^T$
and $v = (v_1, v_2)^T$ of the image stimulus and the preferred
direction of the receptive field is $\theta = 0$,
it follows from Equations~(\ref{eq-rcurve-order1}),  (\ref{eq-rcurve-order2}),
(\ref{eq-rcurve-order3}) and (\ref{eq-rcurve-order4})
in Appendix~\ref{app-anal-simpl-cells} that
the resulting speed selectivity is for $|u| = r \, |v|$ of the form
\begin{equation}
  \label{eq-rcurve-order-all}
  R_{\varphi^m}(r) = A_{\varphi^m,\max}(0, r \, |v|;\; \kappa, |v|)
  = \frac{1}{\left( r^2-2 r+2 \right)^{m/2}}.
\end{equation}
Notably, these idealized models for direction and speed selectivity curves,
that beyond the order $m$ of spatial differentiation
only depend on the following shape parameters of the spatio-temporal
receptive fields:
(i)~the inclination angle $\theta$ between the motion direction
of the stimulus and the preferred direction of the receptive field,
(ii)~the ratio $r = |u|/|v|$ between the speed parameters of the
stimulus and the receptive field, and
(iii)~the degree of elongation $\kappa$ of the spatial component
of the receptive field.
In this respect, we thereby obtain very parameter-compact
idealized models for the shapes of the idealized direction
and speed selectivity curves.

Corresponding combined direction and speed selectivity measures
for the more general case, when the spatial and
the temporal scale parameters $\sigma_{\varphi}$ and $\sigma_t$
are not necessarily coupled in relation to the speed
parameter $|v|$ of the receptive field
according to (\ref{eq-couple-sigmaphi-sigmat}),
are given in
Equations~(\ref{eq-dir-sel-simple-1-general})--(\ref{eq-dir-sel-simple-4-general})
in Figure~\ref{eq-dir-sel-expr-no-rel-sigma-xt}.
In the special case when the inclination angle $\theta = 0$,
these expressions reduce to resulting speed
selectivity curves of the form
\begin{equation}
  \label{eq-rcurve-order-all-noncoupled}  
  R_{\varphi^m}(|u|;\; \sigma_{\varphi}, \sigma_t, |v|)
  = \left(
    \frac{\sigma_{\varphi}}{\sqrt{\sigma_{\varphi}^2+\sigma_t^2 \, (|u|-|v|)^2}}
  \right)^m.
\end{equation}
Graphs of the corresponding direction selectivity curves
with coupled spatial and scale parameters are shown
in Figure~\ref{fig-dir-sel-simple-veladapt-anal-const-u} for
different degrees of elongation and in
Figure~\ref{fig-dir-sel-simple-veladapt-anal-const-kappa} for
different values of the relative image speed $r = |u|/|v|$ between the
image speed $|u|$ of the motion stimulus and the preferred speed
$|v|$ of the spatio-temporal receptive field.
Figure~\ref{fig-dir-sel-simple-veladapt-anal-theta-0} does in turn
show the corresponding speed selectivity curves for the different
idealized models of simple cells of orders $m \in \{ 1, 2, 3, 4 \}$,
again based on coupled values of the spatial and scale parameters,

According to the most idealized version of the normative theory
for visual receptive fields, there should be spatio-temporal
receptive fields for all motion directions in image space,
all motion speeds and all degrees of elongation, while
under the constraint of a finite processing capacity
the image velocities $|v|$ and the degree of elongation $\kappa$
would have to be restricted to some finite ranges $|v| \in [|v|_{\min}, |v|_{\max}]$
and $\kappa \in [\kappa_{\min}, \kappa_{\max}]$, where
the corresponding finite intervals $[|v|_{\min}, |v|_{\max}]$
and $[\kappa_{\min}, \kappa_{\max}]$ then determine the capacity
of the vision system in terms of the variabilities
in the speeds of motion stimuli and how steep slant angles
the vision system can accurately handle
(Lindeberg \citeyear{Lin26-JCompNeuroSci} Sections~6.3.3 and~6.3.5).
Specifically, a logarithmic distribution over each interval constitutes
canonical default choice because of complementary
requirements of self-similarity for the non-negative entities
$|v|$ and $\kappa$ (Jaynes \citeyear{Jay68-SMC}).

The visualizations in
Figures~\ref{fig-dir-sel-simple-veladapt-anal-const-u}--\ref{fig-dir-sel-simple-veladapt-anal-const-kappa}
do from this perspective reflect subsets of the variabilities in the shapes of the
direction and speed selectivity curves that we could expect from an
idealized vision system according to the underlying
normative theory for visual receptive fields.

Regarding qualitative properties of the direction selectivity for the
different orders $m$ of spatial differentiation, by comparing the
direction selectivity graphs for the idealized models
of simple cells, we generally find that:
\begin{itemize}
\item
  The direction selectivity becomes sharper with increasing degree of
  elongation $\kappa$ and for increasing order $m$ of spatial
  differentiation.
\item
  The decrease in speed selectivity with increasing speed of the
  probing motion stimulus becomes more abrupt with increasing order
  of spatial differentiation.
\item
  The lobes in the direction selectivity curves for the direction
  perpendicular to the preferred direction of the receptive field
  become smaller with increasing order of spatial differentiation.
\end{itemize}
In the following section, we will next described that corresponding qualitative
properties do also hold regarding idealized models of velocity-tuned
complex cells of the forms
(\ref{eq-quasi-quad-dir-vel-adapt-spat-temp-12-int}) and
(\ref{eq-quasi-quad-dir-vel-adapt-spat-temp-34-int}).

\begin{figure*}[hbtp]
  \begin{center}
    \begin{tabular}{ccc}
      & {\em\footnotesize Velocity-tuned complex cell (12)\/}
      & {\em\footnotesize Velocity-tuned complex cell (34)\/} \\
      & $A_{{\cal Q}_{12},\max}(\theta, r=1;\; \kappa)$ 
      & $A_{{\cal Q}_{34},\max}(\theta, r=1;\; \kappa)$ \\
      {\footnotesize $\kappa = 1$}
      & \includegraphics[height=0.14\textheight]{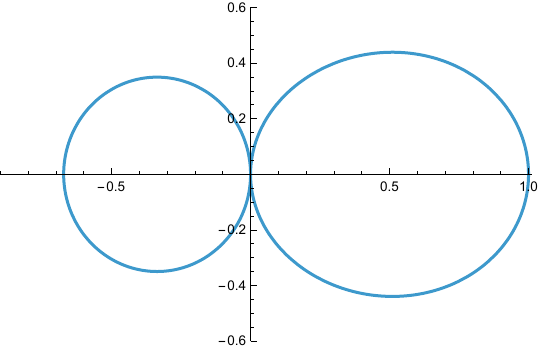}
      &
        \includegraphics[height=0.14\textheight]{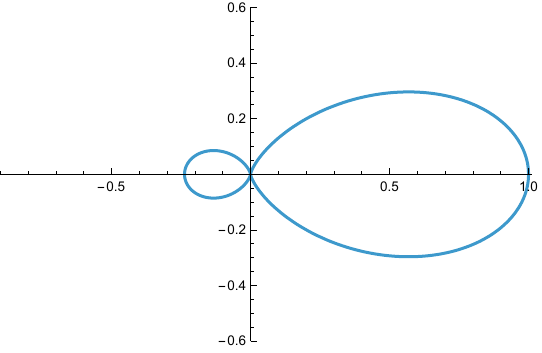} \\
 %     & \includegraphics[height=0.14\textheight]{diradaptphiphiphi-dirsel-omegasel-kappa1-eps-converted-to.pdf}
%      & \includegraphics[height=0.14\textheight]{diradaptphiphiphiphi-dirsel-omegasel-kappa1-eps-converted-to.pdf} \\
      {\footnotesize $\kappa = 2$}
      & \includegraphics[height=0.14\textheight]{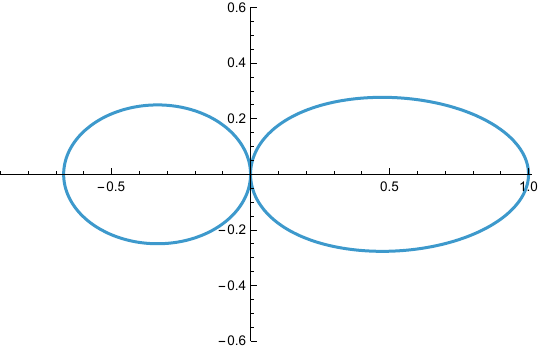}
      & \includegraphics[height=0.14\textheight]{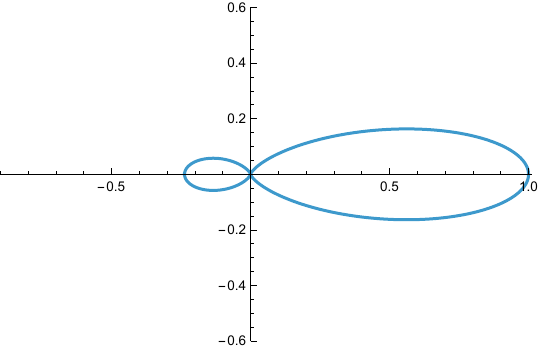}\\
%      & \includegraphics[height=0.14\textheight]{diradaptphiphiphi-dirsel-omegasel-kappa2-eps-converted-to.pdf}
%      & \includegraphics[height=0.14\textheight]{diradaptphiphiphi-dirsel-omegasel-kappa2-eps-converted-to.pdf} \\
      {\footnotesize $\kappa = 4$}
      & \includegraphics[height=0.14\textheight]{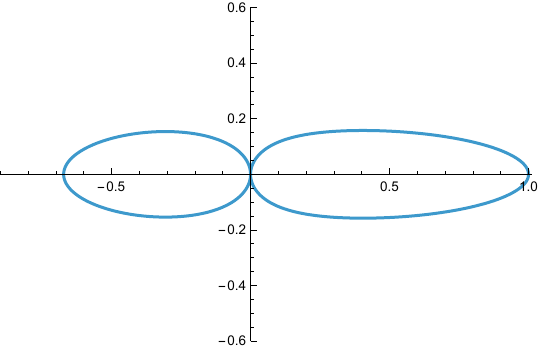}
      & \includegraphics[height=0.14\textheight]{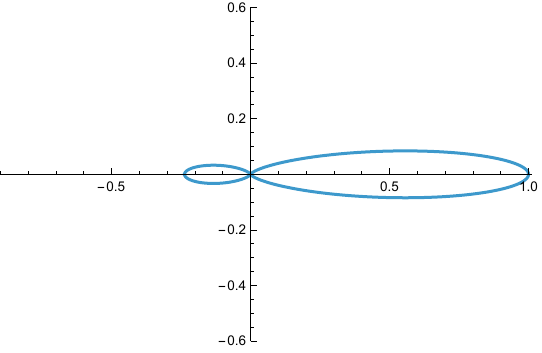}\\
%      & \includegraphics[height=0.14\textheight]{diradaptphiphiphi-dirsel-omegasel-kappa4-eps-converted-to.pdf}
%      & \includegraphics[height=0.14\textheight]{diradaptphiphiphiphi-dirsel-omegasel-kappa4-eps-converted-to.pdf} \\
      {\footnotesize $\kappa = 8$}
      & \includegraphics[height=0.14\textheight]{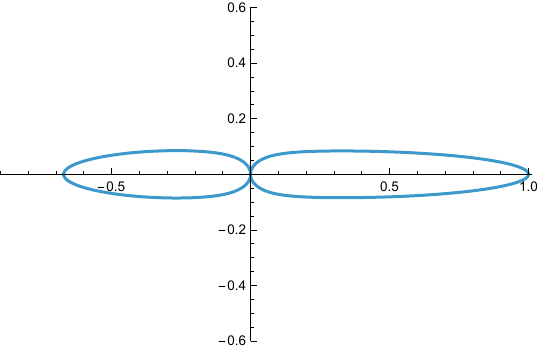}
      & \includegraphics[height=0.14\textheight]{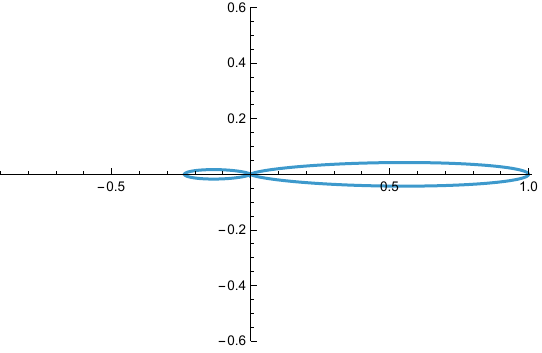}\\
%      & \includegraphics[height=0.14\textheight]{diradaptphiphiphi-dirsel-omegasel-kappa8-eps-converted-to.pdf}
%      & \includegraphics[height=0.14\textheight]{diradaptphiphiphiphi-dirsel-omegasel-kappa8-eps-converted-to.pdf} \\
     \end{tabular}
  \end{center}
  \caption{Graphs of the {\em direction selectivity curves
      of the forms $A_{{\cal Q}_{12},\max}(\theta, r;\; \kappa)$
      according to (\ref{eq-A-Q12-int-compl}) and
      $A_{{\cal Q}_{34},\max}(\theta, r;\; \kappa) $
      according to (\ref{eq-A-Q34-int-compl}) for idealized
      models of velocity-tuned complex cells\/} based on {\bf (left column)} spatial
    integration of a quadratic
    combination of first- and second-order spatial derivatives
    with zero-order temporal Gaussian kernels
    into $\overline{{\cal Q}}_{\varphi,12,\vel,\norm} L$ according
    to (\ref{eq-quasi-quad-dir-vel-adapt-spat-temp-12-int})
    {\bf (right column)} spatial integration of a quadratic
    combination  of third- and fourth-order spatial derivatives
    with zero-order temporal Gaussian kernels
    into $\overline{{\cal Q}}_{\varphi,34,\vel,\norm} L$ according
    to (\ref{eq-quasi-quad-dir-vel-adapt-spat-temp-34-int}),
    {\em shown for different values of the degree of elongation
    $\kappa$\/} between the spatial scale parameters in the vertical
    {\em vs.\/}\ the horizontal directions. Observe how the direction
    selectivity varies strongly depending on the eccentricity
    $\epsilon = 1/\kappa$ of the receptive fields.
    {\bf (top row)} Results for $\kappa = 1$.
    {\bf (second row)} Results for $\kappa = 2$.
    {\bf (third row)} Results for $\kappa = 4$.
    {\bf (bottom row)} Results for $\kappa = 8$.}
  \label{fig-dir-sel-complex-const-u}
\end{figure*}

\section{Direction and speed selectivity properties for idealized
  models of complex cells}
\label{sec-dir-sel-compl-cells}

Given the above analysis for a set of different models of simple
cells, as varying by the order of spatial differentiation, let us
next combine the output from these simple cells into a set of
idealized models of velocity-tuned complex cells.

Appendix~\ref{app-anal-compl-cells} describes the main steps in
performing such an analysis for idealized models of velocity-tuned complex cells
of the forms
(\ref{eq-quasi-quad-dir-vel-adapt-spat-temp-12-int})
and
(\ref{eq-quasi-quad-dir-vel-adapt-spat-temp-34-int}),
based on
spatial derivatives of orders $\{ 1, 2 \}$ and $\{ 3, 4 \}$, respectively,
Using this methodology, explicit expressions can be obtained for
joint direction and speed selectivity properties for both these types
of models of complex cells. These results are, however,
rather complex, as shown in the explicit expressions for the
results in Equations~(\ref{eq-A-Q12-int-compl}) and
(\ref{eq-A-Q34-int-compl}) in
Figure~\ref{fig-eq-ampl-meas-int-compl-cells-12-34}
in Appendix~\ref{app-anal-compl-cells}.

In the special case when the inclination angle $\theta = 0$, and we
parameterize the speed parameters $|u|$ and $|v|$ according to
$|u| = r \, |v|$, the resulting dependencies on the 
parameter $r$ are, however, in a simpler form, as shown in
Equations~(\ref{eq-speed-sel-curve-compl-12})
and (\ref{eq-speed-sel-curve-compl-34}) and
Figure~\ref{fig-alg-expr-R-curves-int-compl-cells}.

Figure~\ref{fig-dir-sel-complex-veladapt-anal-theta-0} shows
corresponding graphs of the speed selectivity curves
$R_{{\cal Q}_{12}}(r)$ according to
(\ref{eq-speed-sel-curve-compl-12}) and
$R_{{\cal Q}_{34}}(r)$ according to
(\ref{eq-speed-sel-curve-compl-34}) for the inclination angle
$\theta = 0$. As can be seen from these results, the speed selectivity
decreases more abruptly towards larger relative speed factors $r$ with
increasing order of spatial differentiation in the idealized models of
the spatio-temporal receptive fields.

Figure~\ref{fig-dir-sel-complex-const-u} shows examples of
direction selectivity curves that arise from the explicit
expressions that we have derived for the  composed amplitude measures
$A_{{\cal Q}_{12},\max}(\theta, r;\; \kappa)$ and
$A_{{\cal Q}_{34},\max}(\theta, r;\; \kappa)$, in the special case
when the ratio $r = |u|/|v|$ between the speed parameter $|u|$ of the motion
stimulus and the speed parameter $|v|$ of the spatio-temporal receptive
field is equal to 1.
As can be seen from these results, the direction selectivity
properties do, as for the previous analysis of simple cells,
become significantly sharper with increasing orders of spatial
differentiation as well as with increasing degree of elongation
$\kappa$ of the receptive fields.
Additionally, the size of the lobe in opposite direction to the
preferred direction of the receptive field becomes smaller
with increasing order of spatial differentiation.

Again, similarly to the previous treatment for simple cells in
the previous section, according to the most idealized version
of the normative theory for visual receptive fields, there should
be complex cells for all combinations of the shape parameters
of the receptive fields in terms of the direction and the speed
parameters as well as with regard to the degree of elongation of the receptive field.
This, in turn, implies that there should in addition to a variability
over the order $m$ of spatial differentiation be corresponding
variabilities in the shapes of the direction and speed selectivity
curves for all combinations of
(i)~the inclination angle between the motion direction of the
stimulus and the preferred direction of the spatio-temporal receptive field,
(ii)~the ratio $r = |u|/|v|$ between the speed parameters of
the stimulus and the receptive field, and
(iii)~the degree of elongation $\kappa$ of the receptive field.

From this perspective,
Figure~\ref{fig-dir-sel-complex-const-u}
illustrates a subset of the variabilities in the shapes of speed and direction
selectivity curves that we could expect from an idealized
vision system according to the underlying normative theory
of visual receptive fields.

Appendix~\ref{app-cart-dir-sel-plots} provides corresponding
Cartesian, as opposed to polar, plots of the idealized models
for direction selectivity curves for simple cells and complex cells.

\begin{figure*}[hbtp]
%    \begin{center}
%      \begin{tabular}{cccc}
%        \includegraphics[width=0.20\textwidth]{Orban86-fig6A.png}
%        & \includegraphics[width=0.22\textwidth]{diradaptphi-rcurve-loglin-eps-converted-to.pdf} 
%        & \includegraphics[width=0.20\textwidth]{Orban86-fig6B.png}
%        & \includegraphics[width=0.22\textwidth]{diradaptphiphi-rcurve-loglin-eps-converted-to.pdf}\\
%        \includegraphics[width=0.20\textwidth]{Orban86-fig3.png}
%        & \includegraphics[width=0.22\textwidth]{intquasi12-rcurve-loglin-eps-converted-to.pdf}  
%        & \includegraphics[width=0.20\textwidth]{Orban86-fig6C.png}
%        & \includegraphics[width=0.22\textwidth]{intquasi34-rcurve-loglin-eps-converted-to.pdf} \\
%      \end{tabular}
%    \end{center}
  \begin{center}
    \begin{subfigure}{0.48\textwidth}
      \begin{tabular}{cc}
        {\em central neuron (6A)} & {\em first-order simple cell $(|v| = 1.4)$} \\ \smallskip \\
        \includegraphics[width=0.40\textwidth]{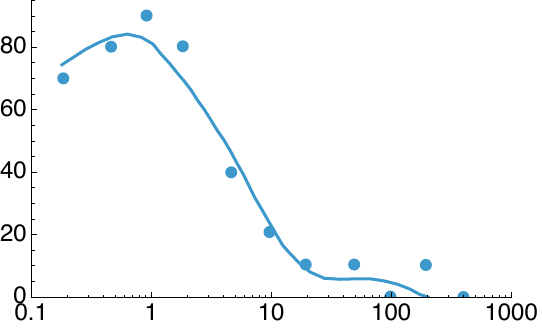}
       & \includegraphics[width=0.40\textwidth]{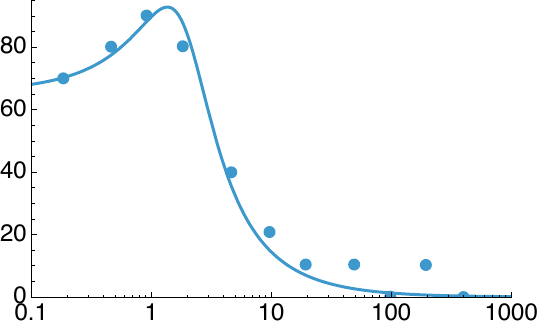}
      \end{tabular}
      \caption{Modelling of neuron 6A}
    \end{subfigure}
    \hfill
    \begin{subfigure}{0.48\textwidth}
      \begin{tabular}{cc}
        {\em central neuron (6B)} & {\em second-order simple cell $(|v| = 3.1)$} \\ \smallskip \\
        \includegraphics[width=0.40\textwidth]{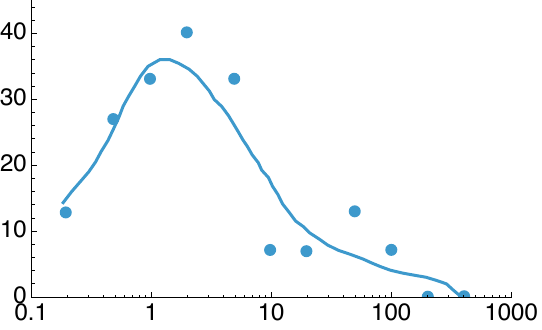}
      & \includegraphics[width=0.40\textwidth]{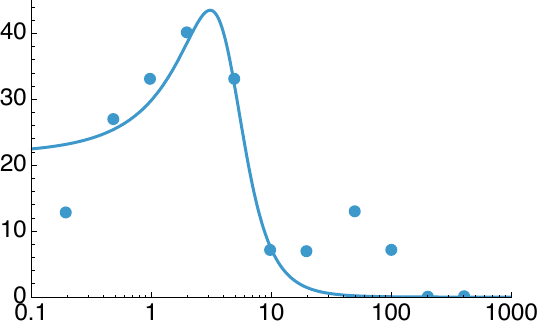}
      \end{tabular}
      \caption{Modelling of neuron 6B}
    \end{subfigure}
  \end{center}
  \medskip
  \begin{center}
    \begin{subfigure}{0.48\textwidth}
      \begin{tabular}{cc}
        {\em peripheral neuron (3)} & {\em complex cell of order $\{ 1, 2 \}$ $(|v|  = 5)$} \\
        \smallskip \\
        \includegraphics[width=0.40\textwidth]{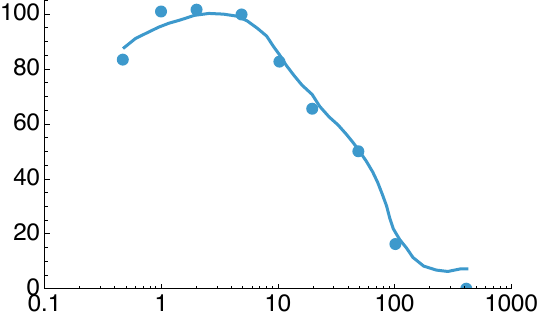}
        & \includegraphics[width=0.40\textwidth]{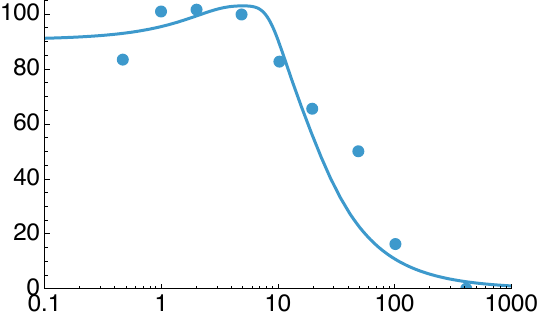}
      \end{tabular}
      \caption{Modelling of neuron 3}
    \end{subfigure}
    \begin{subfigure}{0.48\textwidth}
      \begin{tabular}{cc}
         {\em peripheral neuron (6C)} & {\em complex cell of order $\{ 3, 4 \}$ $(|v| = 48)$} \\ \smallskip \\
        \includegraphics[width=0.40\textwidth]{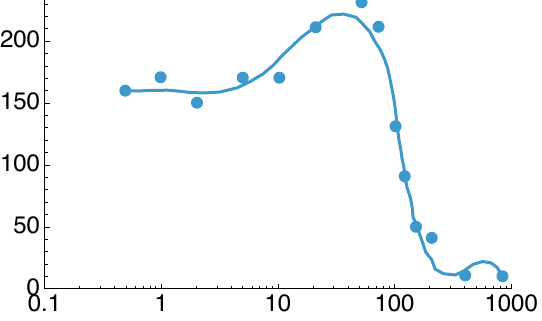}
       & \includegraphics[width=0.40\textwidth]{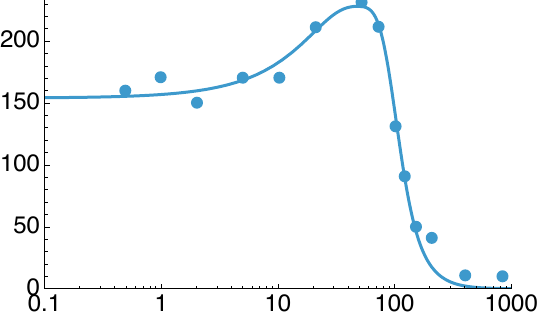} 
      \end{tabular}
      \caption{Modelling of neuron 6C}
    \end{subfigure}            
  \end{center}
%   \begin{center}
%     \begin{tabular}{cccc}
%       {\em central neuron (6A)} & {\em first-order simple cell $(|v| = 1.4)$}
%       & {\em central neuron (6B)} & {\em second-order simple cell $(|v| = 3.1)$} \\ \smallskip \\
%       \includegraphics[width=0.20\textwidth]{Orban86-fig6A-curve+points-eps-converted-to.pdf}
%       & \includegraphics[width=0.20\textwidth]{Orban86-fig6A-model+points-eps-converted-to.pdf} 
%       & \includegraphics[width=0.20\textwidth]{Orban86-fig6B-curve+points-eps-converted-to.pdf}
%       &
%         \includegraphics[width=0.20\textwidth]{Orban86-fig6B-model+points-eps-converted-to.pdf}\\
%       \\ \medskip \\ 
%       {\em peripheral neuron (3)} & {\em complex cell of order $\{ 1, 2 \}$ $(|v|  = 5)$}
%       & {\em peripheral neuron (6C)} & {\em complex cell of order $\{ 3, 4 \}$ $(|v| = 48)$} \\ \smallskip \\
%       \includegraphics[width=0.20\textwidth]{Orban86-fig3-curve+points-eps-converted-to.pdf}
%       & \includegraphics[width=0.20\textwidth]{Orban86-fig3-model+points-eps-converted-to.pdf}  
%       & \includegraphics[width=0.20\textwidth]{Orban86-fig6C-curve+points-eps-converted-to.pdf}
%       & \includegraphics[width=0.20\textwidth]{Orban86-fig6C-model+points-eps-converted-to.pdf} \\
%     \end{tabular}
%   \end{center}
   \caption{Modelling of 4 speed selectivity curves from
     Orban {\em et al.\/}\ (\citeyear{OrbKenNul86-JNeurPhys}) based on
     idealized models of simple cells and complex cells.
    {\bf (a)} A central neuron from Figure~6A modelled as a {\em first-order simple
      cell\/} of the form $A \times R_{\varphi}(r)$ according to (\ref{eq-rcurve-order-all})
    with $m = 1$ for image speed $|v| = 1.4~\mbox{deg/s}$ and amplitude $A = 93$.
    {\bf (b)} A central neuron from Figure~6B modelled as a {\em second-order simple
      cell} of the form $A \times R_{\varphi\varphi}(r)$ according to (\ref{eq-rcurve-order-all})
    with $m = 2$ for image speed $|v| = 3.1~\mbox{deg/s}$ and amplitude $A = 44$.
    {\bf (c)} A peripheral neuron from Figure~3 modelled as a {\em complex cell
      based on first- and second-order spatial derivative responses\/}
    of the form $A \times R_{{\cal Q}_{12}}(r)$ according to (\ref{eq-speed-sel-curve-compl-12})
    for image speed $|v| = 5~\mbox{deg/s}$ and amplitude $A = 103$.
    {\bf (d)} A peripheral neuron from Figure~6C modelled as a {\em complex cell
      based on third- and fourth-order spatial derivative responses\/}
    of the form $A \times R_{{\cal Q}_{34}}(r)$ according to (\ref{eq-speed-sel-curve-compl-34})
    for image speed $|v| = 48~\mbox{deg/s}$ and amplitude $A = 228$.
    For each model, the parameter $r = |u|/|v|$ is the ratio between
    the speed $|u|$ of the stimulus and the speed parameter $|v|$ in the model.
    For each pair of figures, that is for each biological neuron,
    the left subfigure shows the actual measurements
    indicated by points together with a stylized reproduction of the
    the continuous curve drawn by
    Orban {\em et al.\/}\ (\citeyear{OrbKenNul86-JNeurPhys}).
    The right subfigure does in turn show the corresponding idealized
    model fit to the neurophysiological data, again visualized as points.
    From comparisons between these results, we can
    clearly note qualitative similarities between the idealized models
    and the neurophysiological measurements. Please, note, 
    however, that the experimental conditions for obtaining the curves
    are not identical between the idealized models and the
    neurophysiological measurements. The neurophysiological
    measurements have been performed with moving bars while the
    idealized models have been derived from moving sine waves, to enable
    a closed-form theoretical analysis.
    Nevertheless, the qualitative shapes of the fitted idealized
    models agree very well with the qualitative shapes of the
    neurophysiologically recorded speed selectivity curves.
    All figure numbers refer to the figures in Orban {\em et al.\/}\
    (\citeyear{OrbKenNul86-JNeurPhys}) from which the data about the
    neurophysiological measurements have been extracted.
    (Horizontal axes: image speed $|u| \in [0.1, 1000]$ of the
    stimulus in degrees per second.
    Vertical axes: response in spikes per second for the biological
    measurements while showing the rescaled speed selectivity
    functions $R$ multiplied by a complementary amplitude factor
    $A$ for the modelled results.)}
  \label{fig-model-orban-speed-sel-curves}
\end{figure*}

\begin{figure}[hbtp]
  \begin{center}
    \begin{tabular}{cc}
      {\em central neuron (4)} & {\em first-order simple cell $(|v| = 0.68)$} \\
      \\ \smallskip \\
          \includegraphics[width=0.20\textwidth]{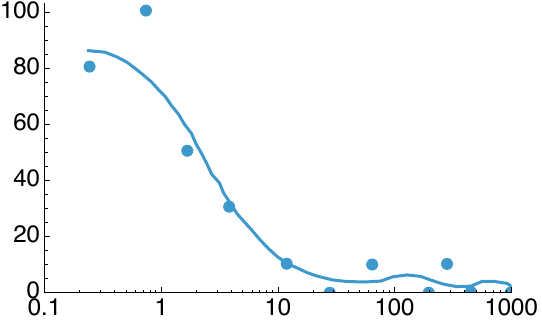}
      &
        \includegraphics[width=0.20\textwidth]{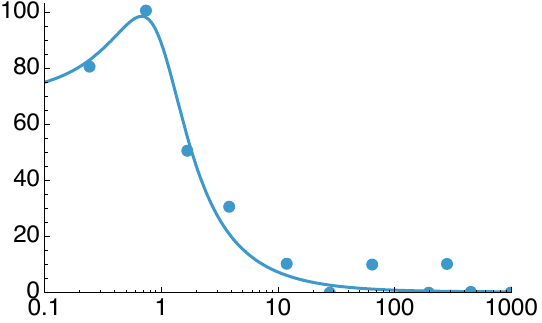}
    \end{tabular}
  \end{center}
  \caption{Illustration of the ability to obtain qualitatively different
    interpretations of the neural measurements by
    Orban {\em et al.\/}\ (\citeyear{OrbKenNul86-JNeurPhys})
    as enabled by the theory presented in this paper:
    {\bf (left)} Original biological measurements of a central neuron
    in V1 together with a stylized reproduction of the
    continuous curve drawn by Orban {\em et al.\/}\
    (\citeyear{OrbKenNul86-JNeurPhys}) in Figure~4 in their original paper.
    {\bf (right)} The result of fitting the idealized model
    corresponding to a first-order simple cell of the form $A \times R_{\varphi}(r)$
    according to (\ref{eq-rcurve-order-all}) with $m = 1$ for image speed
    $|v| = 0.68~\mbox{deg/s}$ and amplitude $A = 98$ to the same measurements.
    The parameter $r = |u|/|v|$ is the ratio between
    the speed $|u|$ of the stimulus and the speed parameter $|v|$ in the model.
    As can be seen from these graphs,
    Orban {\em et al.\/}\ (\citeyear{OrbKenNul86-JNeurPhys})
    interpreted this neuron as having an overall decreasing
    response with increasing image speed $|u|$ of the stimulus, thereby
    essentially ignoring the peak response over the image speeds
    in the second measurement point from the left.
    By fitting the idealized model corresponding to a first-order
    simple cell to the same data, with its prior shape in terms of
    sharp well-localized peaks on a logarithmic speed scale,
    we do on the other hand obtain a clear peak over the
    image speeds for $|v| \approx 0.68~\mbox{deg/s}$.
    (Horizontal axes: image speed $|u| \in [0.1, 1000]$ of the
    stimulus in degrees per second.
    Vertical axes: response in spikes per second for the biological
    measurements while showing rescaled speed selectivity
    function $R$ multiplied by a complementary amplitude factor
    $A$ for the modelled results.)}
  \label{fig-model-Orban-fig4}
\end{figure}

\begin{figure*}[hbtp]
  \begin{center}
    \begin{tabular}{ccc}
      {\em $A_{\varphi}(\theta - \theta_0;\; \kappa)$ for $\theta_0 = 42^{\circ}$, $\kappa = 8$} 
      &  {\em $A_{\varphi}(\theta - \theta_0;\; \kappa)$ for $\theta_0 = 11^{\circ}$, $\kappa = 4$} 
      &  {\em $A_{\varphi\varphi}(\theta - \theta_0;\; \kappa)$ for $\theta_0 = 189^{\circ}$, $\kappa = 4$} \\
      \includegraphics[height=0.30\textwidth]{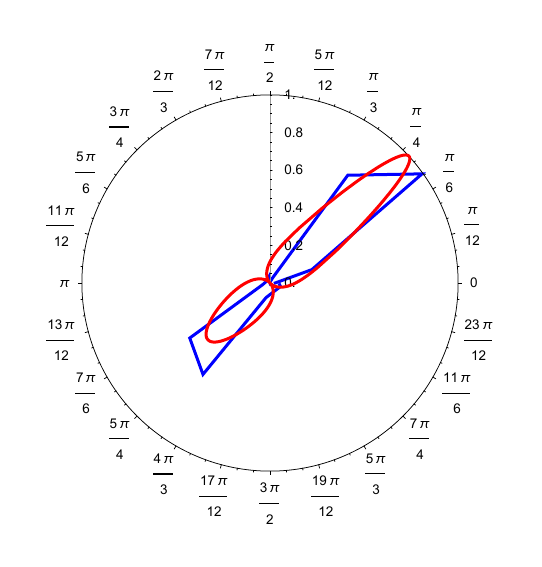}
      &  \includegraphics[height=0.30\textwidth]{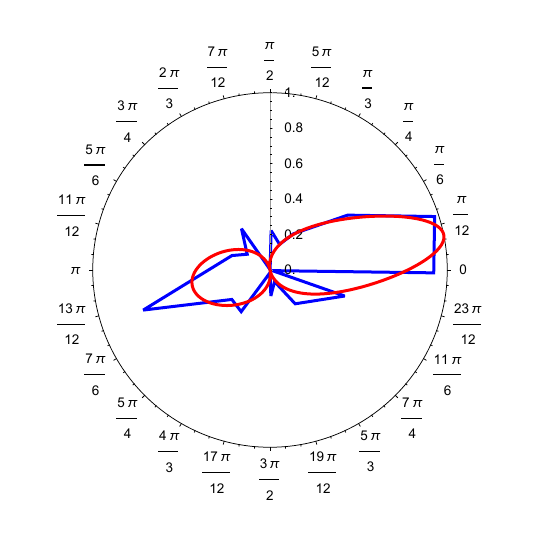}
      &  \includegraphics[height=0.30\textwidth]{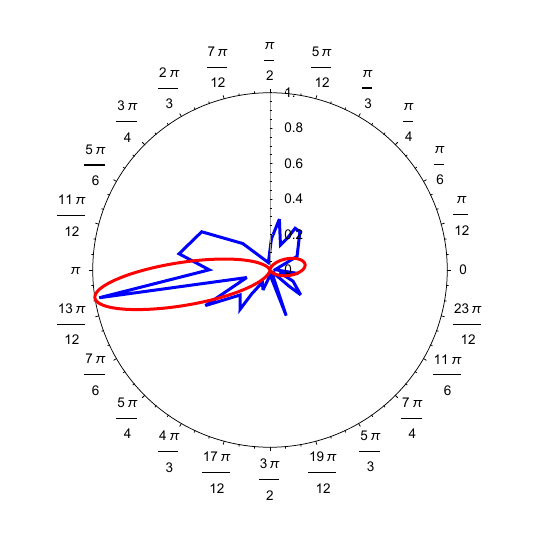} \\
      {\em $A_{\varphi\varphi}(\theta - \theta_0;\; \kappa)$ for $\theta_0 = 113^{\circ}$, $\kappa = 2$} 
      &  {\em $A_{\varphi}(\theta - \theta_0;\; \kappa)$ for $\theta_0 = 338^{\circ}$, $\kappa = 8$} 
      &  {\em $A_{\varphi\varphi}(\theta - \theta_0;\; \kappa)$ for $\theta_0 = 186^{\circ}$, $\kappa = 5$} \\
      \includegraphics[height=0.30\textwidth]{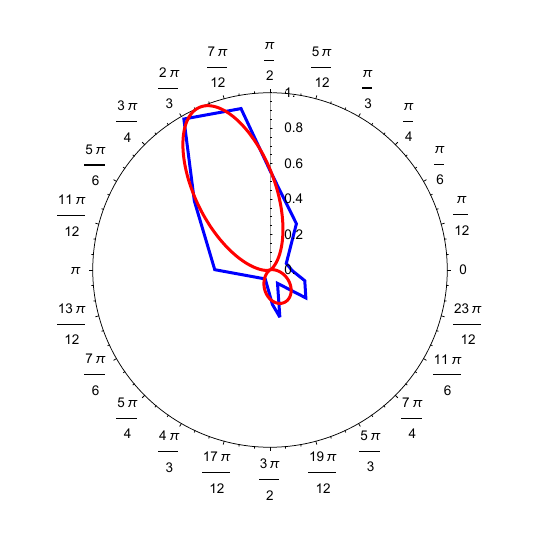}
      &  \includegraphics[height=0.30\textwidth]{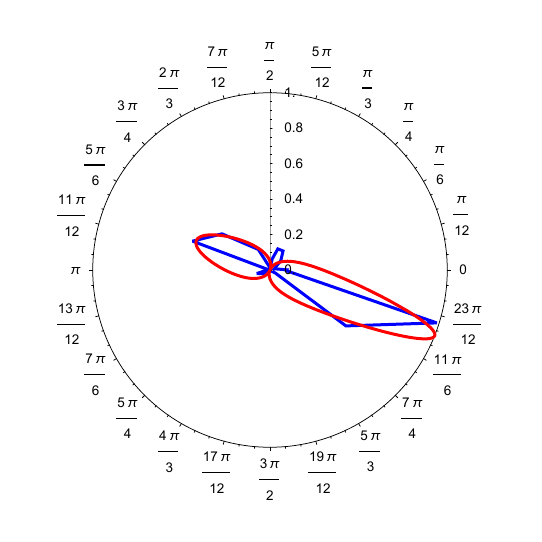}
      &
        \includegraphics[height=0.30\textwidth]{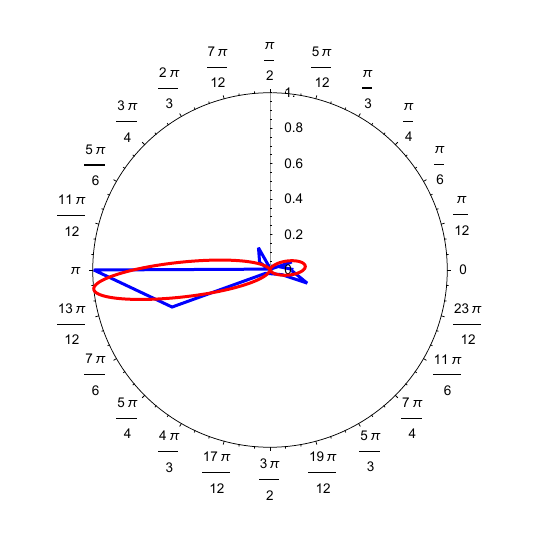}
    \end{tabular}
  \end{center}
  \caption{Modelling of direction selectivity curves for 6 {\em simple cells\/} recorded
    by Livingstone and Conway (\citeyear{LivCon03-JNeuroPhys}) (shown
    as blue data) using idealized models of direction selectivity curves for
    velocity-tuned simple cells of the form
    $A_{\varphi^m}(\theta;\; \kappa)  = A_{\varphi^m,\max}(\theta, |u|;\; \kappa, |v|)$
    for $r = |u|/|v| = 1$ and with $A_{\varphi^m,\max}(\theta, |u|;\; \kappa, |v|)$
    according to (\ref{eq-cond-summ-dir-sel-prop})
    %according to Equation~(\ref{eq-cond-summ-dir-sel-prop})
    (shown as red curves).
    {\bf (top left)} First-order simple cell with $m = 1$ for preferred direction
    $\theta_0 = 42^{\circ}$ and elongation $\kappa = 8$.
    {\bf (top middle)} First-order simple cell with $m = 1$ for preferred direction
    $\theta_0 = 11^{\circ}$ and elongation $\kappa = 4$.
    {\bf (top right}) Second-order simple cell with $m = 2$ for preferred direction
    $\theta_0 = 189^{\circ}$ and elongation $\kappa = 4$.
    {\bf (bottom left)} Second-order simple cell with $m = 2$ for preferred
    direction $\theta_0 = 113^{\circ}$ and elongation $\kappa = 2$.
    {\bf (bottom middle)} First-order simple cell with $m = 1$ for preferred
    direction $\theta_0 = 338^{\circ}$ and elongation $\kappa = 8$.
    {\bf (bottom right}) Second-order simple cell with $m = 2$ for preferred
    direction $\theta_0 = 186^{\circ}$ and elongation $\kappa = 5$.
    (In the polar plots, the increments of $\pi/12$ radians on the circle
    corresponds to increments by $15^{\circ}$ degrees.)}
  \label{fig-model-Liv03-simple}
\end{figure*}

\begin{figure*}[hbtp]
  \begin{center}
    \begin{tabular}{ccc}
      {\em $A_{{\cal Q}_{34}}(\theta - \theta_0;\; \kappa)$ for $\theta_0 = 195^{\circ}$, $\kappa = 0.8$} 
      &  {\em $A_{{\cal Q}_{34}}(\theta - \theta_0;\; \kappa)$ for $\theta_0 = 192^{\circ}$, $\kappa = 0.65$} 
      &  {\em $A_{{\cal Q}_{34}}(\theta - \theta_0;\; \kappa)$ for $\theta_0 = 70^{\circ}$, $\kappa = 0.16$} \\
      \includegraphics[height=0.30\textwidth]{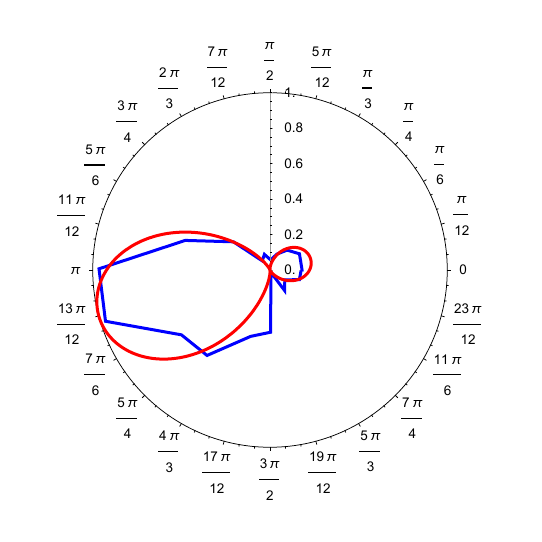}
      &  \includegraphics[height=0.30\textwidth]{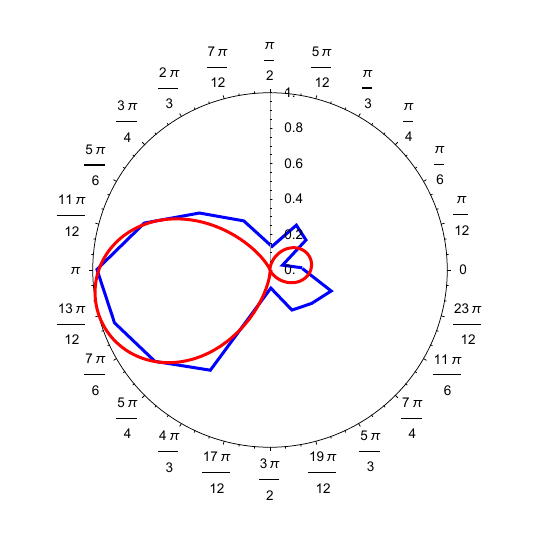}
      &  \includegraphics[height=0.30\textwidth]{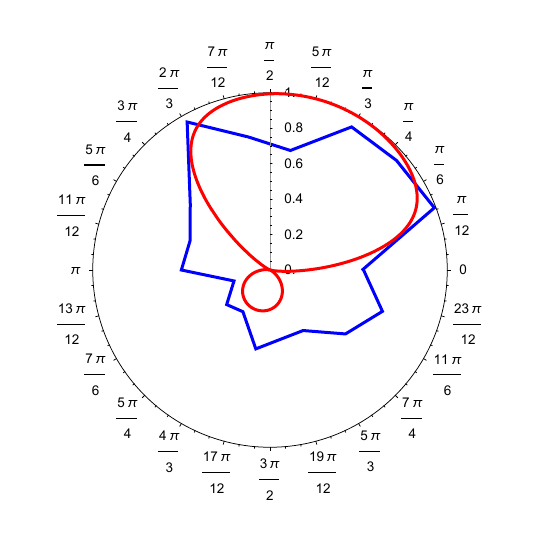} \\
      {\em $A_{{\cal Q}_{34}}(\theta - \theta_0;\; \kappa)$ for $\theta_0 = 350^{\circ}$, $\kappa = 0.1$} 
      &  {\em $A_{{\cal Q}_{34}}(\theta - \theta_0;\; \kappa)$ for $\theta_0 = 344^{\circ}$, $\kappa = 0.7$} 
      &  {\em $A_{{\cal Q}_{34}}(\theta - \theta_0;\; \kappa)$ for $\theta_0 = 193^{\circ}$, $\kappa = 1.2$} \\
      \includegraphics[height=0.30\textwidth]{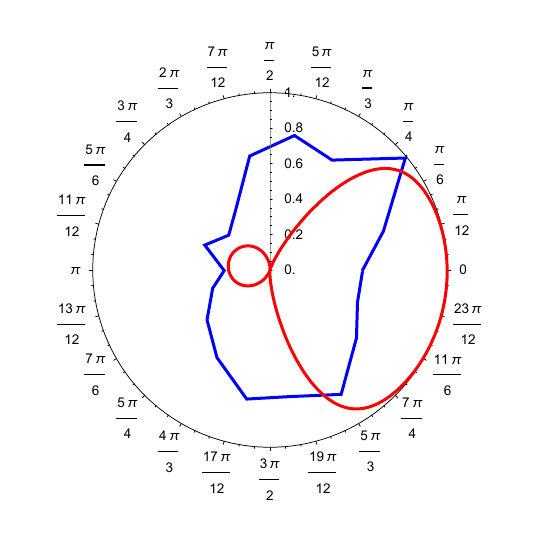}
      &  \includegraphics[height=0.30\textwidth]{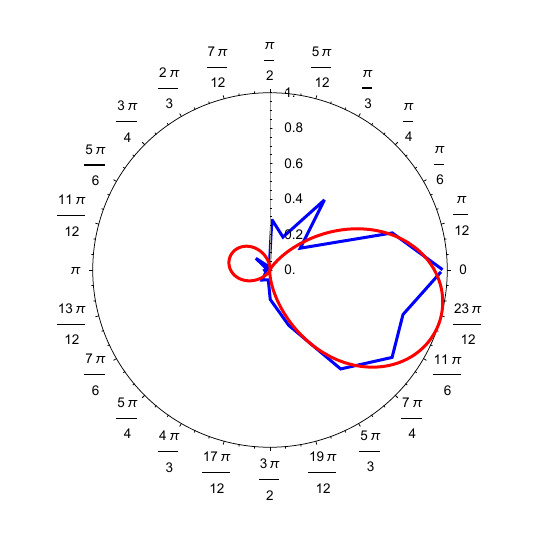}
      &
        \includegraphics[height=0.30\textwidth]{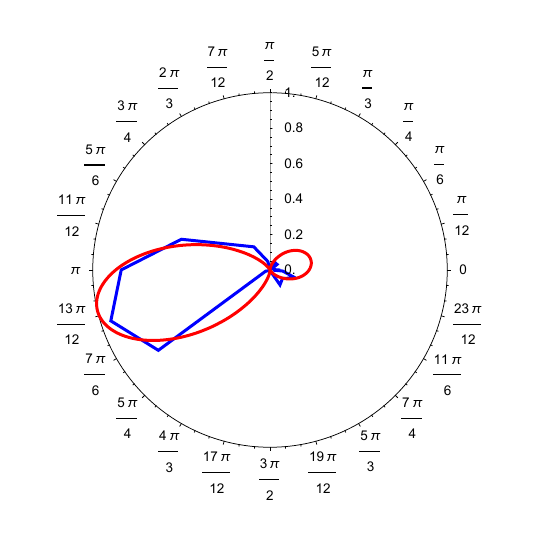}
      \\
       {\em $A_{{\cal Q}_{34}}(\theta - \theta_0;\; \kappa)$ for $\theta_0 = 241^{\circ}$, $\kappa = 2$} 
      &  {\em $A_{{\cal Q}_{34}}(\theta - \theta_0;\; \kappa)$ for $\theta_0 = 281^{\circ}$, $\kappa = 5$} 
      &  {\em $A_{{\cal Q}_{34}}(\theta - \theta_0;\; \kappa)$ for $\theta_0 = 60^{\circ}$, $\kappa = 2.5$} \\
      \includegraphics[height=0.30\textwidth]{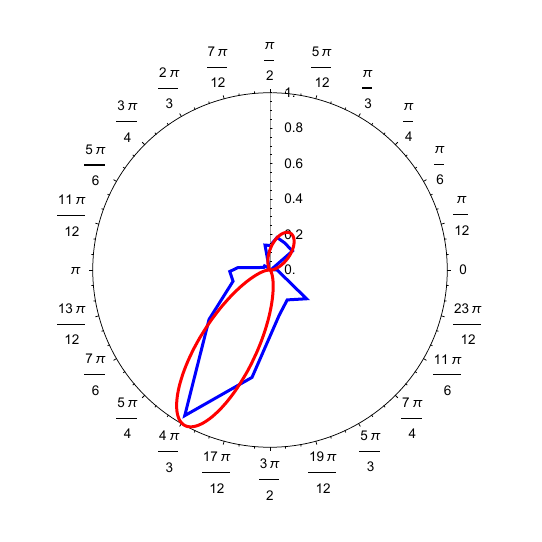}
      &  \includegraphics[height=0.30\textwidth]{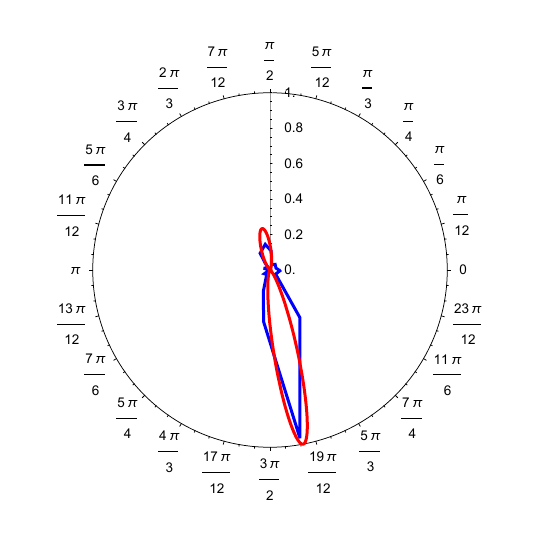}
      &
        \includegraphics[height=0.30\textwidth]{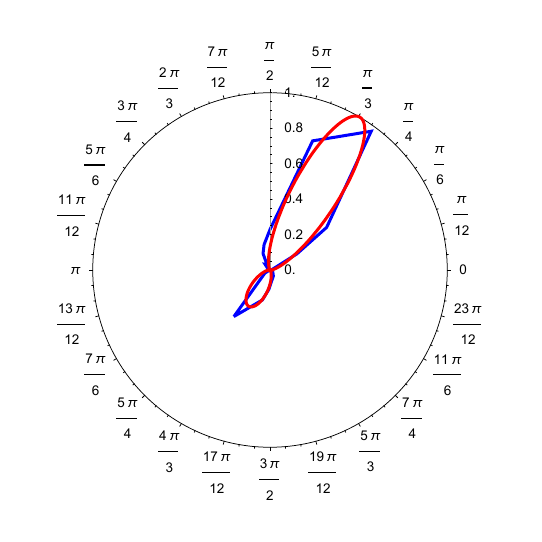}
   \end{tabular}
  \end{center}
  \caption{Modelling of direction selectivity curves for 9 {\em complex cells\/} recorded
    by Livingstone and Conway (\citeyear{LivCon03-JNeuroPhys}) (shown
    as blue data)
    using idealized models of direction selectivity curves for
    velocity-tuned complex cells of the form
    $A_{{\cal Q}_{34}}(\theta;\; \kappa) = A_{{\cal Q}_{34},\max}(\theta, r;\; \kappa)$
    for $r = |u|/|v| = 1$ and with $A_{{\cal Q}_{34},\max}(\theta, r;\; \kappa)$
    according to Equation~(\ref{eq-A-Q34-int-compl}) in
    Figure~\ref{fig-eq-ampl-meas-int-compl-cells-12-34} (shown as red curves).
    {\bf (top left)} Complex cell for preferred direction $\theta_0 = 195^{\circ}$
    and elongation $\kappa = 0.8$.
    {\bf (top middle)} Complex cell for preferred direction $\theta_0 = 192^{\circ}$
    and elongation $\kappa = 0.65$.
    {\bf (top right)} Complex cell for preferred direction $\theta_0 = 70^{\circ}$
    and elongation $\kappa = 0.16$.
    {\bf (middle left)} Complex cell for preferred
    direction $\theta_0 = 350^{\circ}$ and elongation $\kappa = 0.1$.
    {\bf (middle middle)} Complex cell for preferred
    direction $\theta_0 = 344^{\circ}$ and elongation $\kappa = 0.7$.
    {\bf (middle right)} Complex cell for preferred
    direction $\theta_0 = 193^{\circ}$ and elongation $\kappa = 1.2$.
    {\bf (bottom left)} Complex cell for preferred
    direction $\theta_0 = 241^{\circ}$ and elongation $\kappa = 2$.
    {\bf (bottom middle)} Complex cell for preferred
    direction $\theta_0 = 281^{\circ}$ and elongation $\kappa = 5$.
    {\bf (bottom right)} Complex cell for preferred
    direction $\theta_0 = 60^{\circ}$ and elongation $\kappa = 2.5$.
    (In the polar plots, the increments of $\pi/12$ radians on the circle
    corresponds to increments by $15^{\circ}$ degrees.)}
  \label{fig-model-Liv03-complex}
\end{figure*}

\section{Relations to direction and speed selectivity properties of biological neurons}
\label{sec-rels-biol-vision}

In this section, we will relate the above derived theoretical relationships
of the direction selectivity properties of spatio-temporal receptive
fields according to the generalized Gaussian derivative model for
visual receptive fields to available neurophysiological recordings of
motion-sensitive neurons in the primary visual cortex.

\subsection{Relations to neurophysiological measurements of
  motion-sensitive neurons in the primary visual cortex}
\label{sec-rels-neurophys-Orban}
 
In their in-depth analysis of speed sensitivity and direction
selectivity properties of neurons in the primary visual cortex (V1)
in monkeys,
Orban {\em et al.\/}\ (\citeyear{OrbKenNul86-JNeurPhys}) found that%
\footnote{The terminology ``speed'' that we use in this paper means
  the same physical entity as Orban {\em et al.\/}\ (\citeyear{OrbKenNul86-JNeurPhys}) 
  refer to as ``velocity''. The reason why we use speed in this paper,
  is that according to the convention in physics, the notion of
  ``speed'' is without any direction, while the term ``velocity''
  comprises both the speed and the direction of motion.}
  \begin{itemize}
    
\item
  ``Velocity sensitivity of V1 neurons shifts
  to faster velocities with increasing eccentricity.''
\item
  ``All neurons could be classified into three
  categories according to their velocity-response
  curves: velocity low pass, velocity broad band,
  and velocity tuned.''
\item
  ``There is a significant correlation between
  velocity upper cutoff and receptive field width
  among V1 neurons. The change in upper cutoff
  velocity with eccentricity depends both on
  temporal and spatial factors.''
\end{itemize}
Since our way of probing the idealized models of simple cells and
complex cells by sine waves differs from the way that above
neurophysiological results have been obtained using moving bars, we
cannot expect to be able to directly compare%
\footnote{To perform more quantitative comparisons to the neurophysiological
  results, it would furthermore be necessary to get access to raw data for
  a larger population of visual neurons, which we do not have any access to.}
our direction and speed
selectivity curves to the direction and speed selectivity curves
recorded by Orban {\em et al.\/}\ (\citeyear{OrbKenNul86-JNeurPhys}).
We can, however, note that the idealized models of simple cells
in Section~\ref{sec-model-vel-tuned-simple-cells} and that our
idealized models of complex cells in
Section~\ref{sec-model-vel-tuned-complex-cells}, both based on
velocity-adapted Gaussian derivative operators, have qualitative
similarities to the class of velocity-tuned neurons in V1
characterized by Orban {\em et al.\/}\
(\citeyear{OrbKenNul86-JNeurPhys}), in the sense of being selective to
specific directions and speeds of the input stimulus.

\subsubsection{Modelling of speed selectivity curves for
  motion-sensitive neurons in the primary visual cortex recorded by
  Orban et al.\ (\citeyear{OrbKenNul86-JNeurPhys})}
\label{sec-rels-neurophys-spec-Orban}

Figure~\ref{fig-model-orban-speed-sel-curves} shows the result of
modelling the speed selectivity curves for 4 neurons reported
Orban {\em et al.\/}\ (\citeyear{OrbKenNul86-JNeurPhys})
in terms of the idealized models of simple cells and complex cells
as derived in Sections~\ref{sec-dir-sel-simpl-cells}
and~\ref{sec-dir-sel-compl-cells}.
The top row shows two central%
\footnote{According to the terminology in
  Orban {\em et al.\/}\ (\citeyear{OrbKenNul86-JNeurPhys}),
  ``central neurons'' were within $[0^{\circ},2^{\circ}]$ whereas
  ``peripheral neurons'' were within $[15^{\circ}, 25^{\circ}]$.}
neurons that modelled in terms of
velocity-tuned simple cells of orders 1 and 2,
whereas the bottom row shows two peripheral neurons modelled in
terms of velocity-tuned complex cells based on derivatives of
orders $\{1, 2 \}$ or $\{ 3, 4 \}$.

For the central neurons 6A and 6B and for the peripheral neuron 6C,
the two parameters of the models, the amplitude $A$
and the image speed $|v|$, were determined using the
Wolfram Mathematica function FindFit%
\footnote{The Wolfram Mathematica function FindFit is a multi-purpose
  function for model fitting that in its default mode performs
  least-squares fitting using the Levenberg-Marquardt method,
  which interpolates between the Gauss-Newton method and gradient descent,
  with automatic symbolic differentiation with respect to the
  parameters of the model. The FindFit function can, however, also use other numerical
  optimization algorithms and automatically choose between these,
  see the online information at
  \url{https://reference.wolfram.com/language/ref/FindFit.html}
  for further information.}%
\footnote{In practice, concerning the fitting of speed selectivity curves,
  the relative speed ratio $r$ in the idealized models for
  the speed selectivity curves of the form $R(r)$ was for
  each model of the speed selectivity curve replaced
  by its definition $r = |u|/|v|$.
  Then, the model fitting was then performed by inserting the
  discrete values of the speed parameter $|u|$ of the stimulus
  underlying the biological measurements, while the optimization
  was then performed over the speed parameter $|v|$ of the
  underlying spatio-temporal receptive field.}
applied to sampled measurements
extracted from the corresponding figures in the original paper by
Orban {\em et al.\/}\ (\citeyear{OrbKenNul86-JNeurPhys})
using the Webplot digitizer app (https://automeris.io/).
For the peripheral neuron 3, the parameters $A$ and $v$ were
determined manually from correspondingly sampled data from the
underlying original illustration.
(For the idealized models of complex cells, the relative
integration scale $\gamma$ was throughout set to $\gamma = \sqrt{2}$.)
As can be seen from the results, our derived idealized models do
rather well reflect the qualitative shapes of the biologically
measured speed selectivity curves. Specifically, our idealized models
do better reflect the peaks in the neurons 6A, 6B and 6C compared to
the drawn speed selectivity curves in the original paper.

Figure~\ref{fig-model-Orban-fig4} shows corresponding results for a
central neuron reported in Figure~4 in
Orban {\em et al.\/}\ (\citeyear{OrbKenNul86-JNeurPhys}).
By fitting the speed selectivity curve of a velocity-tuned simple cell
of order 1 to the data corresponding to the underlying biological
measurements using the Wolfram Mathematica function FindFit,
we notably obtain a curve model of a qualitatively different type
than the curve drawn by the authors of the original publication.
Notably, Orban {\em et al.\/}\ (\citeyear{OrbKenNul86-JNeurPhys})
ignored the peak of the second speed measurement and drew a curve
that overall decreases with the speed $|u|$ of the motion stimulus.
Using the idealized model of a velocity-tuned simple cell of order 1,
we do on the other hand get a model with qualitatively different
properties, that is with a clear peak over the image speed $|u|$,
which also overall better agrees with the actual underlying biological measurements.

In these respects, the predictions about speed selectivity curves
obtained from the theory presented in
Sections~\ref{sec-dir-sel-simpl-cells}
and~\ref{sec-dir-sel-compl-cells}
are in good agreement with the results of neurophysiological
measurements of motion-sensitive neurons in the primary visual cortex
performed by Orban {\em et al.\/}\ (\citeyear{OrbKenNul86-JNeurPhys}).

For the purpose of the above modelling results, the type of model
(simple cell of order 1, 2, 3 or 4 alternatively complex cell based
on spatial derivatives of orders $\{ 1, 2 \}$ or $\{ 3, 4 \}$) was
chosen based on the overall shape of the variations in the speed
selectivity measurements of the neuron.
Specifically, this determination was based on (i)~the ratio between the
peak value and the value at the leftmost end, as well as on (ii)~the slope
of the flank to the right.

For the idealized models of simple cells,
the ratio between the peak value and the value at the leftmost end
increases with the order of spatial differentiation,
as shown in the bottom row in
Figure~\ref{fig-dir-sel-simple-veladapt-anal-theta-0}.
For the idealized models of complex cells, this ratio is also larger
for the complex cell on spatial derivatives of orders $\{ 3, 4 \}$
than for the complex cell on spatial derivatives of
orders $\{ 1, 2 \}$, as shown in the bottom row in
Figure~\ref{fig-dir-sel-complex-veladapt-anal-theta-0}.
Additionally, the slope of the flank to the right does also gets
steeper with increasing order of spatial differentiation, both for the
idealized models of simple cells and for the idealized models of
complex cells.
These properties together thereby lead us to choose the types of idealized models for
simple cells or complex cells for the neurons 6A, 6B, 3, 6C and 4
as shown in Figures~\ref{fig-model-orban-speed-sel-curves}
and~\ref{fig-model-Orban-fig4}.

Concerning models of visual neurons beyond the velocity-tuned cells
considered as the main topic of this paper,
by choosing the speeds of the motion parameters in the idealized
models of simple cells and complex cells sufficiently low, their
resulting speed selectivity properties can additionally be made to be structurally similar
to the motion-sensitive neurons with speed low pass character,
as described by Orban {\em et al.\/}\
(\citeyear{OrbKenNul86-JNeurPhys}). 
Furthermore, by combining the output from several
such velocity-tuned models of complex
cells over wider ranges of motion speeds, while keeping the direction
of motion the same, it ought to be possible to formulate idealized
models of complex cells with more broadband speed selectivity
properties, as described by Orban {\em et al.\/}\
(\citeyear{OrbKenNul86-JNeurPhys}).

In these ways, we can note good qualitative similarities between
the direction and speed selectivity properties of our idealized models
of simple cells and
the neurophysiological measurements by Orban {\em et al.\/}\
(\citeyear{OrbKenNul86-JNeurPhys}), thus supporting the idea of
expanding theoretical models of motion-sensitive neurons with respect to the
degrees of freedom of local Galilean transformations.

Additionally, our new theoretical models of speed selectivity
curves induce a more theoretically based taxonomy of different
types of velocity-tuned neurons in the primary visual cortex,
as first coarsely categorized into simple cells or complex cells,
and then with a finer categorization in terms of the orders of spatial
differentiation in the corresponding underlying spatio-temporal receptive fields,
here corresponding to a total number of $4+ 2 = 6$ classes.

\subsubsection{Modelling of direction selectivity curves of simple
  cells and complex cells recorded by Livingstone and Conway (\citeyear{LivCon03-JNeuroPhys})}
\label{sec-model-livingstone}

Figures~\ref{fig-model-Liv03-simple}--\ref{fig-model-Liv03-complex}
show the result of modelling of the
direction selectivity curves of biological simple cells and complex
cells in Macaque V1 as recorded
by Livingstone and Conway (\citeyear{LivCon03-JNeuroPhys})
in terms of (i)~our derived direction selectivity curves
$A_{\varphi^m,\max}(\theta, |u|;\; \kappa, |v|)$
for simple cells according to (\ref{eq-cond-summ-dir-sel-prop}),
and (ii)~our derived direction selectivity curves
$A_{{\cal Q}_{12},\max}(\theta, r;\; \kappa)$ and
$A_{{\cal Q}_{34},\max}(\theta, r;\; \kappa)$
for complex cells according to
Equations~(\ref{eq-A-Q12-int-compl})
and~(\ref{eq-A-Q34-int-compl}) in
Figure~\ref{fig-eq-ampl-meas-int-compl-cells-12-34} 
in Appendix~\ref{app-anal-compl-cells}. Here, we have
(i)~determined the order of spatial differentiation from the size of
lobe opposite to the preferred direction of the neuron, and
(ii)~determined the preferred direction $\theta_0$ and the degree of
elongation $\kappa$ in the idealized model of the direction
selectivity curve manually.%
\footnote{Due to the way that the direction selectivity curves are
  presented in the graphs in
  Livingstone and Conway (\citeyear{LivCon03-JNeuroPhys}),
  we have not found it possible to sample the original
  measurement data to a regular uniformly spaced grid over the direction angles
  from those polar plots. Hence, it would not be
  meaningful to aim at determining the parameters of the models
  by an automated model fitting procedure. We do, however, regard the
  results of this manual fitting procedure as sufficient in the
  respect of supporting the claim that
  the shapes of the direction selectivity curves recorded by
  Livingstone and Conway (\citeyear{LivCon03-JNeuroPhys})
  can, for the neurons that reflect the properties of velocity-tuned
  simple cells or velocity-tuned complex cells, be well
  approximated by our idealized models of corresponding direction
  selectivity curves.}

As can be seen from the results, for most of the neurons,
the idealized models of direction selectivity curves,
as derived here based on the normative theory for
visual receptive fields, are able to very well model the qualitative shapes
of the recorded direction selectivity measurements from the biological
neurons. Specifically, the modelling results of the simple cells
comprise both first-order and second-order simple cells,
with the type of model determined by the size of the lobe
in the opposite direction to the preferred direction of the neuron.

Importantly, receptive fields based on first-order or second-order
spatial derivatives reflect different aspects of the spatial
variations in the image data. First-order derivatives respond strongly to
edges, whereas second-order derivatives respond strongly to peaks.
In this respect, first-order and second-order derivatives provide
complementary cues, where both components are necessary to
appropriately characterize the local image structures.
The results of our modelling of direction selectivity curves thereby
demonstrate that the underlying receptive fields reflect both
of these complementary aspects of the local image structures.

\begin{figure*}[hbtp]
  \begin{center}
    \begin{tabular}{cccc}
      {\em $A_{\varphi}(\theta, r;\; \kappa)$ for $\kappa = 1$}
      &  {\em $A_{\varphi}(\theta, r;\; \kappa)$ for $\kappa = 2$}
      &  {\em $A_{\varphi}(\theta, r;\; \kappa)$ for $\kappa = 4$}
      & {\em $A_{\varphi}(\theta, r;\; \kappa)$ for $\kappa = 8$} \\
      \includegraphics[width=0.22\textwidth]{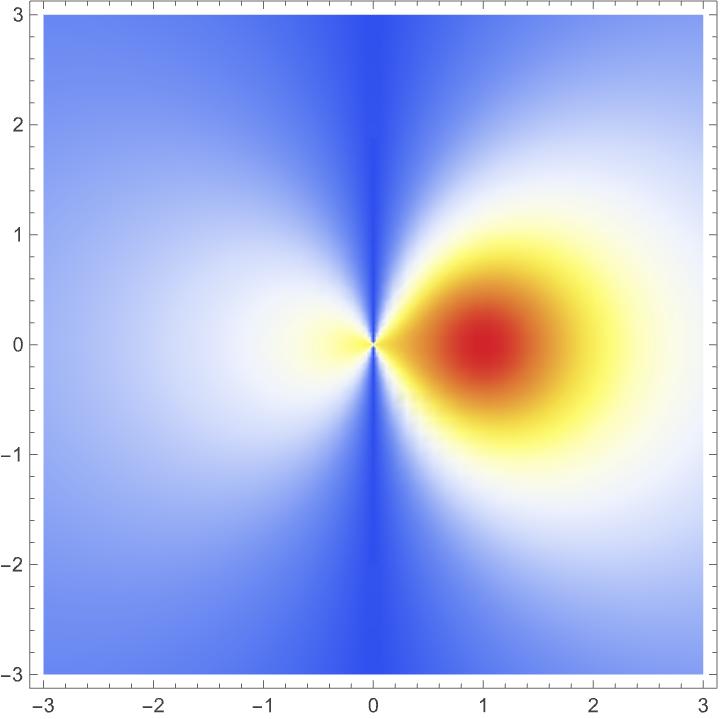}
      & \includegraphics[width=0.22\textwidth]{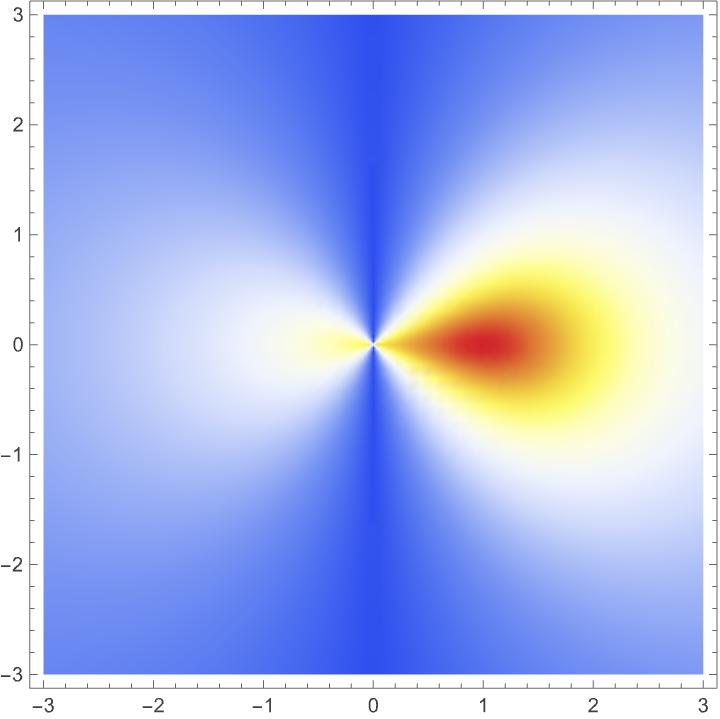}
      & \includegraphics[width=0.22\textwidth]{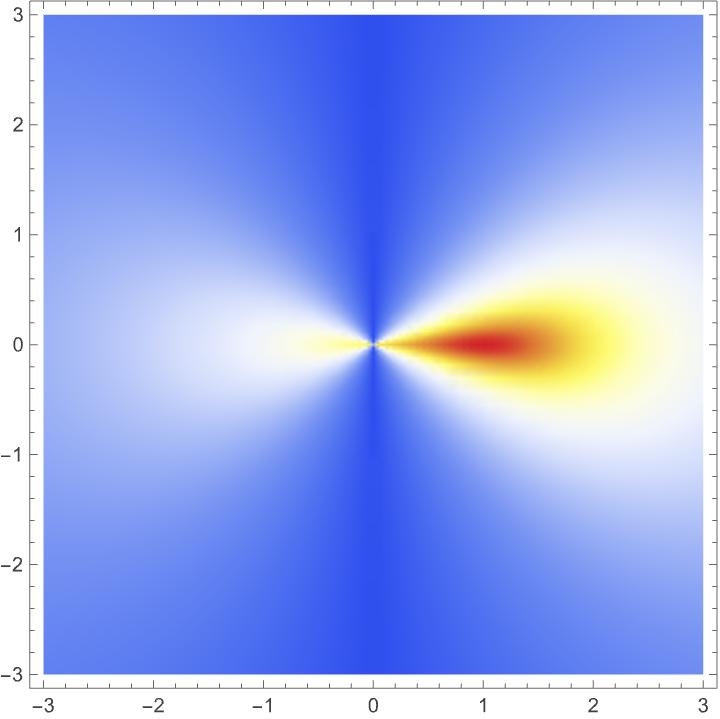}
      & \includegraphics[width=0.22\textwidth]{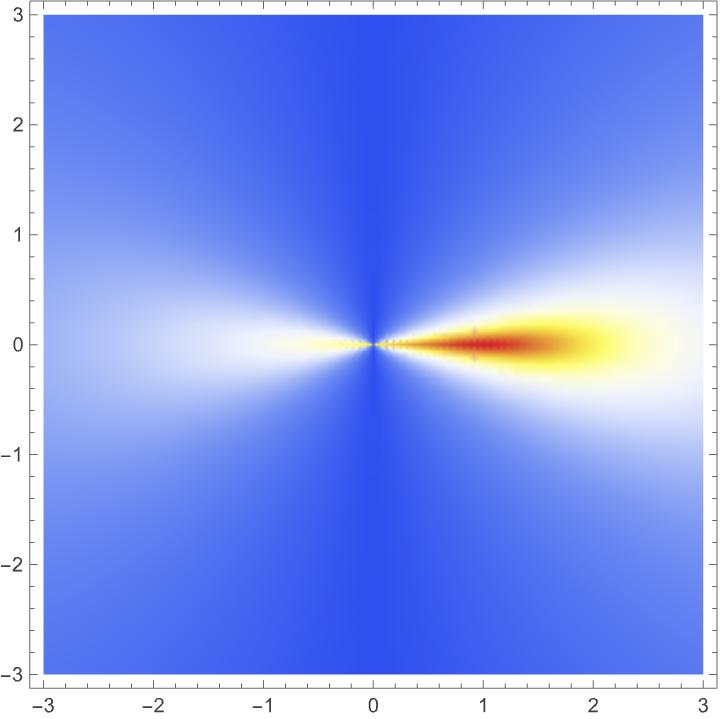} \\
      {\em $A_{\varphi\varphi}(\theta, r;\; \kappa)$ for $\kappa = 1$}
      &  {\em $A_{\varphi\varphi}(\theta, r;\; \kappa)$ for $\kappa = 2$}
      &  {\em $A_{\varphi\varphi}(\theta, r;\; \kappa)$ for $\kappa = 4$}
      & {\em $A_{\varphi\varphi}(\theta, r;\; \kappa)$ for $\kappa = 8$} \\
      \includegraphics[width=0.22\textwidth]{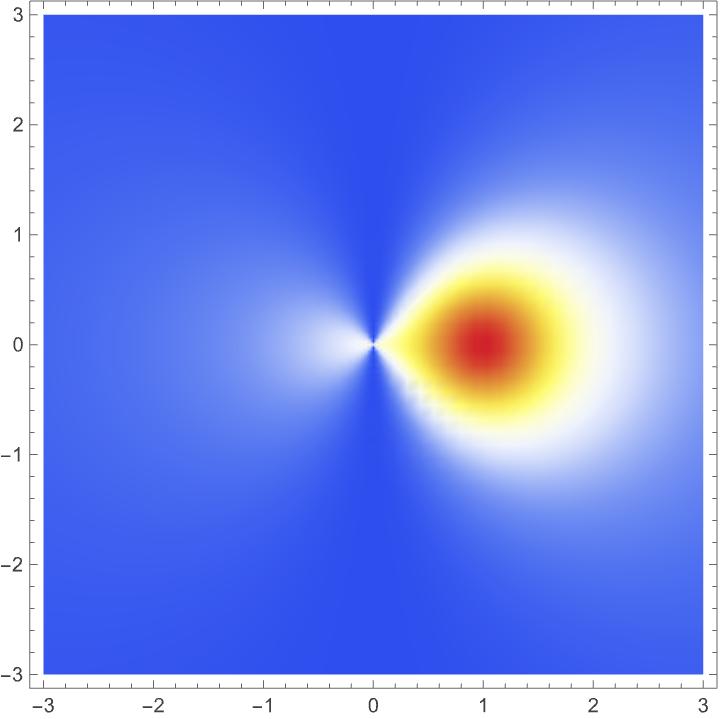}
      & \includegraphics[width=0.22\textwidth]{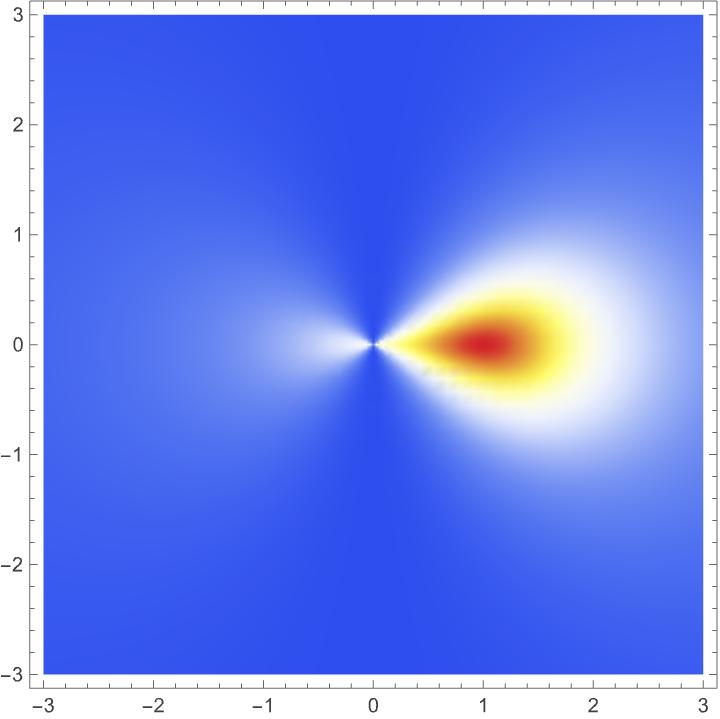}
      & \includegraphics[width=0.22\textwidth]{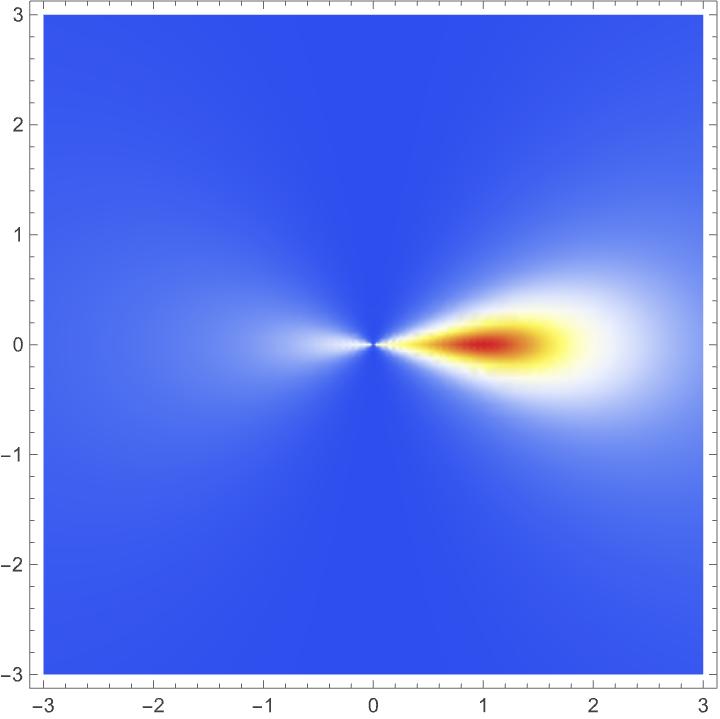}
      & \includegraphics[width=0.22\textwidth]{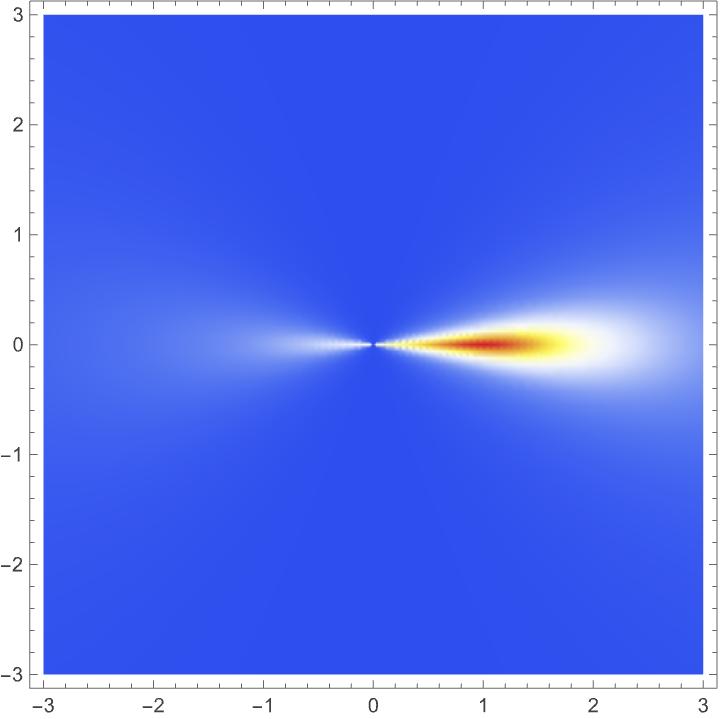} \\
     {\em $A_{\varphi\varphi\varphi}(\theta, r;\; \kappa)$ for $\kappa = 1$}
      &  {\em $A_{\varphi\varphi\varphi}(\theta, r;\; \kappa)$ for $\kappa = 2$}
      &  {\em $A_{\varphi\varphi\varphi}(\theta, r;\; \kappa)$ for $\kappa = 4$}
      & {\em $A_{\varphi\varphi\varphi}(\theta, r;\; \kappa)$ for $\kappa = 8$} \\
      \includegraphics[width=0.22\textwidth]{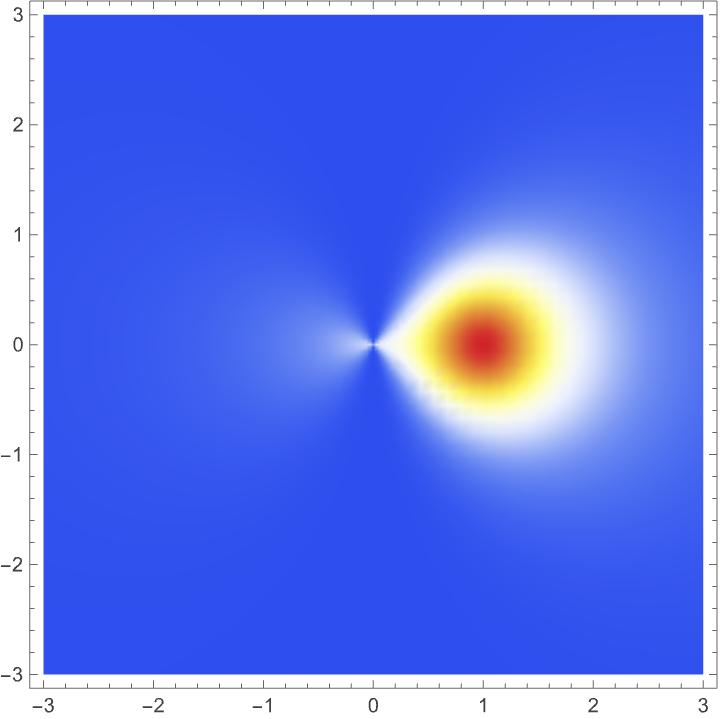}
      & \includegraphics[width=0.22\textwidth]{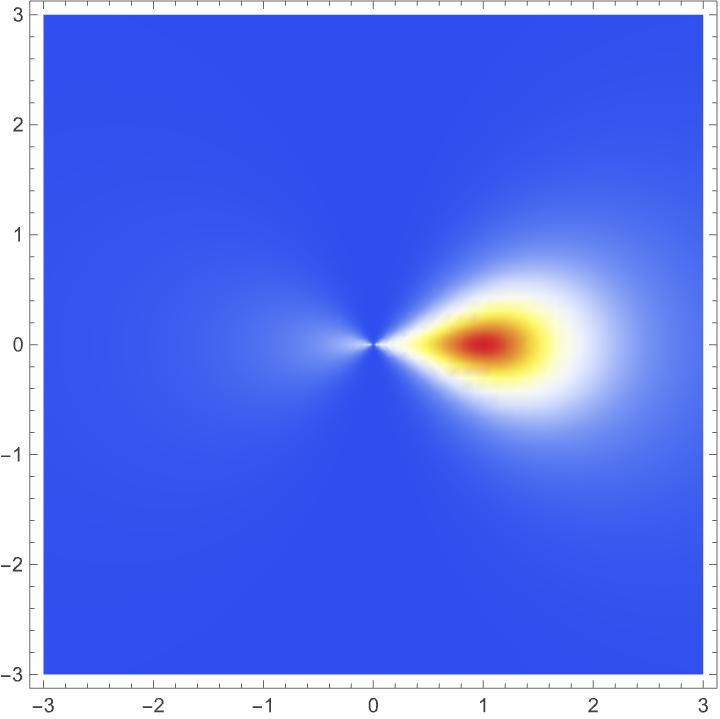}
      & \includegraphics[width=0.22\textwidth]{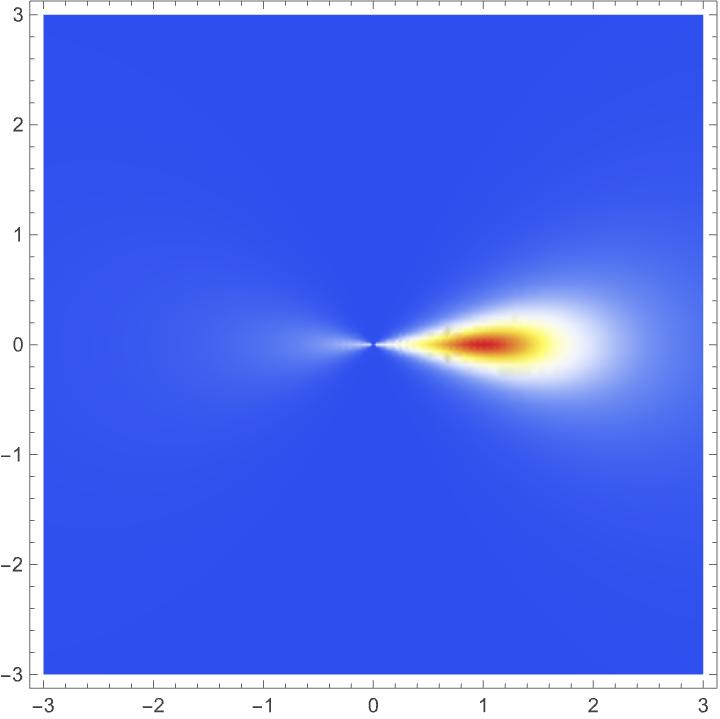}
      & \includegraphics[width=0.22\textwidth]{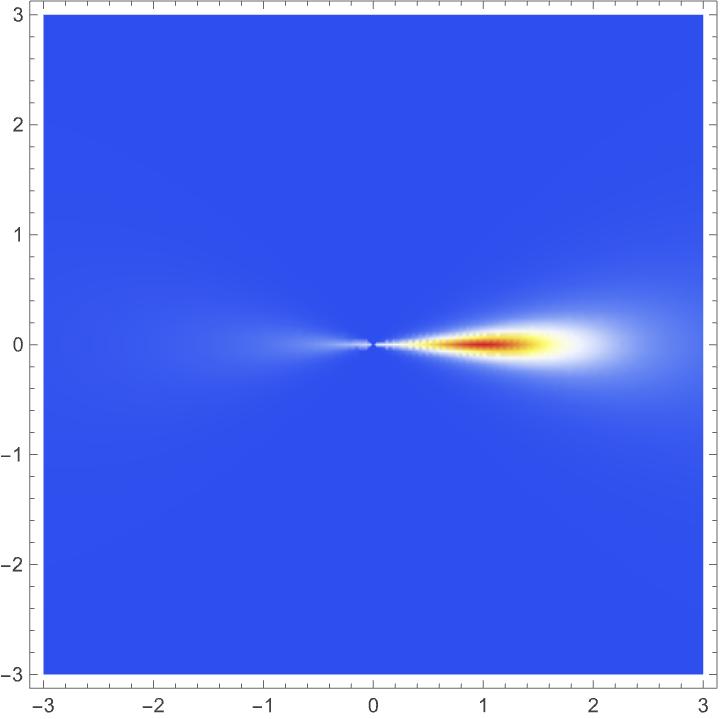} \\
     {\em $A_{\varphi\varphi\varphi\varphi}(\theta, r;\; \kappa)$ for $\kappa = 1$}
      &  {\em $A_{\varphi\varphi\varphi\varphi}(\theta, r;\; \kappa)$ for $\kappa = 2$}
      &  {\em $A_{\varphi\varphi\varphi\varphi}(\theta, r;\; \kappa)$ for $\kappa = 4$}
      & {\em $A_{\varphi\varphi\varphi\varphi}(\theta, r;\; \kappa)$ for $\kappa = 8$} \\
      \includegraphics[width=0.22\textwidth]{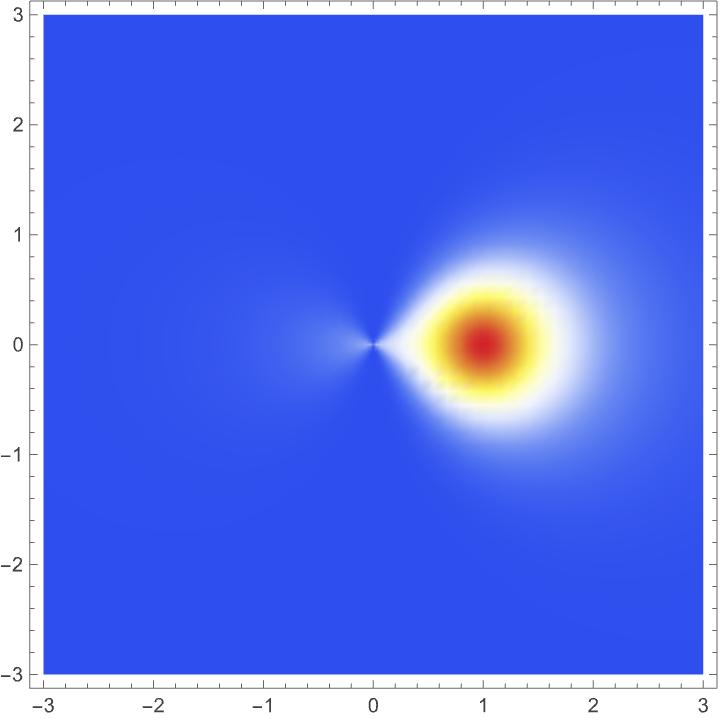}
      & \includegraphics[width=0.22\textwidth]{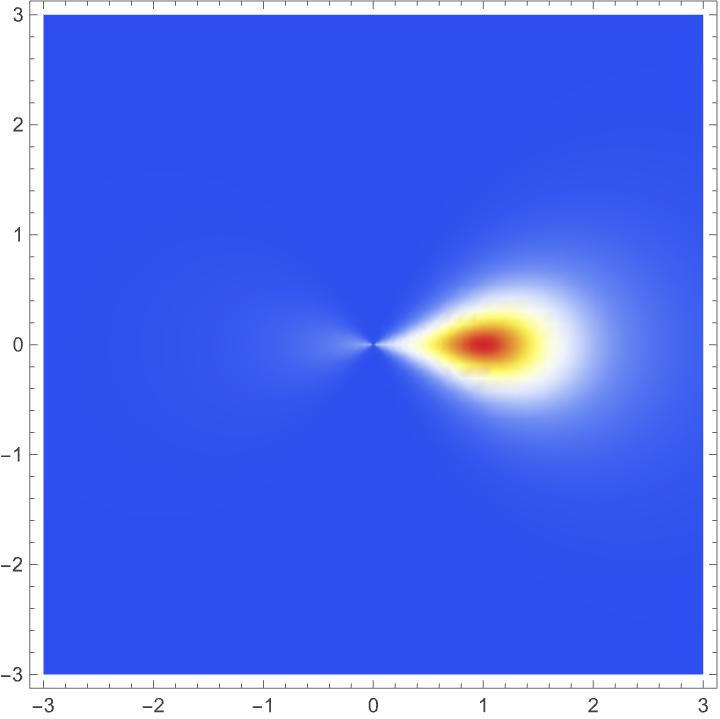}
      & \includegraphics[width=0.22\textwidth]{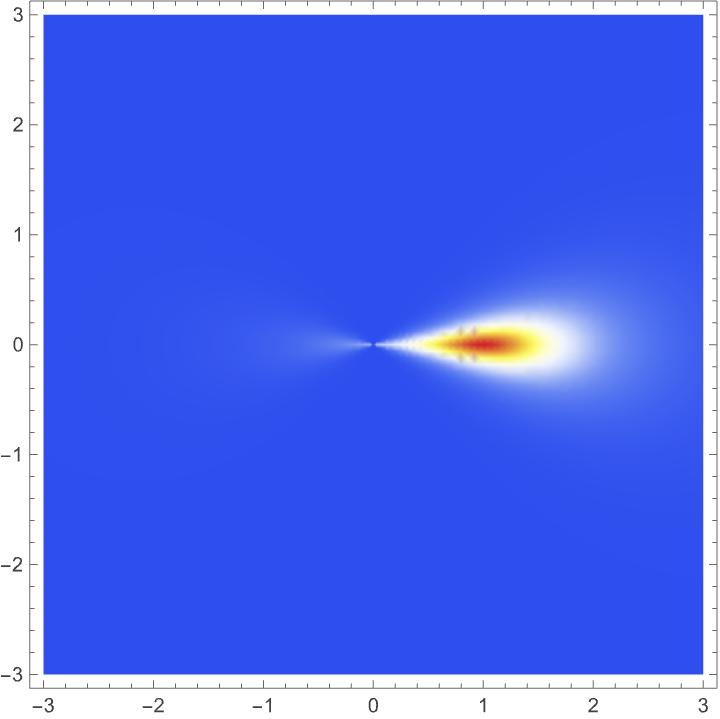}
      & \includegraphics[width=0.22\textwidth]{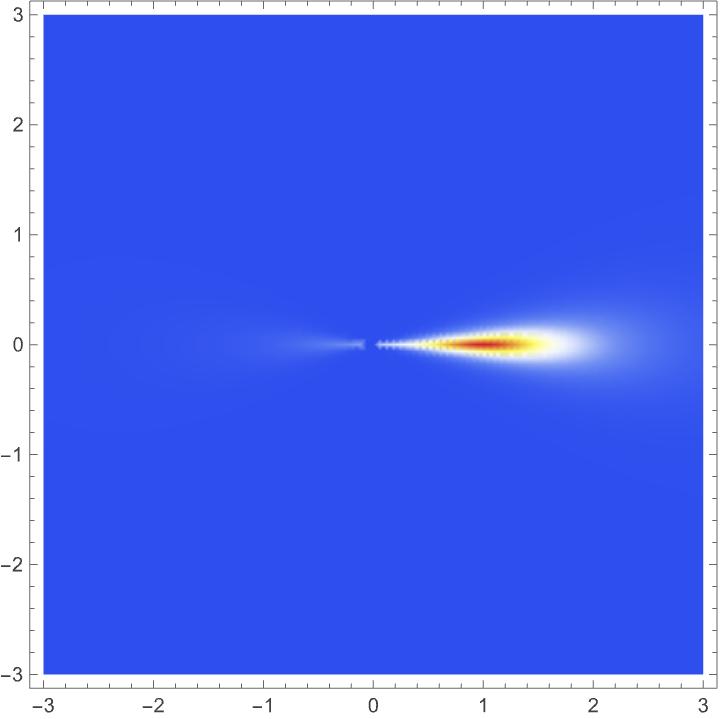} 
    \end{tabular}
  \end{center}
  \caption{Density plots of the {\em joint selectivity maps\/} over directions
    and speeds for the idealized models of {\em velocity-tuned simple cells\/} of orders
    $m \in \{ 1, 2, 3, 4 \}$ for different values of the elongation
    $\kappa$, with the preferred direction of the receptive field
    corresponding to the horizontal direction with $\theta_0 = 0$.
    Here, the relative speed ratio $r$ and the inclination
    angle $\theta$ are treated as polar coordinates, which are then
    mapped to a Cartesian grid. As can be seen from these
    visualizations, the direction selectivity properties become more
    narrow both with increasing order $m$ of spatial differentiation,
    and with increasing elongation $\kappa$.
    (Horizontal axes: $r \cos \theta \in [-3, 3]$.
    Vertical axes: $r \sin \theta \in [-3, 3]$.)
    (In these coloured maps, red represents high values, while blue
    represents low values.)}
  \label{fig-joint-sel-maps-simple}
\end{figure*}

\begin{figure*}[hbtp]
  \begin{center}
    \begin{tabular}{cccc}
      {\em $A_{{\cal Q}_{12}}(\theta, r;\; \kappa)$ for $\kappa = 1$}
      &  {\em $A_{{\cal Q}_{12}}(\theta, r;\; \kappa)$ for $\kappa = 2$}
      &  {\em $A_{{\cal Q}_{12}}(\theta, r;\; \kappa)$ for $\kappa = 4$}
      & {\em $A_{{\cal Q}_{12}}(\theta, r;\; \kappa)$ for $\kappa = 8$} \\
      \includegraphics[width=0.22\textwidth]{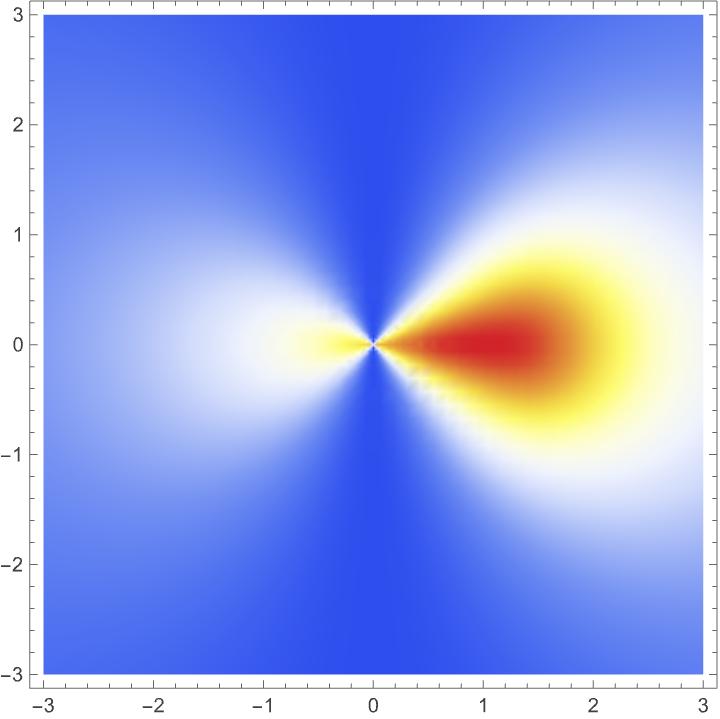}
      & \includegraphics[width=0.22\textwidth]{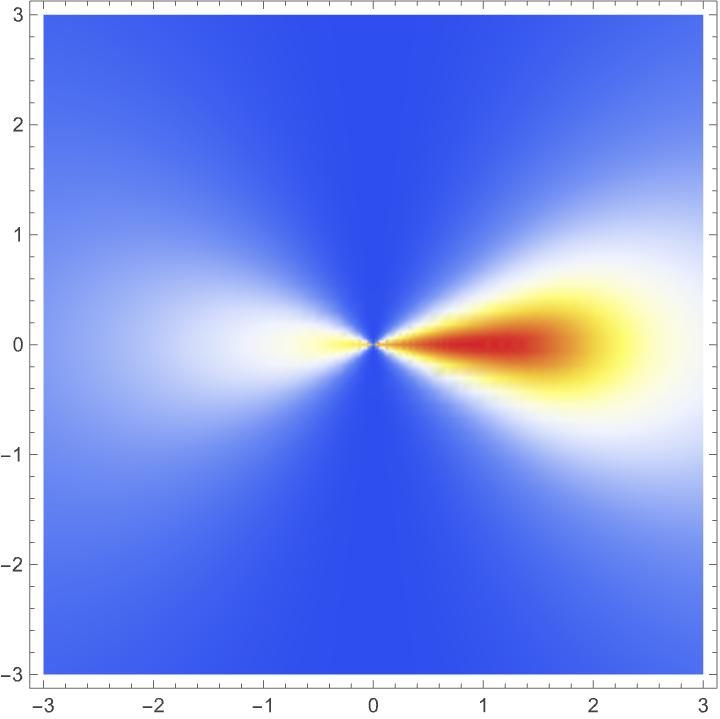}
      & \includegraphics[width=0.22\textwidth]{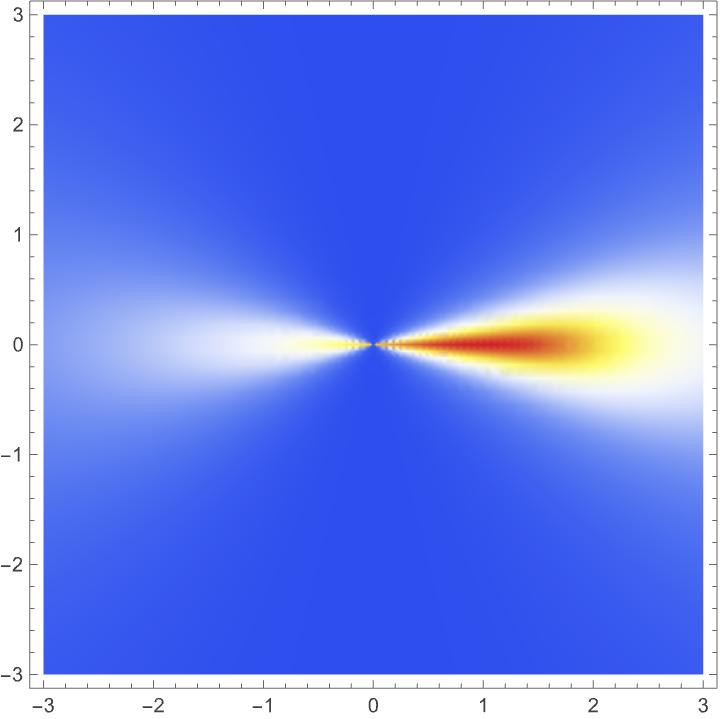}
      & \includegraphics[width=0.22\textwidth]{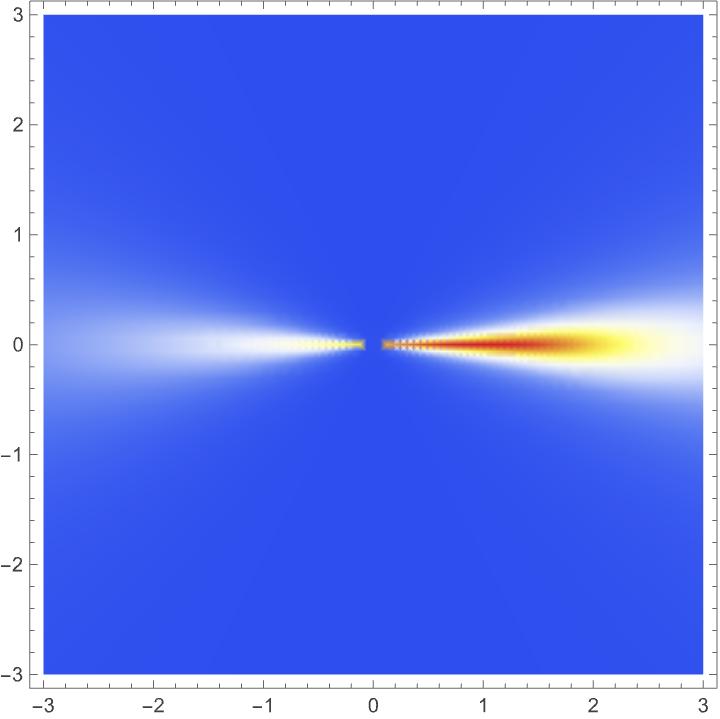} \\
     {\em $A_{{\cal Q}_{34}}(\theta, r;\; \kappa)$ for $\kappa = 1$}
      &  {\em $A_{{\cal Q}_{34}}(\theta, r;\; \kappa)$ for $\kappa = 2$}
      &  {\em $A_{{\cal Q}_{34}}(\theta, r;\; \kappa)$ for $\kappa = 4$}
      & {\em $A_{{\cal Q}_{34}}(\theta, r;\; \kappa)$ for $\kappa = 8$} \\
      \includegraphics[width=0.22\textwidth]{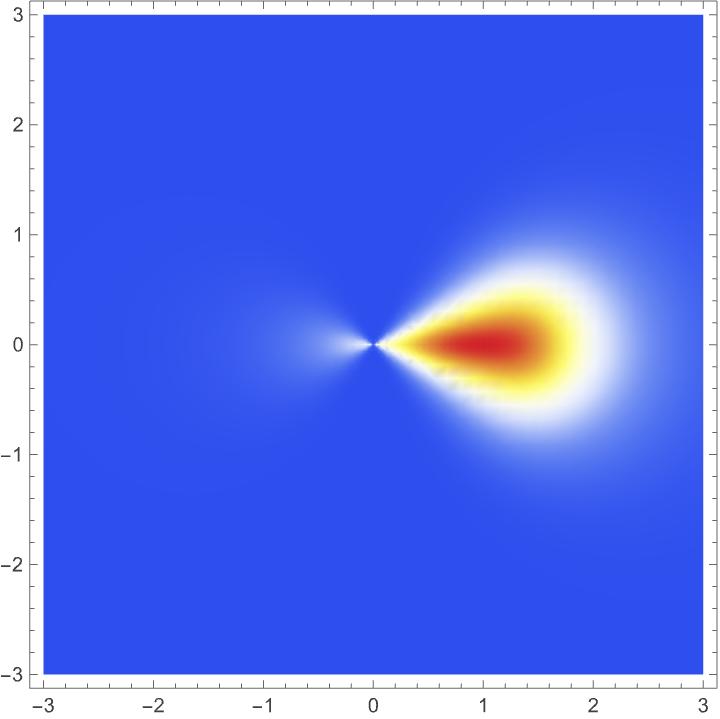}
      & \includegraphics[width=0.22\textwidth]{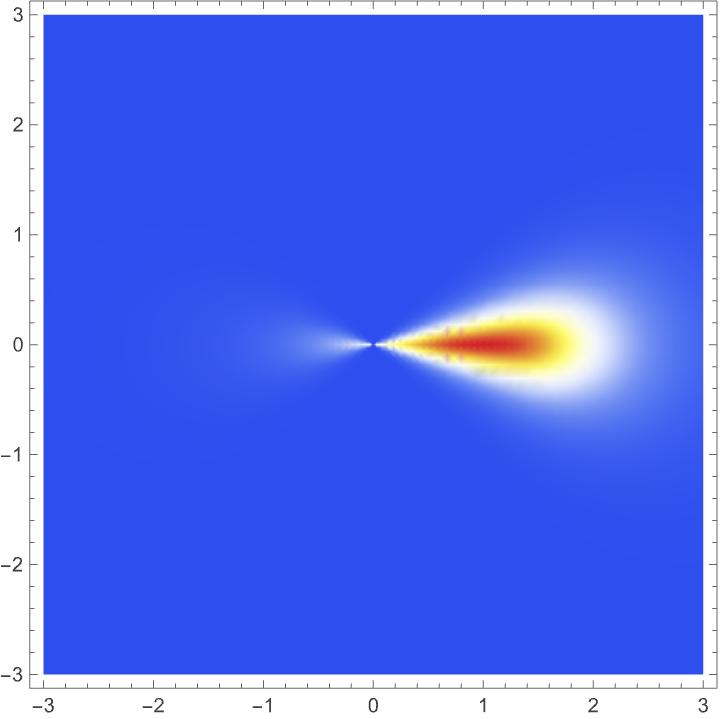}
      & \includegraphics[width=0.22\textwidth]{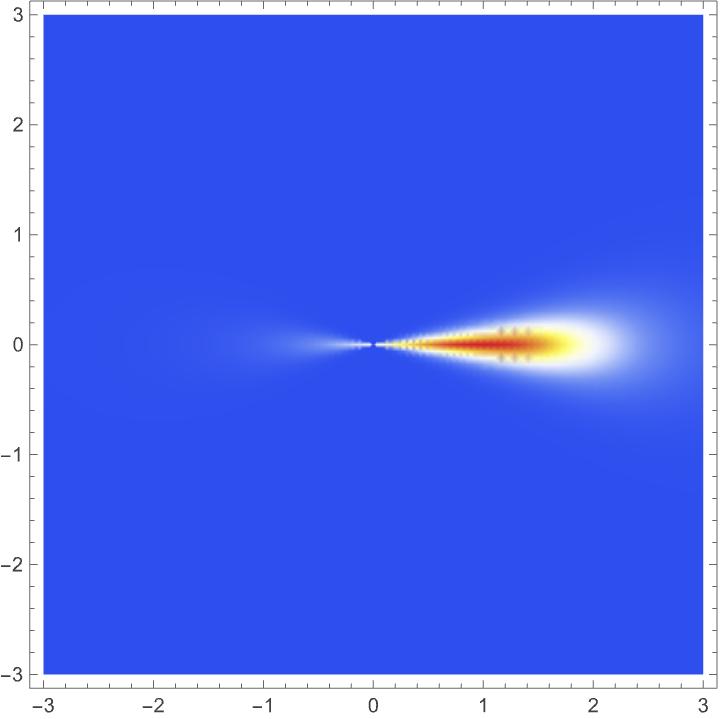}
      & \includegraphics[width=0.22\textwidth]{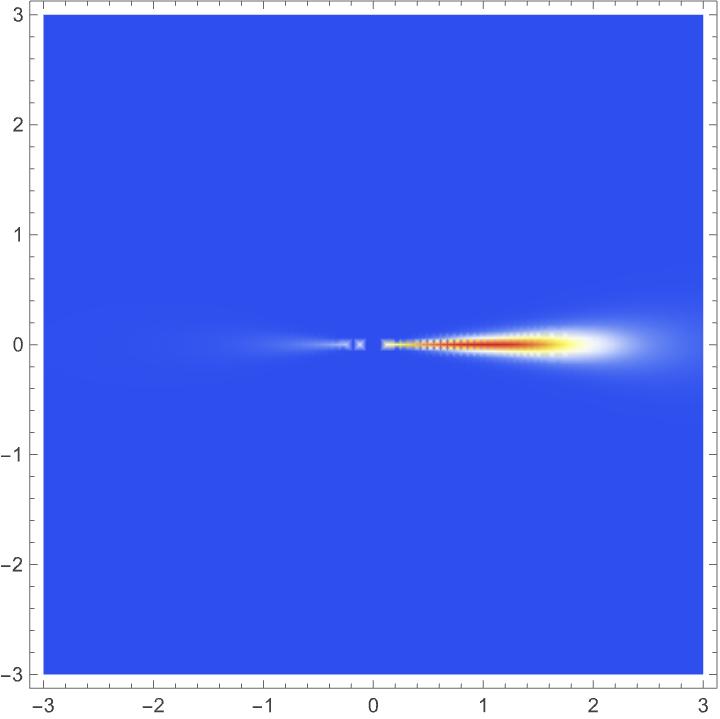} 
     \end{tabular}
  \end{center}
  \caption{Density plots of the {\em joint selectivity maps\/} over directions
    and speeds for the idealized models of {\em velocity-tuned complex cells\/} of orders
    $\{ 1, 2 \}$ or $\{ 3, 4 \}$ for different values of the elongation
    $\kappa$, with the preferred direction of the receptive field
    corresponding to the horizontal direction with $\theta_0 = 0$.
    Here, the relative speed ratio $r$ and the inclination
    angle $\theta$ are treated as polar coordinates, which are then
    mapped to a Cartesian grid. As can be seen from these
    visualizations, the direction selectivity properties become more
    narrow both with increasing order of spatial differentiation,
    and with increasing elongation $\kappa$.
    (Horizontal axes: $r \cos \theta \in [-3, 3]$.
    Vertical axes: $r \sin \theta \in [-3, 3]$.)
    (In these coloured maps, red represents high values, while blue
    represents low values.)}
  \label{fig-joint-sel-maps-complex}
\end{figure*}

Experimentally, one has previously wondered why motion-sensitive
neurons may also respond strongly in the opposite direction of the
preferred direction of the neuron
(Orban {\em et al.\/} \citeyear{OrbKenNul86-JNeurPhys}).
With our theoretically derived models for the direction selectivity
properties of simple cells in terms of directional derivatives of
velocity-adapted spatio-temporal Gaussian kernels, we can provide a possible
explanation for why such strong responses in the opposite direction
may occur. The lobe in the opposite direction of the preferred
direction of neuron arises as a direct consequence of the response
properties of the underlying spatio-temporal receptive field to motion
stimuli in different directions, specifically as the result of processing input
stimuli with different motion directions using the same
computational machinery.

The idealized models for the spatio-temporal receptive fields
of simple cells have been derived from a set of symmetry properties
regarding the environment in combination with a set of internal consistency
requirements to guarantee internally consistent treatment of
spatio-temporal input data over multiple spatio-temporal scales.
The idealized models for complex cells have additionally been complemented with
quadratic non-linear operations, to lead to response properties
that less depend on the local phase information in the input data.
Then, these computational models for spatio-temporal receptive
fields have been probed with moving sine waves for different
inclination angles and speeds, which results in a strong
peak in the direction of the preferred direction of each spatio-temporal
receptive fields, but also in a less strong peak in the opposite
direction to the preferred direction of the spatio-temporal receptive field.

How this response then varies with the inclination
angle is then merely a direct consequence of the mathematical model
for the spatio-temporal receptive field, where one may then interestingly
note that this gives rise to a lobe in the opposite direction, and the
size of the lobe then strongly depends on the order of differentiation
and whether the model is a model for a simple cell or a complex cell.
In this respect, there is specifically no particular mechanism
beyond the overall computational function of the spatio-temporal
receptive that underlies the properties of the lobe in the opposite
direction.

Other possible explanations can also be stated regarding this
secondary response lobe in the anti-preferred direction of
biological neurons terms of:
(i)~underlying orientation selectivity combined with incomplete
suppression of opposite-direction motion,
(ii)~shared spatial geometry in the sense that a moving stimulus
to either the left or the has exactly the same spatial orientation,
which naturally stimulates the receptive field in both directions
(Albright \citeyear{Alb84-JNeuroPhys}),
%W{\"o}rg{\"o}tter and Eysel \citeyear{WorEys87-BICY}),
(iii)~spatio-temporal inseparability in the sense that in motion energy
models, the Fourier energy spectrum overlaps with
the motion in the opposite direction
(Adelson and Bergen \citeyear{AdeBer85-JOSA},
Emerson {\em et al.\/} \citeyear{EmeBerAde92-VR}), or
(iv)~incomplete asymmetric inhibition, due to the property
that direction selectivity may rely upon delayed, spatially
asymmetric inhibition to block responses when a stimulus moves;
if that veto is only partially effective, then some excitation leaks
through, producing a smaller secondary lobe in the opposite direction
(Livingstone \citeyear{Liv98-Neuron}), see, however, also
Priebe and Ferster (\citeyear{PriFer08-Neuron}) for a contrary
wider view.

Let us, however, in this context emphasize that our aim here is not
in any way to delve deeply into the topic of explaining the secondary lobe in the
opposite direction to the preferred direction of a motion-sensitive
neuron. It is rather just to demonstrate that the secondary lobe in the
opposite direction is a direct consequence of a pure feed-forward
model for the spatio-temporal receptive field.

For two of the complex cells (top right and middle left
in Figure~\ref{fig-model-Liv03-complex}) there are notable
qualitative differences between the shapes of the biologically
recorded direction selectivity curves and what can be spanned by
the two types of idealized models for velocity-tuned complex cells considered here.
Those two complex cells respond to a much wider span of motion
directions than the other complex cells, and specifically beyond what
is compatible with and can be captured with the
more narrow response properties of the proposed model for
velocity-tuned complex cells.
One may therefore consider if those direction selectivity curves
should be categorized as due to other types of complex cells,
not explicitly modelled here.
Notably, in his work on motion-sensitive neurons in the primary visual
cortex, Orban {\em et al.\/} (\citeyear{OrbKenNul86-JNeurPhys})
also considered different and other types of neurons than the class of
velocity-tuned neurons.

Another specific aspect of these results is that the biologically recorded
direction selectivity curves are in this modelling better explained by considering both the
simple cells and the complex cells as having a variability in the
degree of elongation $\kappa$ of the receptive fields.
In this respect, these results are consistent with the prediction about a
variability in the degree of elongation of the receptive fields in the
primary visual cortex, as predicted in
Lindeberg (\citeyear{Lin25-JCompNeurSci-spanelong},
\citeyear{Lin25-PONE}), and do additionally extend this prediction to
also comprise a variability in the degree of elongation for
genuine spatio-temporal receptive fields.

\subsection{Joint selectivity maps over directions and speeds}
\label{sec-joint-selectivity-maps}

For the idealized models of simple cells of order $m$, the {\em joint\/}
  selectivity over
the motion direction $\theta$ and the relative speed factor $r = |u|/|v|$
is given by the compact expression (\ref{eq-cond-summ-dir-sel-prop})
\begin{multline}
  A_{\varphi^m,\max}(\theta, r;\; \kappa) = \\
  = \alpha_m \left( \frac{\cos (\theta )}{\sqrt{\kappa ^2 \sin ^2(\theta )-2 r \cos (\theta
   )+2 \cos ^2(\theta )+r^2}} \right)^m,
\end{multline}
while for the idealized models of complex cells based on spatial
derivatives of orders $\{ 1, 2 \}$ or $\{ 3, 4 \}$, the corresponding
joint selectivities $A_{{\cal Q}_{12},\max}(\theta, r;\; \kappa)$
and $A_{{\cal Q}_{34},\max}(\theta, r;\; \kappa)$ are given
by Equations~(\ref{eq-A-Q12-int-compl})
and~(\ref{eq-A-Q34-int-compl}), respectively.

Figures~\ref{fig-joint-sel-maps-simple}--\ref{fig-joint-sel-maps-complex}
show two-dimensional maps of these joint
selectivity maps in terms of polar plots over the logarithmic
relative speed factor $r$ for different values of the degree of
elongation $\kappa$.
As can be seen from these
visualizations, the direction selectivity properties become more
narrow both with increasing order of spatial differentiation,
and with increasing elongation $\kappa$.
Specifically, these joint direction and speed selectivity maps
characterise how the notions of direction selectivity and
speed selectivity are intimately coupled. 

From a theoretical viewpoint, such joint selectivity
properties ought to constitute a more natural concept to consider,
compared to just studying the projections to either pure speed selectivity
along the preferred direction of the neuron or the
direction selectivity for the preferred speed of the neuron,
as have so far been measured in biological experiments.

An interesting question for future research does hence concern
how well these idealized models for joint selectivity maps predict
corresponding joint selectivity maps for biological neurons.
To answer this question, new biological measurements would, however,
need to be performed.

\subsection{Relationships to the hypothesis about Galilean covariant
  spatio-temporal receptive fields in the primary visual cortex}
\label{sec-biol-rel-gal-cov}

In relation to covariance properties of visual receptive fields with
regard to basic geometric image transformations,
Lindeberg
(\citeyear{Lin23-FrontCompNeuroSci,Lin26-JCompNeuroSci})
proposed the working hypothesis that simple cells in primary visual
cortex ought to be covariant with respect to Galilean transformations of
the form
\begin{equation}
  \left\{
    \begin{array}{l}
      x_1' = x_1 + v_1 \, t, \\
      x_2' = x_2 + v_2 \, t.
    \end{array}
  \right.
\end{equation}
This would then imply that the shapes of the spatio-temporal
receptive fields of simple cells in the primary visual cortex ought to
be expanded over the degrees of freedom of Galilean transformations.
Concretely, this would imply that spatio-temporal receptive fields
of the form (\ref{eq-spat-temp-RF-model-der-norm-caus}) would be
present for a range of image velocities $v = (v_1, v_2)^T \in \bbbr^2$.

One may therefore ask if there would be neurophysiological support for
this hypothesis. The purpose with this section is to evaluate the
hypothesis about Galilean covariance in relation to existing
neurophysiological evidence. With regard to the theoretically
derived idealized models for direction selectivity curves,
this topic is essential, since the variabilities in the shapes
of the underlying spatio-temporal receptive fields with
regard to the parameters of local Galilean transformations
give rise to variabilities in the shapes of the idealized
direction and speed selectivity curves.

First of all, the the comparison between our theoretical results regarding
direction and speed selectivity properties of our idealized models of
simple cells and complex cells to the neurophysiological results by
Orban {\em et al.\/}\ (\citeyear{OrbKenNul86-JNeurPhys})
and Livingstone and Conway (\citeyear{LivCon03-JNeuroPhys})
in Section~\ref{sec-rels-neurophys-Orban} 
are consistent with an
expansion of the spatio-temporal receptive field shapes over the
degrees of freedom of local Galilean transformations.

Furthermore, regarding the processing of time-dependent image data in the primary
visual cortex and the middle temporal visual area MT, we have the following facts:
\begin{itemize}
\item
  the ability of the simple cells in the visual system of higher
    mammals to compute spatio-temporal receptive
    field responses similar to velocity-adapted temporal derivatives
    (DeAngelis {\em et al.\/}\
    \citeyear{DeAngOhzFre95-TINS,deAngAnz04-VisNeuroSci};
    Lindeberg \citeyear{Lin21-Heliyon} Figure~18 bottom part),
  \item
    visual neurons in the primate visual cortex being known
    to be direction selective
    (Hubel \citeyear{Hub59-JPhys}, Orban {\em et al.\/} \citeyear{OrbKenNul86-JNeurPhys},
    Churchland {\em et al.\/}\ \citeyear{ChuPriLis05-JNeuroPhys}),
    and ``... organized into subcolumns within an iso-orientation
    column, with each subcolumn preferring motion in
    a different direction'' (Elstrott and Feller \citeyear{ElsFel09-CurrOpNeurBiol}),
  \item
    the receptive
    fields in the middle temporal visual area MT being able to compute
    to a rich variety of direction-selective responses
    (Orban \citeyear{Orb97-ExtrStriCortPrim},
    Born and Bradley \citeyear{BorBra05-AnnRevNeurSci}),
  \item
    with the output from the primary visual cortex providing 
    input to the middle temporal visual area MT (Movshon and Newsome \citeyear{MovNew98-JNeuroSci})
    and
  \item
    with the direction-selective neurons in MT organized into direction
    columns (Albright {\em et al.\/}\
    \citeyear{AlbDesGro84-JNeuroPhys}).
  \end{itemize}
  From these results, it appears as if we can regard the primary visual
  cortex V1 and the middle temporal visual area MT as performing expansions of the visual
  representations over the directions and speeds of local Galilean motions.
  Thereby, it seems plausible that:
  \begin{itemize}
  \item
    the visual system should be able to compute Galilean-covariant
    receptive field responses, so as to be able to process visual
    information over a wide range of motion speeds
    and motion directions,
  \item
    this motion processing hierarchy
    could then be based on
    Galilean-covariant receptive fields in the primary visual cortex.
\end{itemize}
Given such an expansion of the shapes of the
spatio-temporal receptive fields in the primary visual cortex over the
degrees of freedom of local Galilean transformations, a consequence of
that would then be variabilities in the direction selectivity
properties and the speed selectivity properties of the visual
neurons, as recorded in neurophysiological measurements.

\section{Proposals for further directions of neurophysiological experiments}
\label{sec-sugg-neuro-exps}

Given the above theoretical results, with their qualitative
relationships to existing results from neurophysiological
measurements, a highly interesting topic would be to aim at additional
more quantitative comparisons.

The publicly available data regarding
explicitly reconstructed spatio-temporal receptive fields from
neurophysiological recordings is, however, very limited.
For this reason, it would be highly interesting if complementary
neurophysiological recordings%
\footnote{Such experiments should preferably be done for higher
  mammals with general purpose vision systems, like primates or cats.
  Recordings from mice may specifically not be appropriate, since the
  vision system of mice is far less developed compared the visual
  systems of higher mammals, see {\em e.g.\/}\ Huberman and
  Niell (\citeyear{HubNie11-TINS}). Specifically, recent results by
  Fu {\em et al.\/} (\citeyear{FuPieWilBasMuhDiaFroResPonDenSinTolFra24-CellRep})
  of reconstructing most excitatory inputs (MEIs) for early visual receptive
  fields in mice suggest that complex spatial features emerge earlier in the
  visual pathway of mice compared to primates, and that the receptive
  fields of mice are therefore more complex than a Gaussian
  derivative model, as used in the theoretical model for performing
  the theoretical analysis in this paper.}
could be performed, if possible with
recordings of both direction and speed selectivity properties of
visual neurons with explicitly reconstructed spatio-temporal receptive
fields from the same neurons, to answer the
following theoretically motivated questions:
\begin{itemize}
\item
  How well can the spatio-temporal receptive fields of simple cells in
  the primary visual cortex be modelled by velocity-adapted
  spatio-temporal affine-Gaussian-derivative-based receptive fields
  of the form (\ref{eq-spat-temp-RF-model-der-norm-caus}), if the
  entire 2+1-D spatio-temporal receptive field is reconstructed, and
  not only visualized in a 1+1-D cross-section, as done by
  DeAngelis {\em et al.\/}\
  (\citeyear{DeAngOhzFre95-TINS,deAngAnz04-VisNeuroSci}) and
  de~Valois {\em et al.\/}\ (\citeyear{ValCotMahElfWil00-VisRes}),
  and for which the match is qualitatively very good, see
  Lindeberg \citeyear{Lin21-Heliyon} Figure~18 bottom part?
\item
  Which orders $m$ of spatial differentiation describe the
  variabilities in spatio-temporal receptive fields of simple cells in the primary
  visual cortex?
\item
  How wide is the variability in the degree of elongation $\kappa$ in
  such models of the spatio-temporal receptive fields of simple cells
  in the primary visual cortex?
\item
  How wide are the variabilities in the spatial scales $\sigma_{\varphi}$ and
  the temporal scales $\sigma_t$, as depending on the distance from the
  center of the fovea? Does the vision system only implement a set of
  finest spatial scales $\sigma_{\varphi}$, or does it comprise a substantial
  variability? Similarly, does the vision system only implement a set
  of finest temporal scales $\sigma_t$?
  See Lindeberg (\citeyear{Lin26-JMIV}) for a further theoretical
  motivation to how a vision system could choose to only implement a
  set of finest spatial and temporal scales.
\item
  How is the ratio $\sigma_{\varphi}/\sigma_t$ related to the speed parameter
  $|v|$ of the local Galilean motion in the velocity-adapted model of
  the spatio-temporal receptive fields?
\item
  How well do the neurophysiological recordings of the direction and speed
  selectivities of the visual neurons agree with the theoretical
  predictions obtained from (i)~first determining the shape parameters
  $\varphi$, $\sigma_{\varphi}$, $\kappa$, $\sigma_t$ and $v$ in an idealized
  model of the spatio-temporal receptive field fitted to a
  reconstructed biological receptive field, and then (ii)~using the
  predictions about direction and speed selectivity properties
  from the here derived theoretical results concerning the speed and
  direction selectivity properties of idealized models of
  spatio-temporal receptive fields in terms of affine Gaussian
  derivatives?
\end{itemize}
Answering these theoretically motivated questions would provide further cues towards
understanding the computational function of the spatio-temporal
receptive fields in the primary visual cortex, and concerning to what
extent we can regard the influence of geometric image transformations
on the receptive field responses as a primary factor for the
development of the visual receptive fields.
Such more detailed comparisons between the properties of the
theoretical model and the properties of the biological neurons could
also provide cues to possible ways of extending the theoretical model
for spatio-temporal receptive fields with additional%
\footnote{The current model for linear spatio-temporal receptive
  fields, that we base this work upon, has
  been derived under minimal assumptions regarding the structure of
  the environment in combination with internal consistency
  requirements to guarantee theoretically well-founded treatment of
  image structures over multiple spatial and temporal scales
  (Lindeberg \citeyear{Lin10-JMIV,Lin21-Heliyon}).}
{\em a priori\/} information,
if motivated from results from additional neurophysiological experiments.

\section{Summary and discussion}
\label{sec-summ-disc}

When looking at neurophysiological measurements of neuronal
responses in the visual cortex to visual stimulation, 
one may ask why they look the way they do.
In this paper, we have presented a theory with associated
closed-form expressions for how the direction
and speed selectivity properties of idealized models of
simple cells and complex cells in the primary visual cortex
are determined by inherent shape parameters of the
underlying spatio-temporal receptive fields in such
idealized models of motion-sensitive visual neurons.
Additionally, we have shown how variabilities in the shape
parameters of the underlying spatio-temporal receptive fields
give rise to corresponding variabilities in the shapes and tuning widths
of the resulting direction and speed selectivity curves,
and compared these types of theoretically based predictions to the
results of neurophysiological recordings.

The underlying receptive field models have been obtained from a normative
theory of visual receptive fields, derived from axiomatic assumptions
regarding structural properties of the environment in combination with
internal consistency requirements to guarantee theoretically
well-founded processing of image structures over multiple spatial and
temporal scales. This theory states that a canonical way to model
linear receptive fields, corresponding to simple cells in the primary
visual cortex, is in terms of
spatial and temporal derivatives of affine Gaussian kernels, combined with a
mechanism of velocity adaptation to handle the influence on image data
from local image motions. Based on these idealized models of simple
cells, we have formulated idealized models of complex cells, based on
local quadratic energy models of simple cells, combined with regional spatial integration.

Here, the shapes of the underlying spatio-temporal receptive fields
in the idealized models for simple cells are derived by axiomatically
determined uniqueness.
The idealized models for complex cells are, however, not unique.
Instead, the idealized models of velocity-tuned complex cells
considered here should
rather be seen as the simplest possible types of models of complex cells
that produce approximately phase-independent responses to a sine wave
with a frequency matching the scale parameter of the model,
by combining the output from multiple simple cells in an
approximation of quadrature referred to as quasi quadrature, and then
additionally integrating those squared receptive field responses over a region in the image domain with its
size proportional to the size of the support regions of the simple
cells. Hence, regarding the modelling of complex cells, there is a
potential for complementing the here considered idealized models of velocity-tuned complex cells
with idealized models of other types of complex cells.
Additionally, there could be a potential in complementing these models
of complex cells with a more fine-grained modelling of how local
information within the support region of the complex cells is
both processed and combined in a non-linear manner.

By subjecting these idealized models of motion-sensitive neurons in the
primary visual cortex to a structurally similar probing method, as used
for probing the direction and speed selectivity properties of visual
neurons in neurophysiological experiments,
we have derived closed form expressions for the direction
and speed selectivity properties of our idealized models of simple
cells, as summarized in
Equations~(\ref{eq-cond-summ-dir-sel-prop})
and~(\ref{eq-rcurve-order-all}) for the special case when spatial and
temporal scale parameters $\sigma_{\varphi}$ and $\sigma_t$ are coupled to the
speed parameter $|v|$ of the spatio-temporal receptive field according
to $\sigma_{\varphi}/\sigma_t = |v|$. We have also derived more general results
for such combined direction and speed selectivity properties, reproduced in
Figure~\ref{eq-dir-sel-expr-no-rel-sigma-xt}
and Equation~(\ref{eq-rcurve-order-all-noncoupled})
for the more general
case, when there is no such coupling between the scale parameters in
relation to the speed parameter of the receptive field.

In essence, these results show that the direction selectivity
properties become more narrow with increasing order of spatial
differentiation as well as with increasing degree of elongation in the
underlying spatial components of the receptive fields.
The speed selectivity properties also become more narrow with
increasing order of spatial differentiation, but are for the
inclination angle $\theta = 0$ independent of
the degree of elongation of the receptive fields.

For our idealized models of complex cells, the resulting closed form
expressions of the combined direction and speed selectivity properties are given
in Equations~(\ref{eq-A-Q12-int-compl}) and~(\ref{eq-A-Q34-int-compl})
in Figure~\ref{fig-eq-ampl-meas-int-compl-cells-12-34}
in Appendix~\ref{app-anal-compl-cells}. 
The qualitative results regarding more narrow
direction selectivity properties with increasing order of spatial
integration and increasing degree of elongation hold for the idealized
models of complex cells in a similar manner as for the idealized
models of simple cells. The speed selectivity properties for our idealized models of
complex cells for the inclination angle $\theta = 0$ are given in
Equations~(\ref{eq-speed-sel-curve-compl-12})
and~(\ref{eq-speed-sel-curve-compl-34}) in
Figure~\ref{fig-alg-expr-R-curves-int-compl-cells}.

By comparisons with the results of neurophysiological recordings of
motion-sensitive neurons in the primary visual cortex,
the presented
theoretical results are consistent with the previously
reported class of velocity-tuned motion-sensitive neurons
reported by Orban {\em et al.\/}\ (\citeyear{OrbKenNul86-JNeurPhys}),
in the sense
that the models of the neurons respond maximally to the combination of
a preferred motion direction and a preferred motion speed, and that
the magnitude of the response decreases when changing the motion
direction and/or the motion speed from the preferred value.

Specifically, we have used our theoretically derived models for speed
selectivity curves to model the speed selectivity measurements
reported in Figures~3, 6A, 6B, 6C and 4 in
Orban {\em et al.\/}\ (\citeyear{OrbKenNul86-JNeurPhys}).
By our results, reported in
Figures~\ref{fig-model-orban-speed-sel-curves}
and~\ref{fig-model-Orban-fig4} in this paper, we have found that our idealized
models of simple cells and complex cells do well capture the
qualitative shapes of biological speed selectivity measurements for
both central and peripheral visual neurons in the primary visual
cortex. Thereby, we have demonstrated that it is possible to
predict properties of motion-sensitive visual neurons by analysing the
properties of a theoretically principled (axiomatically derived)
idealized model for spatio-temporal receptive fields.

In this way, our catalogue of $4+2=6$ different models of simple cells
and complex cells provide a complementary taxonomy that could be used
to categorize further biological measurements of speed selectivity
properties of neurons in the primary visual cortex. Specifically, by
our theoretically principled low-parameter models, we have also
provided a novel way of reinterpreting previous biological
measurements, as demonstrated in Figure~\ref{fig-model-Orban-fig4},
where we showed that our theoretically derived model does both provide
a qualitatively different interpretation of the overall speed
selectivity properties of the neuron reported in Figure~4 in
Orban {\em et al.\/}\ (\citeyear{OrbKenNul86-JNeurPhys})
and also leads to better agreement with the underlying biological
measurements. From our results in
Figure~\ref{fig-model-orban-speed-sel-curves} in this paper,
for the neurons reported in Figures~6A, 6C and~4
in Orban {\em et al.\/}\ (\citeyear{OrbKenNul86-JNeurPhys}),
we also show that the results of fitting our theoretically derived low-parameter
speed selectivity models to the neurophysiological measurements do
also provide seemingly better fits to the biological data than
the curves drawn in the original publication, as based on the
derived prior concerning the comparably sharp peaks
in the idealized speed selectivity curves.

Regarding the experimentally recorded direction selectivity curves of
biological simple cells and complex cells recorded by
Livingstone and Conway (\citeyear{LivCon03-JNeuroPhys}),
we also, for most of the neurons, obtained qualitatively very good matches between
the biological direction selectivity curves and our idealized
models for most of the neurons reported in this study.
For a few of the neurons, however, the qualitative shapes of the
biological direction selectivity curves differ from what can be
spanned by our classes of idealized velocity-tuned models.
One may therefore consider if those neurons should be categorized as
belonging to a different category, beyond the velocity-tuned models
that we consider here.

Notably, while Larisch and Hamker (\citeyear{LarHam25-NeurNetw})
in their recent work aim at explaining direction selectivity
properties of the neurons in the primary visual cortex by simulations of
complex neural networks aimed at mimicking the early visual pathway.
In this paper, we instead derived closed-form expressions for both
direction and speed selectivity properties by closed-form
theoretical analysis. In this way, our underlying aim is to uncover
essential computational mechanisms in the spatio-temporal receptive
fields, which in turn determine the direction and speed selectivity
properties of the motion-sensitive neurons in the primary visual cortex.

It should further be emphasized that the presented theoretical
analysis is performed as an almost direct
consequence of the underlying axiomatically determined model for
covariant spatio-temporal receptive fields over multiple spatial and
temporal scales with explicit associated variabilities over image orientations,
the degree of elongation of the receptive fields and Galilean transformations.
The theoretical models, that are derived,
specifically have very few parameters, thereby providing a different
form of explanatory power compared to the results of simulating complex neural
networks, which by necessity will contain a very large number of
parameters, for which it may be far from clear how to tune those
parameters in a principled way, since many of those parameters may not
be directly accessible to be observable from neurophysiological
measurements.

Hence, besides the determination of the rather few inherent parameters in the
idealized models for visual receptive fields to the experimental
measurements from a specific neuron, no actual adaptions have
been performed of the underlying theoretical model to make it fit to the biological data.
Instead, the underlying idealized models for the spatio-temporal receptive fields
have been derived in a principled way, based on symmetry properties
imposed on an idealized vision system. Then, the idealized models
for the direction selectivity and speed selectivity properties
follow as {\em direct consequences of the underlying normative
  theory\/}.

Compared to computationally more fine-grained models for direction selectivity
properties of neural computations based on networks of explicit
models of neurons, the derived idealized models of direction
and speed proposed theoretical predictions are based on purely
{\em functional models\/} of the computational functions of the
spatio-temporal receptive fields, and are derived in closed form.
Thereby, the derived models are very parameter compact.

Compared to explicit computational models based on networks
of models of neurons, in approach proposed here
there is additionally no need for setting
the large number of parameters usually needed to fine-tune such models,
and for which it may not be clear how those parameters should
be determined in a principled way. Notably, when using 
explicit neural models of visual computations, there may be
parameters in those models that are not directly observable
in relation to neurophysiological measurements.

In general, if one has a sufficiently flexible model with a
sufficiently large number of internal parameters,
it ought to be possible to fit such a model
to a given dataset. While one may then possibly get a quantitatively
very good fit to the measurement data, the explanatory value of
such a modelling approach could, however, then be questioned,
depending on the nature of the model and on to what extent
the internal parameters can be determined in a principled manner.
The important property of the proposed idealized models for the direction
and speed selectivity curves derived here in this respect
is that qualitatively very good fits to the biological data can be
obtained with very few internal and then also functionally directly interpretable
parameters in the idealized models.

Beyond the order of spatial differentiation, the only parameters
in the idealized models
are the shape parameters of the receptive fields in terms of
(i)~the inclination angle $\theta$ between
the stimulus and the preferred direction of the receptive field,
(ii)~the ratio $r = |u|/|v|$ between the speed parameters of the
stimulus and the receptive field, and
(iii)~the degree of elongation $\kappa$ of the spatial component
of the receptive field.
This implies that properties of the model can be analysed by
pure theoretical analysis, without any need for resorting to
extensive numerical simulations, and also that the model fitting can
be performed with established model fitting schemes,
such as the both powerful and robust Levenberg-Marquardt method,
over a very small set of model parameters. This implies
that the determination of initial seeds for the numerical
optimization will be less critical.

Additionally, we have also described how the shape parameters
of the receptive fields directly relate to the influence of geometric image
transformations on the incoming spatio-temporal image data to
the vision system, and how the predicted variabilities in the shape
parameters of the spatio-temporal receptive fields, as predicted
from the underlying normative theory of visual receptive fields,
directly correspond to corresponding variabilities in the shapes
of the direction and speed selectivity curves for the idealized
models of simple cells and complex cells.

Complementing with the fact that the shape parameters
of the receptive fields directly connect to the influence of
geometric image transformations according to the
unified theory for this in Lindeberg (\citeyear{Lin25-JMIV}),
it then follows that the variabilities in the shapes of speed and direction
selectivity curves, as have been recorded in neurophysiological
experiments, can be explained from a requirement that the
visual system should have the ability to handle the influence of
geometric image transformations on the incoming image data,
based on covariance properties that lead to
variabilities in the shapes of the receptive fields according
to the degrees of freedom of geometric image transformations.

In this way, the presented methodology provides a way to
explain the results that we get from neurophysiological
measurements of visual neurons, by anchoring those to
direct consequences of a normative theory for
visual receptive fields, that derives the receptive field profiles by necessity.
The very good qualitative similarities that we have obtained from
these idealized predictions and the biological data could in this
sense can also be regarded as strong support for the underlying
normative theory of visual receptive fields based on the generalized
Gaussian derivative model.

By comparisons with overall results concerning direction selective
neurons in the primary visual cortex V1 and the middle temporal area
MT in Section~\ref  {sec-biol-rel-gal-cov},
as reported by
Albright {\em et al.\/}\ (\citeyear{AlbDesGro84-JNeuroPhys}),
see Born and Bradley (\citeyear{BorBra05-AnnRevNeurSci}),
Elstrott and Feller (\citeyear{ElsFel09-CurrOpNeurBiol})
Orban {\em et al.\/}\ (\citeyear{OrbKenNul86-JNeurPhys}),
Orban (\citeyear{Orb97-ExtrStriCortPrim}),
Movshon and Newsome (\citeyear{MovNew98-JNeuroSci}) and
Churchland {\em et al.\/}\ (\citeyear{ChuPriLis05-JNeuroPhys}),
and as summarized 
in Section~\ref{sec-rels-neurophys-Orban},
with their organization
into subcolumns preferring motion in different directions, we have
also found this organization consistent with the wider hypothesis proposed
in Lindeberg
(\citeyear{Lin23-FrontCompNeuroSci,Lin26-JCompNeuroSci}),
that the population of receptive fields of the simple cells in the
primary visual cortex ought to be covariant under local Galilean
transformations.

This would then specifically imply that the
spatio-temporal receptive fields ought to have their shapes expanded over the
degrees of freedom of local transformations. In practice, this would
imply that there would be multiple copies of similar types of
spatio-temporal receptive fields for different values of the velocity
parameter $v = (v_1, v_2)^T$ in the idealized model
(\ref{eq-spat-temp-RF-model-der-norm-caus}) of simple cells, as illustrated in
Figures~\ref{fig-noncaus-strfs} and~\ref{fig-timecaus-strfs}.

Unfortunately, the amount of publicly available data regarding the
neurophysiological properties of spatio-temporal receptive fields in
the primary visual cortex is very limited. To address further
questions regarding the quantitative modelling of the spatio-temporal
receptive fields of simple cells and complex cells, with their
direction and speed selectivity properties, we have therefore proposed
a set of outlines for further neurophysiological measurements in
Section~\ref{sec-sugg-neuro-exps}.

Notwithstanding such potential opportunities for additional more detailed
quantitative modelling and comparisons, if further neurophysiological
measurements would become available, we have in
this treatment theoretically analyzed how the direction and speed
selectivity properties of idealized models of simple cells and complex
cells can be related to inherent properties of the their underlying
spatio-temporal receptive fields, and derived explicit relationships
for how the direction and speed selectivity properties can for these
models be directly related to intrinsic shape parameters of the receptive
fields. Our intention is that these results should contribute to a
better theoretical understanding regarding the computational
mechanisms underlying the processing of motion stimuli
in the primary visual cortex.

Overall, these results are also consistent with the wider hypothesis
in Lindeberg
(\citeyear{Lin21-Heliyon,Lin23-FrontCompNeuroSci,Lin26-JCompNeuroSci})
that the influence of geometric image transformations on the receptive
field responses may constitute a primary factor in the development of the
receptive fields in the primary visual cortex.

\subsection{Directions for future work}

\subsubsection{Theoretical analysis for stimuli based on moving bars}
\label{app-theor-anal-moving-bars}

In the theoretical analysis performed in the paper, we have throughout
used stimuli based on moving sine waves, which allows for closed-form
theoretical analysis and also leads to closed-form predictions about
idealized direction and speed selectivity curves.

In the comparisons to biological results, we have, however, compared
to the results of neurophysiological measurements based on moving bars.
From this perspective, the theoretical model does not fully reflect
the conditions in the biological experiments, but does instead constitute
an approximation.%
\footnote{Notably, however, the theoretical analysis was performed
  and written up (Lindeberg \citeyear{Lin25-arXiv-dirsel})
  {\em before\/} I started to look for biological results
  that I could the compare theoretical predictions to. In this respect,
  the theoretical results can be regarded as true predictions, and not
  merely the result of adapting a theoretical model to a specific data set.}
The very good similarity between the theoretical predictions from
the model and the biological results, specifically in view of the
very small number of parameters in the idealized models for
the selectivity curves, do, on the other hand,
support that this approximation ought to be quite good.

If the primary goal would instead be to as accurately as possible
model precisely the biological result that we have compared to in the paper,
then it could be interesting to try to carry out a corresponding
theoretical analysis based on moving bars.
For moving bars of a finite width, such a theoretical analysis
would, however, probably not be possible to carry out
in closed form. Hence, one would instead probably have
to instead resort to numerical simulations, which would
then not make it possible to obtain as compact closed-form
predictions as derived in
Sections~\ref{sec-dir-sel-simpl-cells}--\ref{sec-dir-sel-compl-cells}
with the corresponding technical details in
Appendices~\ref{app-anal-simpl-cells}--\ref{app-anal-compl-cells}.

Alternatively, one could possibly conceive to instead
investigate if a corresponding theoretical analysis could
be performed with infinitely thin moving bars
(corresponding to the extension of pointwise delta functions
of infinite high to both infinitely thin and infinitely high bars
with extended extent along straight lines).
The results from such a theoretical analysis, would, however,
then again merely constitute an approximation, then, however,
in relation to a non-infinitesimal width of the stimulus.

One may, however, consider if the results from such theoretical
or numerical predictions from moving bars would be radically
different from the theoretical predictions derived here based
on moving sine waves.

In relation to this, one could conceive that, as seen from the viewpoint of a local neuron,
that reacts to the visual stimuli from an either moving bar or moving
sine wave that passes by its corresponding spatio-temporal receptive
fields, the overall effect on the neuron could be largely similar
for these two types of stimuli, at least with regard to the overall
variabilities with respect to variations in the direction and the
speed of the stimulus. Hence, one may conceive that the theoretical
predictions derived from moving sine waves could also largely
reflect the effect of varying the direction and the speed of the
stimulus for moving bars.

A solid theoretical and/or numerical analysis could then evaluate
to how high degree of accuracy such an approximation would be valid.

\subsubsection{Theoretical and/or numerical analysis for Gabor-type models of spatio-temporal receptive fields}

It could additionally also be of interest to investigate if it would
be possible to perform a corresponding analysis for Gabor-type
models of spatio-temporal receptive fields.
Notably, however, the degrees of freedom for Gabor-type models
would then be somewhat different than for the generalized
Gaussian derivative model considered here. Specifically, explicit Gabor-type
models have not previously been formulated with joint variabilities
over the degree of elongation and local Galilean motion, why 
extensions of the Gabor-type models would then be needed in
this respect. Additionally, it is also not clear concerning to what
extent such an analysis could be performed in closed form---our
previous theoretical results regarding spatial orientation selectivity
properties in Lindeberg (\citeyear{Lin25-JCompNeurSci-orisel}) were
possible to perform in closed form for the generalized Gaussian
derivative model but not for a corresponding Gabor model.

By performing such a theoretical and/or numerical analysis
for Gabor-type models of spatio-temporal receptive fields,
one could then possibly evaluate the quality of the predictions
generated by the generalized Gaussian derivative model for
visual receptive fields in relation to the quality of corresponding
predictions based on Gabor-type models of visual receptive fields.

\subsubsection{More extensive quantitative evaluation of the idealized models for direction and speed selectivity properties for simple cells and complex cells}

Given the very good agreement between the qualitative shapes of
the idealized models for speed and direction selectivity curves
to neurophysiological measurements as reported in
Section~\ref{sec-rels-neurophys-Orban}, a complementary
question concerns how well these models fit in a quantitative
manner in relation to corresponding biological measurements
for different species.

Notably, if such more quantitative measurements are to be
performed, then ideally:
\begin{itemize}
\item
  the speed selectivity measurements ought to be acquired
  with a denser sampling of the velocity of the stimulus
  than used in the neurophysiological recordings by
  Orban {\em et al.\/}\ (\citeyear{OrbKenNul86-JNeurPhys}),
\item
  the direction selectivity measurements ought to be performed
  with a denser sampling of the inclination angle of the stimulus
  than used in the neurophysiological recordings by
  Livingstone and Conway (\citeyear{LivCon03-JNeuroPhys}),
\item
  the neurophysiological measurements ought have associated
  error bars, to reflect the accuracy of the neurophysiological
  measurements, and thereby make it possible to in a quantitative
  manner relate the degree of approximation of the theoretical
  in relation to the inherent accuracy in the biological data.
\end{itemize}
Additionally, it would be very interesting to also have data acquired under
joint variations of the direction and speed parameters, to evaluate
the predictions about joint direction and speed selectivity maps
described in Section~\ref{sec-joint-selectivity-maps}.

Depending on the quality of the biological data, and on whether the
biological data have been acquired using stimuli in terms of either moving
sine waves or moving bars, it could in such context also be interesting
to perform a theoretical analysis of the form outlined in
Section~\ref{app-theor-anal-moving-bars}, to evaluate the
degree of approximation of using the here derived theoretical predictions
derived based on moving sine waves in relation to corresponding
predictions based on moving bars, and to evaluate these
two lines of theoretical predictions in relation to corresponding
biological measurements.

Depending on what specific circumstances the biological
data have been acquired under, there could also be a
potential in more closely adapting the computational
and selection mechanism in the theoretical model
to more closely reflect the actual experimental conditions.
The intermediate results in the derivations in
Appendices~\ref{app-anal-simpl-cells}--\ref{app-anal-compl-cells}
could then constitute a starting point for performing
such more closely adapted theoretical analysis.

Beyond more detailed quantitative modelling of direction and speed
selectivity curves, such a more quantitative experimental evaluation
could also more generally be used to further evaluate the predictive
nature of the underlying normative theory for visual receptive fields,
according to conceptual framework outlined in the introduction.

\section*{Acknowledgements}

I would like to thank the anonymous reviewers for valuable comments
that have improved the presentation.

\appendix

\section*{Appendix}

Remark: In the the calculations in
Appendices~\ref{app-anal-simpl-cells}--\ref{app-anal-compl-cells},
where the preferred direction
of the receptive field is throughout assumed to correspond to the
horizontal direction $\varphi = 0$, we, for simplicity, denote the
magnitude of the motion parameter of the receptive field
by $v$ (as opposed to $|v|$ in the main body of the paper).
Similarly, we denote the magnitude of the motion parameter
of the stimulus by $u$ (as opposed to $|u|$ in the main body of the
paper).

The angle between the motion parameters of the stimulus and the
receptive field is throughout denoted by $\theta$, which now
thereby is defined relative to the horizontal axis.

Additionally, we denote the scale parameter $\sigma_{\varphi}$
along the preferred direction of the receptive field,
which now coincides with the first coordinate axis, as $\sigma_1$,
and the scale parameter $\sigma_{\orth}$ in the orthogonal direction,
which now coincides with the second coordinate axis, as $\sigma_2$.

\section{Analysis of the direction- and speed selectivity properties
  of velocity-tuned simple cells of orders between 1 and 4}
\label{app-anal-simpl-cells}

\subsection{Analysis for first-order simple cell without temporal
  differentiation}

Consider a velocity-adapted receptive field
corresponding to a {\em first-order\/} scale-normalized
Gaussian derivative with scale parameter $\sigma_1$ and speed $v$ in the horizontal
$x_1$-direction, a zero-order Gaussian kernel with scale parameter
$\sigma_2$ in the vertical $x_2$-direction, and a zero-order
Gaussian derivative with scale parameter $\sigma_t$
in the temporal direction, corresponding to $\varphi = 0$, $v = 0$,
$\Sigma_0 = \diag(\sigma_1^2, \sigma_2^2)$,
$m = 1$ and $n = 0$ in (\ref{eq-spat-temp-RF-model-der-norm-caus}):
\begin{align}
  \begin{split}
    & T_{0,\norm}(x_1, x_2, t;\; \sigma_1, \sigma_2, \sigma_t) =
  \end{split}\nonumber\\
  \begin{split}
     & = \frac{\sigma_1}{(2 \pi)^{3/2} \, \sigma_1 \sigma_2 \sigma_t} \,
            \partial_{x_1} 
            \left. \left(
                e^{-x_1^2/2\sigma_1^2 - x_2^2/2 \sigma_2^2 -  t^2/2\sigma_t^2}
            \right) \right|_{x_1 \rightarrow x_1-v t}
  \end{split}\nonumber\\
  \begin{split}
    & = \frac{(x_1 - v t)}{(2 \pi)^{3/2} \, \sigma_1^2 \sigma_2 \sigma_t} \,
               e^{-(x_1- vt)^2/2\sigma_1^2 - x_2^2/2 \sigma_2^2 - t^2/2\sigma_t^2}.
  \end{split}
\end{align}
The corresponding receptive field response,
when subjecting this receptive field to a moving sine wave of the form
(\ref{eq-sine-wave-model-vel-adapt-anal}), is then, after solving the
convolution integral in Wolfram Mathematica,
\begin{align}
   \begin{split}
     L_{0,\norm}(x_1, x_2, t;\; \sigma_1, \sigma_2, \sigma_t) =
  \end{split}\nonumber\\
  \begin{split}
    & = \int_{\xi_1 = -\infty}^{\infty}  \int_{\xi_2 = -\infty}^{\infty} \int_{\zeta = -\infty}^{\infty}
             T_{0,\norm}(\xi_1, \xi_2, \zeta;\; \sigma_1, \sigma_2, \sigma_t)
  \end{split}\nonumber\\
  \begin{split}
    & \phantom{= = \int_{\xi_1 = -\infty}^{\infty}  \int_{\xi_2 = -\infty}^{\infty}}
             \times f(x_1 - \xi_1, x_2 - \xi_2, t - \zeta) \, d \xi_1 \xi_2 d\zeta
  \end{split}\nonumber\\
  \begin{split}
    & = \omega  \, \sigma_1 \cos \theta \, 
  \end{split}\nonumber\\
  \begin{split}
   & \phantom{= =}
       \times e^{-\frac{\omega^2}{2} 
           \left( \left(\sigma_1^2+\sigma_t^2 v^2\right) \cos^2 (\theta)
             +\sigma_2^2 \sin ^2 \theta  -2 \sigma_t^2 u v \cos \theta +\sigma_t^2 u^2\right)}
  \end{split}\nonumber\\
  \begin{split}
    \label{eq-L0-vel-adapt-anal}
    & \phantom{= =}
           \times \cos
             \left(
               \cos (\theta) \, x_1 + \sin (\theta) \, x_2 -\omega  \, u \, t  + \beta
             \right),
   \end{split}         
\end{align}
{\em i.e.\/},\ a cosine wave with amplitude
\begin{align}
   \begin{split}
      & A_{\varphi}(\theta, u, \omega;\; \sigma_1, \sigma_2, \sigma_t,
      v) = 
   \end{split}\nonumber\\
   \begin{split}
     & = \omega \, \sigma_1 \left| \cos \theta \right| \, 
   \end{split}\nonumber\\
  \begin{split}
    \label{eq-A-varphi-spat-temp-anal}
      & \phantom{= =}         
          \times
          e^{-\frac{\omega^2}{2} 
               \left(\cos^2 \theta  \left(\sigma_1^2+\sigma_t^2 v^2\right)
               +\sigma_2^2 \sin ^2 \theta  -2 \sigma_t^2 u v \cos \theta +\sigma_t^2 u^2\right)}.
  \end{split}
\end{align}
Assume that a biological experiment regarding the response properties of the
receptive field is performed by varying the angular frequency
$\omega$ to get the maximum value of the
response over these parameters. Differentiating the amplitude $A_{\varphi}$
with respect to $\omega$ and setting this derivative to
zero, while also setting $\sigma_2 = \kappa \, \sigma_1$
and $v = \sigma_1/\sigma_t$, then gives
\begin{equation}
  \label{eq-omega1-vel-adapt}  
  \hat{\omega}_{\varphi} =
  \frac{1}{\sqrt{\frac{\sigma_1^2 \left(\kappa ^2 v^2 \sin ^2(\theta )-2 u v \cos
          (\theta )+2 v^2 \cos ^2(\theta )+u^2\right)}{v^2}}}
\end{equation}
Inserting this value into
$A_{\varphi}(\theta, u, \omega;\; \sigma_1, \sigma_2, \sigma_t, v)$,
then gives the following combined direction and speed selectivity measure
\begin{multline}
  \label{eq-Aphi-max}
  A_{\varphi,\max}(\theta, u;\; \kappa, v) = \\
  = \frac{\cos (\theta )}{\sqrt{e} \sqrt{\kappa ^2 \sin ^2(\theta )-\frac{2 u \cos (\theta
   )}{v}+2 \cos ^2(\theta )+\frac{u^2}{v^2}}},
\end{multline}
which for the inclination angle $\theta = 0$ and for a coupling of the
speed values of the form $u = r \, v$, reduces to the following
speed dependency, notably independent of the degree of elongation $\kappa$:
\begin{equation}
  \label{eq-rcurve-order1}
  R_{\varphi}(r) = A_{\varphi,\max}(0, r \, v;\; \kappa, r \, v)
  = \frac{1}{\sqrt{r^2-2 r+2}}.
\end{equation}
The left column in Figure~\ref{fig-dir-sel-simple-veladapt-anal-const-u} shows
the result of plotting the measure $A_{\varphi,\max}(\theta, u;\; \kappa, v)$ of
the direction selectivity as function of the inclination angle $\theta$
for a few values of the scale parameter ratio $\kappa$
for the special case when the speed of the stimulus is equal to
the preferred speed of the receptive field $u = v$, with the values rescaled
such that the peak value for each graph is equal to 1. As we can see
from the graphs, the direction selectivity becomes sharper with increasing
values of the degree of elongation $\kappa$.
The left column in
Figure~\ref{fig-dir-sel-simple-veladapt-anal-const-kappa}
shows corresponding results of
varying the magnitude of the motion according to $u = r \, v$
for different values of $r$, while keeping the degree of elongation
fixed to $\kappa = 2$. As we can see from the results, the direction
selectivity of the simple cells prefers motion speeds $u$ that are
near the preferred speed $v$ of the receptive field.
The left graph in
Figure~\ref{fig-dir-sel-simple-veladapt-anal-theta-0} finally
shows the relative speed selectivity curve $R_{\varphi}(r)$.

\subsection{Analysis for second-order simple cell without temporal
  differentiation}

Consider next a velocity-adapted receptive field
corresponding to a {\em second-order\/} scale-normalized
Gaussian derivative with scale parameter $\sigma_1$ and speed $v$ in the horizontal
$x_1$-direction, a zero-order Gaussian kernel with scale parameter
$\sigma_2$ in the vertical $x_2$-direction, and a zero-order
Gaussian derivative with scale parameter $\sigma_t$
in the temporal direction, corresponding to $\varphi = 0$, $v = 0$,
$\Sigma_0 = \diag(\sigma_1^2, \sigma_2^2)$,
$m = 2$ and $n = 0$ in (\ref{eq-spat-temp-RF-model-der-norm-caus}):
\begin{align}
  \begin{split}
    & T_{00,\norm}(x_1, x_2, t;\; \sigma_1, \sigma_2, \sigma_t) =
  \end{split}\nonumber\\
  \begin{split}
     & = \frac{\sigma_1^2}{(2 \pi)^{3/2} \, \sigma_1 \sigma_2 \sigma_t} \,
            \partial_{x_1 x_1} 
            \left. \left(
                e^{-x_1^2/2\sigma_1^2 - x_2^2/2 \sigma_2^2 - t^2/2\sigma_t^2}
            \right) \right|_{x_1 \rightarrow x_1-v t}
  \end{split}\nonumber\\
  \begin{split}
    & = \frac{((x_1 - v t)^2 - \sigma_1^2)}{(2 \pi)^{3/2} \, \sigma_1^3 \sigma_2 \sigma_t} \,
    e^{-(x_1 - v t)^2/2\sigma_1^2 - x_2^2/2 \sigma_2^2 - t^2/2\sigma_t^2}.
  \end{split}
\end{align}
The corresponding receptive field response,
when subjecting this receptive field to a moving sine wave of the form
(\ref{eq-sine-wave-model-vel-adapt-anal}), is then, after solving the
convolution integral in Wolfram Mathematica,
\begin{align}
   \begin{split}
     L_{00,\norm}(x_1, x_2, t;\; \sigma_1, \sigma_2, \sigma_t) =
  \end{split}\nonumber\\
  \begin{split}
    & = \int_{\xi_1 = -\infty}^{\infty}  \int_{\xi_2 = -\infty}^{\infty} \int_{\zeta = -\infty}^{\infty}
             T_{00,\norm}(\xi_1, \xi_2, \zeta;\; \sigma_1, \sigma_2, \sigma_t)
  \end{split}\nonumber\\
  \begin{split}
    & \phantom{= = \int_{\xi_1 = -\infty}^{\infty}  \int_{\xi_2 = -\infty}^{\infty}}
             \times f(x_1 - \xi_1, x_2 - \xi_2, t - \zeta) \, d \xi_1 \xi_2 d\zeta
  \end{split}\nonumber\\
  \begin{split}
    & = -\omega^2  \sigma_1^2 \cos^2 \theta \, 
 \end{split}\nonumber\\
  \begin{split}
    & \phantom{= =}
    \times
       e^{-\frac{\omega^2}{2} 
           \left( \left(\sigma_1^2+\sigma_t^2 v^2\right) \cos^2 \theta
             +\sigma_2^2 \sin ^2 \theta  -2 \sigma_t^2 u v \cos \theta +\sigma_t^2 u^2\right)}
  \end{split}\nonumber\\
  \begin{split}
    \label{eq-L00-vel-adapt-anal}
    & \phantom{= =}
           \times \sin
             \left(
               \cos (\theta) \, x_1 + \sin (\theta) \, x_2 - \omega \, u \, t  + \beta
             \right),
   \end{split}         
\end{align}
{\em i.e.\/},\ a sine wave with amplitude
\begin{align}
   \begin{split}
      & A_{\varphi\varphi}(\theta, u, \omega;\; \sigma_1, \sigma_2,
      \sigma_t, v) = \\
  \end{split}\nonumber\\
  \begin{split}
      & = \omega^2  \sigma_1^2 \cos^2 \theta \, 
   \end{split}\nonumber\\
  \begin{split}
    \label{eq-A-varphivarphi-spat-temp-anal}
    & \phantom{= =}    
        \times
        e^{-\frac{\omega^2}{2} 
             \left(\cos^2 \theta  \left(\sigma_1^2+\sigma_t^2 v^2\right)
              +\sigma_2^2 \sin ^2 \theta  -2 \sigma_t^2 u v \cos \theta +\sigma_t^2 u^2\right)}.
  \end{split}         
\end{align}
Assume that a biological experiment regarding the response properties of the
receptive field is performed by varying the angular frequency
$\omega$ to get the maximum value of the
response over this parameter. Differentiating the amplitude $A_{\varphi\varphi}$
with respect to $\omega$ and setting this derivative to
zero, while also setting $\sigma_2 = \kappa \, \sigma_1$
and $v = \sigma_1/\sigma_t$,
then gives
\begin{equation}
  \label{eq-omega2-vel-adapt}
   \hat{\omega}_{\varphi\varphi}
   = \frac{\sqrt{2}}{\sqrt{\frac{\sigma_1^2 \left(\kappa ^2 v^2 \sin ^2(\theta )-2 u v
   \cos (\theta )+2 v^2 \cos ^2(\theta )+u^2\right)}{v^2}}}.
\end{equation}
Inserting this value into
$A_{\varphi\varphi}(\theta, u, \omega;\; \sigma_1, \sigma_2,
\sigma_t, v)$,
then gives the following combined direction and speed selectivity measure
\begin{multline}
  \label{eq-Aphiphi-max}
  A_{\varphi\varphi,\max}(\theta, u, \; \kappa, v) = \\
= \frac{2 v^2 \cos ^2(\theta )}{e \left(\kappa ^2 v^2 \sin ^2(\theta )-2 u v \cos (\theta
   )+2 v^2 \cos ^2(\theta )+u^2\right)},
\end{multline}
which for the inclination angle $\theta = 0$ and for a coupling of the
speed values of the form $u = r \, v$, reduces to the following
speed dependency, notably again independent of the degree of elongation $\kappa$:
\begin{equation}
  \label{eq-rcurve-order2}
  R_{\varphi\varphi}(r) = A_{\varphi\varphi,\max}(0, r \, v;\; \kappa, r \, v)
  = \frac{1}{r^2-2 r+2}.
\end{equation}
The middle left column in Figure~\ref{fig-dir-sel-simple-veladapt-anal-const-u} shows
the result of plotting the measure $A_{\varphi\varphi,\max}(\theta, u;\; \kappa, v)$ of
the direction selectivity as function of the inclination angle $\theta$
for a few values of the scale parameter ratio $\kappa$
for the special case when the speed of the stimulus is equal to
the preferred speed of the receptive field $u = v$, with the values rescaled
such that the peak value for each graph is equal to 1. As we can see
from the graphs, the direction selectivity becomes sharper with increasing
values of the degree of elongation $\kappa$.
The middle left column in
Figure~\ref{fig-dir-sel-simple-veladapt-anal-const-kappa}
shows corresponding results of
varying the magnitude of the motion according to $u = r \, v$
for different values of $r$, while keeping the degree of elongation
fixed to $\kappa = 2$. As we can see from the results, again the direction
selectivity of the simple cells prefers motion speeds $u$ that are
near the preferred speed $v$ of the receptive field.
The middle left graph in
Figure~\ref{fig-dir-sel-simple-veladapt-anal-theta-0} finally
shows the relative speed selectivity curve $R_{\varphi\varphi}(r)$.

\subsection{Analysis for third-order simple cell without temporal
  differentiation}

Let us next consider a velocity-adapted receptive field
corresponding to a {\em third-order\/} scale-normalized
Gaussian derivative with scale parameter $\sigma_1$ and speed $v$ in the horizontal
$x_1$-direction, a zero-order Gaussian kernel with scale parameter
$\sigma_2$ in the vertical $x_2$-direction, and a zero-order
Gaussian derivative with scale parameter $\sigma_t$
in the temporal direction, corresponding to $\varphi = 0$, $v = 0$,
$\Sigma_0 = \diag(\sigma_1^2, \sigma_2^2)$,
$m = 3$ and $n = 0$ in (\ref{eq-spat-temp-RF-model-der-norm-caus}):
\begin{align}
  \begin{split}
    & T_{000,\norm}(x_1, x_2, t;\; \sigma_1, \sigma_2, \sigma_t) =
  \end{split}\nonumber\\
  \begin{split}
     & = \frac{\sigma_1^3}{(2 \pi)^{3/2} \, \sigma_1 \sigma_2 \sigma_t} \,
            \partial_{x_1 x_1 x_1} 
            \left. \left(
                e^{-x_1^2/2\sigma_1^2 - x_2^2/2 \sigma_2^2 - t^2/2\sigma_t^2}
            \right) \right|_{x_1 \rightarrow x_1-v t}
  \end{split}\nonumber\\
  \begin{split}
    & = \frac{-((x_1 - v t)^3 - 3 \sigma_1^2 x)}{(2 \pi)^{3/2} \, \sigma_1^4 \sigma_2 \sigma_t} \,
               e^{-(x_1 - v t)^2/2\sigma_1^2 - x_2^2/2 \sigma_2^2 - t^2/2\sigma_t^2}.
  \end{split}
\end{align}
The corresponding receptive field response,
when subjecting this receptive field to a moving sine wave of the form
(\ref{eq-sine-wave-model-vel-adapt-anal}),
can then be expressed as, after solving the
convolution integral in Wolfram Mathematica,
\begin{align}
   \begin{split}
     L_{000,\norm}(x_1, x_2;\; \sigma_1, \sigma_2) =
  \end{split}\nonumber\\
  \begin{split}
    & = \int_{\xi_1 = -\infty}^{\infty}  \int_{\xi_2 = -\infty}^{\infty}
             T_{000,\norm}(\xi_1, \xi_2;\; \sigma_1, \sigma_2)
  \end{split}\nonumber\\
  \begin{split}
    & \phantom{= = \int_{\xi_1 = -\infty}^{\infty}  \int_{\xi_2 = -\infty}^{\infty}}
             \times f(x_1 - \xi_1, x_2 - \xi_2) \, d \xi_1 \xi_2
  \end{split}\nonumber\\
  \begin{split}
    & = - \omega^3 \, \sigma_1^3 \cos^3 (\theta) \
\end{split}\nonumber\\
  \begin{split}
    & \phantom{= =}
    \times
       e^{-\frac{\omega^2}{2} 
           \left( \left(\sigma_1^2+\sigma_t^2 v^2\right) \cos^2 \theta
             +\sigma_2^2 \sin ^2 \theta  -2 \sigma_t^2 u v \cos \theta +\sigma_t^2 u^2\right)}
  \end{split}\nonumber\\
  \begin{split}
    \label{eq-L000-vel-adapt-anal}
    & \phantom{= =}
           \times \cos
             (
                \cos(\theta) \, x_1 + \sin (\theta) \, x_2 - \omega \, u \, t  + \beta
             ),
   \end{split}         
\end{align}
{\em i.e.\/},\ it corresponds to sine wave with amplitude
\begin{multline}
  A_{\varphi\varphi\varphi}(\theta, u, \omega;\; \sigma_1, \sigma_2, v) 
  = \omega^3 \, \sigma_1^3 \cos^3 (\theta) \, \times \\
      \times e^{-\frac{\omega^2}{2} 
           \left( \left(\sigma_1^2+\sigma_t^2 v^2\right) \cos^2 \theta
             +\sigma_2^2 \sin ^2 \theta  -2 \sigma_t^2 u v \cos \theta +\sigma_t^2 u^2\right)}
\end{multline}
To determine the value $\hat{\omega}$ of $\omega$ that gives the
strongest response, let us 
differentiate $A_{\varphi}(\theta, u, \omega;\; \sigma_1, \sigma_2, v)$ with
respect to $\omega$, set the derivative to zero,
while also setting $\sigma_2 = \kappa \, \sigma_1$
and $v = \sigma_1/\sigma_t$,
which gives:
\begin{equation}
  \label{eq-omega3-vel-adapt}
  \hat{\omega}_{\varphi\varphi\varphi}
  = \frac{\sqrt{3}}{\sqrt{\frac{\sigma_1^2 \left(\kappa ^2 v^2 \sin ^2(\theta )-2 u v
   \cos (\theta )+2 v^2 \cos ^2(\theta )+u^2\right)}{v^2}}}.
\end{equation}
Inserting this value into $A_{\varphi\varphi\varphi}(\theta, u, \omega;\; \sigma_1, \sigma_2, v)$
then gives rise to a combined direction and speed selectivity measure of the form
\begin{multline}
    \label{eq-ori-sel-simple-3der}
  A_{\varphi\varphi\varphi,\max}(\theta, u;\; \kappa, v) = \\
  = \frac{3 \sqrt{3} \cos ^3(\theta )}{e^{3/2} \left(\kappa ^2 \sin ^2(\theta )-\frac{2 u
   \cos (\theta )}{v}+2 \cos ^2(\theta )+\frac{u^2}{v^2}\right)^{3/2}},
\end{multline}
which for the inclination angle $\theta = 0$ and for a coupling of the
speed values of the form $u = r \, v$, reduces to the following
speed dependency, notably again independent of the degree of elongation $\kappa$:
\begin{equation}
  \label{eq-rcurve-order3}
  R_{\varphi\varphi\varphi}(r) = A_{\varphi\varphi\varphi,\max}(0, r \, v;\; \kappa, r \, v)
  = \frac{1}{\left(r^2-2 r+2\right)^{3/2}}.
\end{equation}
The middle right column in Figure~\ref{fig-dir-sel-simple-veladapt-anal-const-u} shows
the result of plotting the measure
$A_{\varphi\varphi\varphi,\max}(\theta, u;\; \kappa, v)$ of
the direction selectivity as function of the inclination angle $\theta$
for a few values of the scale parameter ratio $\kappa$
for the special case when the speed of the stimulus is equal to
the preferred speed of the receptive field $u = v$, with the values rescaled
such that the peak value for each graph is equal to 1. As we can see
from the graphs, the direction selectivity becomes sharper with increasing
values of the degree of elongation $\kappa$.
The middle right column in
Figure~\ref{fig-dir-sel-simple-veladapt-anal-const-kappa}
shows corresponding results of
varying the magnitude of the motion according to $u = r \, v$
for different values of $r$, while keeping the degree of elongation
fixed to $\kappa = 2$. As we can see from the results, again the direction
selectivity of the simple cells prefers motion speeds $u$ that are
near the preferred speed $v$ of the receptive field.
The middle right graph in
Figure~\ref{fig-dir-sel-simple-veladapt-anal-theta-0} finally
shows the relative speed selectivity curve $R_{\varphi\varphi\varphi}(r)$.

\subsection{Analysis for fourth-order simple cell without temporal
  differentiation}

Let us next consider a fourth-order idealized model of a simple cell 
with with scale parameter $\sigma_1$ and speed $v$ in the horizontal
$x_1$-direction, a zero-order Gaussian kernel with scale parameter
$\sigma_2$ in the vertical $x_2$-direction, and a zero-order
Gaussian derivative with scale parameter $\sigma_t$
in the temporal direction, corresponding to $\varphi = 0$, $v = 0$,
$\Sigma_0 = \diag(\sigma_1^2, \sigma_2^2)$,
$m = 4$ and $n = 0$ in (\ref{eq-spat-temp-RF-model-der-norm-caus}):
\begin{align}
  \begin{split}
    & T_{0000,\norm}(x_1, x_2, t;\; \sigma_1, \sigma_2, \sigma_t) =
  \end{split}\nonumber\\
  \begin{split}
     & = \frac{\sigma_4^3}{(2 \pi)^{3/2} \, \sigma_1 \sigma_2 \sigma_t} \,
            \partial_{x_1 x_1 x_1 x_1} 
            \left. \left(
                e^{-x_1^2/2\sigma_1^2 - x_2^2/2 \sigma_2^2 - t^2/2\sigma_t^2}
            \right) \right|_{x_1 \rightarrow x_1-v t}
  \end{split}\nonumber\\
  \begin{split}
    & = \frac{-((x_1 - v t)^4 - 6 \sigma_1^2 (x_1 - v t)^2 +3 \sigma_1^4)}{(2 \pi)^{3/2}
      \, \sigma_1^4 \sigma_2 \sigma_t} 
 \end{split}\nonumber\\
  \begin{split}    
     &  \phantom{=} \quad \times e^{-(x_1 - v t)^2/2\sigma_1^2 - x_2^2/2 \sigma_2^2 - t^2/2\sigma_t^2}.
  \end{split}
\end{align}
After solving the convolution integral in Wolfram Mathematica,
the corresponding receptive field response,
when subjecting this receptive field to a moving sine wave of the form
(\ref{eq-sine-wave-model-vel-adapt-anal}), is then of the form
\begin{align}
   \begin{split}
     L_{0000,\norm}(x_1, x_2;\; \sigma_1, \sigma_2) =
  \end{split}\nonumber\\
  \begin{split}
    & = \int_{\xi_1 = -\infty}^{\infty}  \int_{\xi_2 = -\infty}^{\infty}
             T_{0000,\norm}(\xi_1, \xi_2;\; \sigma_1, \sigma_2)
  \end{split}\nonumber\\
  \begin{split}
    & \phantom{= = \int_{\xi_1 = -\infty}^{\infty}  \int_{\xi_2 = -\infty}^{\infty}}
             \times f(x_1 - \xi_1, x_2 - \xi_2) \, d \xi_1 \xi_2
  \end{split}\nonumber\\
  \begin{split}
    & = \omega^4 \, \sigma_1^4 \cos^4 (\theta) \
\end{split}\nonumber\\
  \begin{split}
    & \phantom{= =}
    \times
       e^{-\frac{\omega^2}{2} 
           \left( \left(\sigma_1^2+\sigma_t^2 v^2\right) \cos^2 \theta
             +\sigma_2^2 \sin ^2 \theta  -2 \sigma_t^2 u v \cos \theta +\sigma_t^2 u^2\right)}
  \end{split}\nonumber\\
  \begin{split}
    \label{eq-L0000-vel-adapt-anal}
    & \phantom{= =}
           \times \sin
             (
                \cos (\theta) \, x_1 + \sin (\theta) \, x_2 - \omega \, u \, t  + \beta
             ),
   \end{split}         
\end{align}
{\em i.e.\/},\ a sine wave with amplitude
\begin{multline}
  A_{\varphi\varphi\varphi\varphi}(\theta, u, \omega;\; \sigma_1,
  \sigma_2, v) = \\
  = \omega^4 \, \sigma_1^4 \cos^4 (\theta) \,
      e^{-\frac{1}{2} \omega^2 (\sigma_1^2 \cos^2 \theta + \sigma_2^2 \sin^2 \theta)}.
\end{multline}
Again selecting the value of $\hat{\omega}$ at which the amplitude
assumes its maximum over $\omega$ gives
\begin{equation}
  \label{eq-omega4-vel-adapt}  
  \hat{\omega}_{\varphi\varphi\varphi\varphi}
  = \frac{2}{\sqrt{\frac{\sigma_1^2 \left(\kappa ^2 v^2 \sin ^2(\theta )-2 u v \cos
          (\theta )+2 v^2 \cos ^2(\theta )+u^2\right)}{v^2}}},
\end{equation}
which implies that the maximum amplitude over spatial scales as a
function of the inclination angle $\theta$ and the scale parameter
ratio $\kappa$ can be written
\begin{multline}
  \label{eq-ori-sel-simple-4der}
  A_{\varphi\varphi\varphi\varphi,\max}(\theta, u;\; \kappa, v) = \\
  = \frac{16 v^4 \cos ^4(\theta )}{e^2 \left(\kappa ^2 v^2 \sin ^2(\theta )-2 u v \cos
   (\theta )+2 v^2 \cos ^2(\theta )+u^2\right)^2},
\end{multline}
which for the inclination angle $\theta = 0$ and for a coupling of the
speed values of the form $u = r \, v$, reduces to the following
speed dependency, notably again independent of the degree of elongation $\kappa$:
\begin{equation}
  \label{eq-rcurve-order4}
  R_{\varphi\varphi\varphi\varphi}(r) =
  A_{\varphi\varphi\varphi,\max}(0, r \, v;\; \kappa, v)
  = \frac{1}{\left(r^2-2 r+2\right)^2}.
\end{equation}
The right column in Figure~\ref{fig-dir-sel-simple-veladapt-anal-const-u} shows
the result of plotting the measure
$A_{\varphi\varphi\varphi\varphi,\max}(\theta, u;\; \kappa, v)$ of
the direction selectivity as function of the inclination angle $\theta$
for a few values of the scale parameter ratio $\kappa$
for the special case when the speed of the stimulus is equal to
the preferred speed of the receptive field $u = v$, with the values rescaled
such that the peak value for each graph is equal to 1. As we can see
from the graphs, the direction selectivity becomes sharper with increasing
values of the degree of elongation $\kappa$.
The right column in
Figure~\ref{fig-dir-sel-simple-veladapt-anal-const-kappa}
shows corresponding results of
varying the magnitude of the motion according to $u = r \, v$
for different values of $r$, while keeping the degree of elongation
fixed to $\kappa = 2$. As we can see from the results, again the direction
selectivity of the simple cells prefers motion speeds $u$ that are
near the preferred speed $v$ of the receptive field.
The right graph in
Figure~\ref{fig-dir-sel-simple-veladapt-anal-theta-0} finally
shows the relative speed selectivity curve
$R_{\varphi\varphi\varphi\varphi}(r)$.

\section{Analysis for integrated velocity-tuned models of complex cells}
\label{app-anal-compl-cells}

To model the spatial response of a complex cell
according to the spatio-temporal quasi-quadrature measure
(\ref{eq-quasi-quad-dir-vel-adapt-spat-temp-12})
based on velocity-adapted spatio-temporal
receptive fields,
to a moving sine wave of the form
(\ref{eq-sine-wave-model-vel-adapt-anal}),
we combine either the responses of the first- and
second-order simple cells
\begin{equation}
  \label{eq-quasi-quad-dir-vel-adapt-spat-temp-anal-12}
  ({\cal Q}_{0,12,\vel,\norm} L)^2
  = L_{0,\norm}^2 + \, C_{\varphi} \, L_{00,\norm}^2
\end{equation}
with $L_{0,\norm}$ according to (\ref{eq-L0-vel-adapt-anal}) and
$L_{00,\norm}$ according to (\ref{eq-L00-vel-adapt-anal}).
Similarly, we model the spatial response of a complex cell
according to the spatio-temporal quasi-quadrature measure
(\ref{eq-quasi-quad-dir-vel-adapt-spat-temp-34})
based on the responses of the third- and
fourth-order simple cells as
\begin{multline}
  \label{eq-quasi-quad-dir-vel-adapt-spat-temp-anal-34}
  ({\cal Q}_{0,34,\vel,\norm} L)^2
  = L_{000,\norm}^2 
              + \, C_{\varphi\varphi\varphi} \, L_{0000,\norm}^2,
\end{multline}
with $L_{000,\norm}$ according to (\ref{eq-L000-vel-adapt-anal}) and
$L_{0000,\norm}$ according to (\ref{eq-L0000-vel-adapt-anal}).

Selecting the angular frequency as the geometric average of the angular
frequency values at which the above spatio-temporal simple cell models
assume their maxima over angular frequencies, gives either
\begin{multline}
   \label{eq-omegaQ-vel-adapt12}
   \hat{\omega}_{{\cal Q}_{12}}
   = \sqrt{\hat{\omega}_{\varphi} \, \hat{\omega}_{\varphi\varphi}} \\
   = \sqrt[4]{2} \times \\ \sqrt{\frac{v^2}{\sigma_1^2 \left(\kappa ^2 v^2 \sin ^2(\theta )-2 u v
   \cos (\theta )+2 v^2 \cos ^2(\theta )+u^2\right)}}
 \end{multline}
with $\hat{\omega}_{\varphi}$ according to (\ref{eq-omega1-vel-adapt}) and
$\hat{\omega}_{\varphi\varphi}$ according to
(\ref{eq-omega2-vel-adapt}), or
\begin{multline}
   \label{eq-omegaQ-vel-adapt34}
   \hat{\omega}_{{\cal Q}_{34}}
   = \sqrt{\hat{\omega}_{\varphi\varphi\varphi} \,
     \hat{\omega}_{\varphi\varphi\varphi\varphi}} = \\
   = \sqrt{2} \sqrt[4]{3} \times \\ \sqrt{\frac{v^2}{\sigma_1^2 \left(\kappa ^2 v^2 \sin ^2(\theta
         )-2 u v \cos (\theta )+2 v^2 \cos ^2(\theta )+u^2\right)}}
\end{multline}
with $\hat{\omega}_{\varphi\varphi\varphi}$ according to (\ref{eq-omega3-vel-adapt}) and
$\hat{\omega}_{\varphi\varphi\varphi\varphi}$ according to
(\ref{eq-omega4-vel-adapt}).

\begin{figure*}[hbtp]
  \begin{dmath}
    \label{eq-A-Q12-int-compl}
    A_{{\cal Q}_{12},\max}^2(\theta, r;\; \kappa) =
    \cos ^2(\theta ) \exp \left(-\frac{\sqrt{2} \left(5 \kappa ^2 \sin ^2(\theta
          )+\cos ^2(\theta )+2 \cos (2 \theta )+2\right)}{\kappa ^2 \sin ^2(\theta )+r^2-2 r
        \cos (\theta )+2 \cos ^2(\theta )}\right) \left(\cos ^2(\theta ) \left(3 \exp
        \left(\frac{2 \sqrt{2} \left(2 \kappa ^2 \sin ^2(\theta )+\cos (2 \theta
              )+1\right)}{\kappa ^2 \sin ^2(\theta )+r^2-2 r \cos (\theta )+2 \cos ^2(\theta
            )}\right)+1\right)-2 r \cos (\theta ) \left(\exp \left(\frac{2 \sqrt{2} \left(2
              \kappa ^2 \sin ^2(\theta )+\cos (2 \theta )+1\right)}{\kappa ^2 \sin ^2(\theta
            )+r^2-2 r \cos (\theta )+2 \cos ^2(\theta )}\right)+1\right)+\left(\kappa ^2 \sin
        ^2(\theta )+r^2\right) \left(\exp \left(\frac{2 \sqrt{2} \left(2 \kappa ^2 \sin
              ^2(\theta )+\cos (2 \theta )+1\right)}{\kappa ^2 \sin ^2(\theta )+r^2-2 r \cos
            (\theta )+2 \cos ^2(\theta )}\right)+1\right)\right)/
    (\sqrt{2} \left(\kappa ^2 \sin
      ^2(\theta )+r^2-2 r \cos (\theta )+2 \cos ^2(\theta )\right)^2)
  \end{dmath}

  \begin{dmath}
    \label{eq-A-Q34-int-compl}    
    A_{{\cal Q}_{34},\max}^2(\theta, r;\; \kappa) =
    12 \sqrt{3} \cos ^6(\theta ) \exp \left(-\frac{2 \sqrt{3} \left(5 \kappa ^2 \sin
          ^2(\theta )+\cos ^2(\theta )+2 \cos (2 \theta )+2\right)}{\kappa ^2 \sin ^2(\theta
        )+r^2-2 r \cos (\theta )+2 \cos ^2(\theta )}\right) \left(\cos ^2(\theta )
      \left(\left(2+\sqrt{6}\right) \exp \left(\frac{4 \sqrt{3} \left(2 \kappa ^2 \sin
              ^2(\theta )+\cos (2 \theta )+1\right)}{\kappa ^2 \sin ^2(\theta )+r^2-2 r \cos
            (\theta )+2 \cos ^2(\theta )}\right)+2-\sqrt{6}\right)-2 r \cos (\theta ) \left(\exp
        \left(\frac{4 \sqrt{3} \left(2 \kappa ^2 \sin ^2(\theta )+\cos (2 \theta
              )+1\right)}{\kappa ^2 \sin ^2(\theta )+r^2-2 r \cos (\theta )+2 \cos ^2(\theta
            )}\right)+1\right)+\left(\kappa ^2 \sin ^2(\theta )+r^2\right) \left(\exp
        \left(\frac{4 \sqrt{3} \left(2 \kappa ^2 \sin ^2(\theta )+\cos (2 \theta
              )+1\right)}{\kappa ^2 \sin ^2(\theta )+r^2-2 r \cos (\theta )+2 \cos ^2(\theta
            )}\right)+1\right)\right)/
    \left(\kappa ^2 \sin ^2(\theta )+r^2-2 r \cos (\theta )+2
      \cos ^2(\theta )\right)^4
  \end{dmath}
  \caption{Explicit expressions for the squares of the amplitude measures
    $A_{{\cal Q}_{12},\max}(\theta, r;\; \kappa)$
    and $A_{{\cal Q}_{34},\max}(\theta, r;\; \kappa)$ to a
    probing sine wave with inclination angle $\theta$
    for the idealized models of velocity-tuned complex cells
    based on integrated quasi quadrature measures according
    to (\ref{eq-quasi-quad-dir-vel-adapt-spat-temp-12-int}) and
    (\ref{eq-quasi-quad-dir-vel-adapt-spat-temp-34-int}), respectively,
    with $r = u/v$ denoting the ratio between the image speed $u$ of
    the stimulus and the preferred speed $v$ of the receptive field,
    and $\kappa$ denoting the degree of elongation of the receptive field.}
  \label{fig-eq-ampl-meas-int-compl-cells-12-34}
\end{figure*}

Integrating the resulting expressions in Wolfram Mathematica,
with an affine Gaussian window
function according to
(\ref{eq-quasi-quad-dir-vel-adapt-spat-temp-12-int}) or
(\ref{eq-quasi-quad-dir-vel-adapt-spat-temp-34-int}),
and setting the relative weights between first- and second-order
information or between third- and fourth-order information
to $C_{\varphi} = 1/\sqrt{2}$ and $C_{\varphi\varphi\varphi} =
1/\sqrt{2}$,
while also setting $\sigma_2 = \kappa \, \sigma_1$
and $v = \sigma_1/\sigma_t$, 
then leads to the resulting composed amplitude measures
$A_{{\cal Q}_{12},\max}(\theta, r;\; \kappa)$ and
$A_{{\cal Q}_{34},\max}(\theta, r;\; \kappa)$
as functions of the inclination angle $\theta$,
the ratio $r = u/v$ between the speed $u$ of the input stimulus
and the preferred speed $v$ of the receptive
field, and the degree of elongation $\kappa$ of the receptive field
in closed form as
shown in Equations~(\ref{eq-A-Q12-int-compl})
and~(\ref{eq-A-Q34-int-compl}) in
Figure~\ref{fig-eq-ampl-meas-int-compl-cells-12-34}.
%Unfortunately, these expressions are, however, too complex to
%be reproduced here.

In the special case when the inclination angle $\theta$ is zero, the resulting
expressions are, however, more compact and are given in
Figure~\ref{fig-alg-expr-R-curves-int-compl-cells} in
Section~\ref{sec-dir-sel-compl-cells}. Additionally,
Figure~\ref{fig-dir-sel-complex-const-u}  in
Section~\ref{sec-dir-sel-compl-cells} shows graphs of
the corresponding direction selectivity curves in the special case
when the image speed $u$ of the stimulus is equal to the preferred
speed $v$ of the spatio-temporal receptive field.

\begin{figure*}[hbtp]
  \begin{center}
    \begin{tabular}{cc}
      {\em first-order simple cell} & \hspace{-17mm} {\em second-order simple cell} \\
      $A_{\varphi}(\theta;\; r=1, \kappa)$ & \hspace{-17mm} $A_{\varphi\varphi}(\theta;\; r=1, \kappa)$ \\ 
      \includegraphics[height=0.16\textheight]{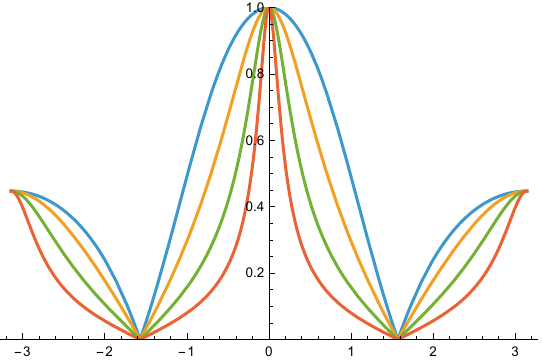}
      & \includegraphics[height=0.16\textheight]{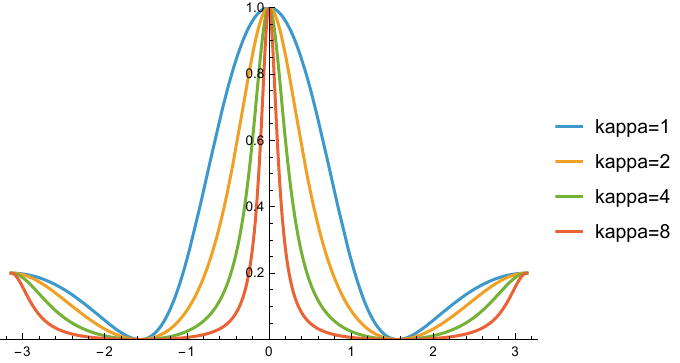} \\
      $\,$ \\
      {\em third-order simple cell} & \hspace{-17mm} {\em fourth-order simple cell} \\
      $A_{\varphi\varphi\varphi}(\theta;\; r=1, \kappa)$ & \hspace{-17mm} $A_{\varphi\varphi\varphi\varphi}(\theta;\; r=1, \kappa)$ \\ 
      \includegraphics[height=0.16\textheight]{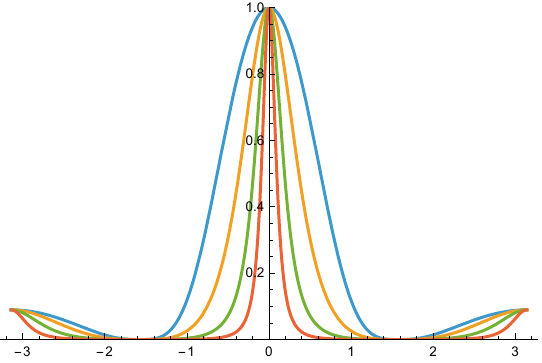}
      & \includegraphics[height=0.16\textheight]{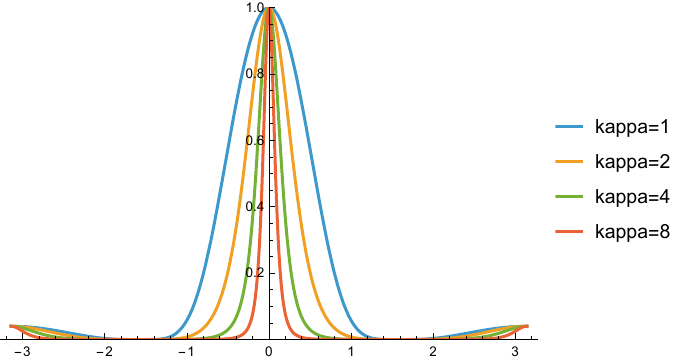} \\
      $\,$ \\
      {\em complex cell (12)} & \hspace{-17mm} {\em complex cell (34)} \\
      $A_{{\cal Q}_{12},\max}(\theta, r=1;\; \kappa)$ & \hspace{-17mm} $A_{{\cal Q}_{34},\max}(\theta, r=1;\; \kappa)$ \\
      \includegraphics[height=0.16\textheight]{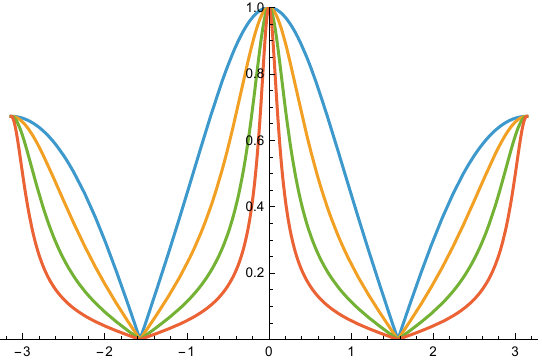}
      &
        \includegraphics[height=0.16\textheight]{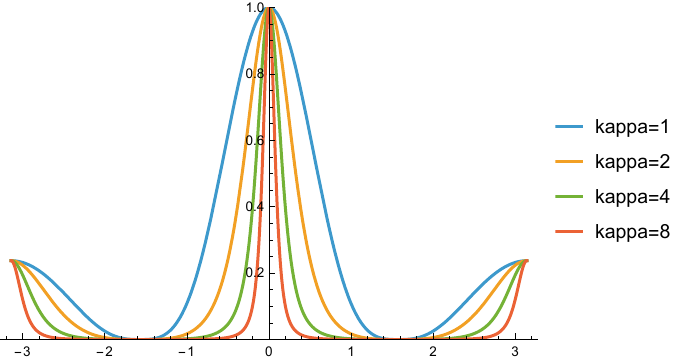}
    \end{tabular}
  \end{center}
  \caption{Cartesian plots of the idealized models for
     the direction selectivity curves for
    {\bf (top and middle rows)} velocity-tuned simple cells of the form
    $A_{\varphi^m}(\theta;\; r, \kappa)  = A_{\varphi^m,\max}(\theta, |u|;\; \kappa, |v|)$
    for $r = |u|/|v|$ and with $A_{\varphi^m,\max}(\theta, |u|;\; \kappa, |v|)$
    according to (\ref{eq-cond-summ-dir-sel-prop})
    for the order spatial differentiation $m \in \{ 1, 2, 3, 4 \}$, and
   {\bf (bottom row)} velocity-tuned complex cells of the forms
   $A_{{\cal Q}_{12},\max}(\theta, r;\; \kappa)$
   according to (\ref{eq-A-Q12-int-compl}) and
   $A_{{\cal Q}_{34},\max}(\theta, r;\; \kappa) $
   according to (\ref{eq-A-Q34-int-compl}),
   for the relative speed ratio $r = |u|/|v| = 1$ and
   different degrees of elongation $\kappa$.
   (Horizontal axes: inclination angle $\theta \in [-\pi, \pi]$.
   Vertical axes: response relative to the peak response for
   $\theta = 0$.)}
 \label{fig-cart-plots-dir-sel}
\end{figure*}

\section{Cartesian plots of the idealized models for direction selectivity curves}
\label{app-cart-dir-sel-plots}

With regard to the notion of direction selectivity, where the
directions are defined over the range $[-\pi, \pi]$,
the notion of polar plots, as used for the neurophysiological
measurements by Livingstone and Conway \citeyear{LivCon03-JNeuroPhys}
and also in Sections~\ref{sec-dir-sel-simpl-cells},
\ref{sec-dir-sel-compl-cells} and~\ref{sec-model-livingstone},
constitute a natural choice, since the plots then naturally reflect
the 2$\pi$-periodicity of the direction selectivity curves.
To enable comparisons to other work, where direction
selectivity may instead have been shown with Cartesian
plots, we provide such visualizations for the idealized
models of simple cells and complex cells, respectively,
in Figure~\ref{fig-cart-plots-dir-sel}.

\begin{figure*}[hbtp]
  \begin{center}
    \begin{subfigure}{0.99\textwidth}
      \begin{tabular}{cccc}
        $\hat{T}_{\varphi}(\omega_1, \omega_2;\; \sigma_{\varphi}, \sigma_{\orth})$
        & $\hat{T}_{\varphi\varphi}(\omega_1, \omega_2;\; \sigma_{\varphi}, \sigma_{\orth})$
        & $\hat{T}_{\varphi\varphi\varphi}(\omega_1, \omega_2;\; \sigma_{\varphi}, \sigma_{\orth})$
        & $\hat{T}_{\varphi\varphi\varphi\varphi}(\omega_1, \omega_2;\;  \sigma_{\varphi}, \sigma_{\orth})$ \\
        \includegraphics[width=0.22\textwidth]{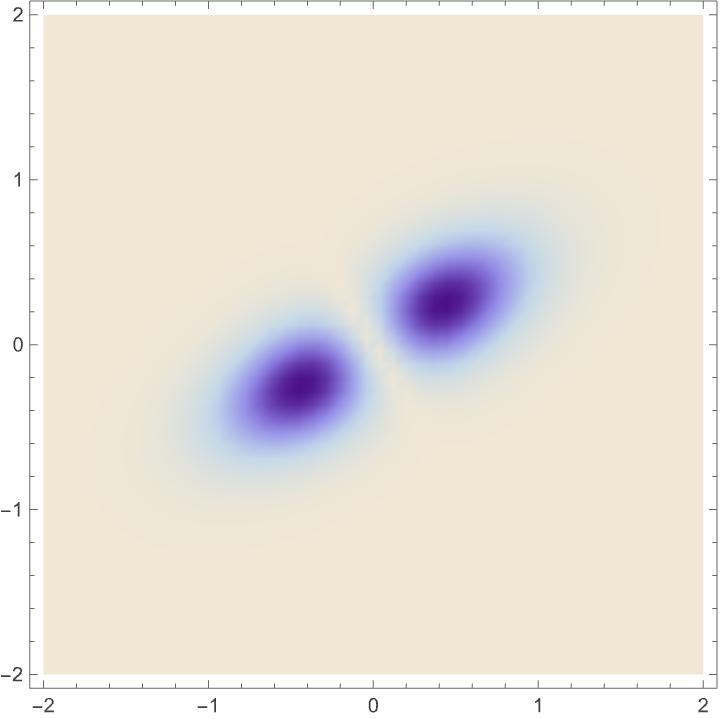}
        & \includegraphics[width=0.22\textwidth]{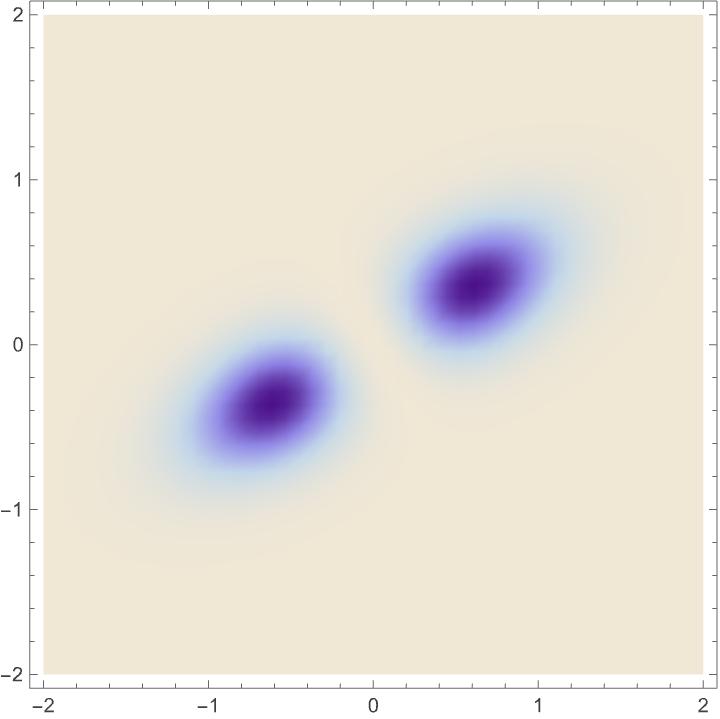}
        & \includegraphics[width=0.22\textwidth]{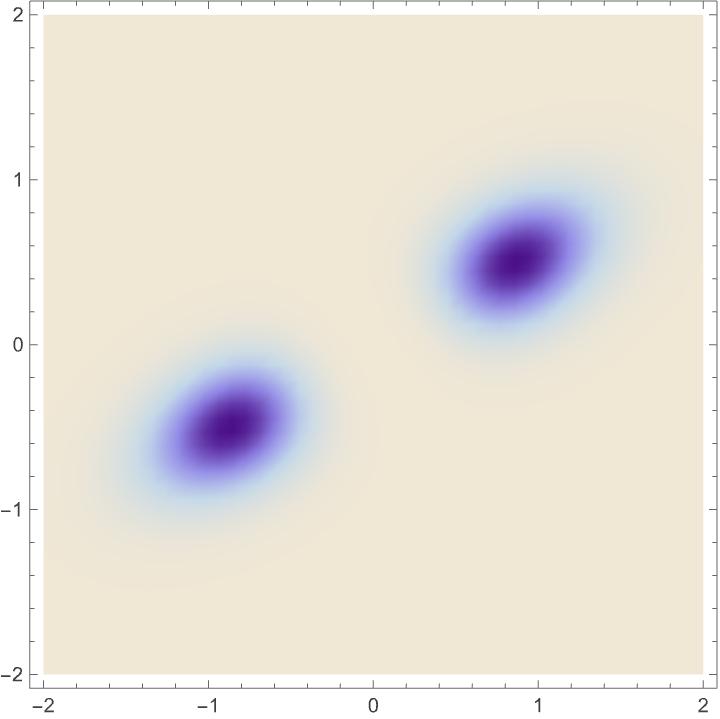}
        & \includegraphics[width=0.22\textwidth]{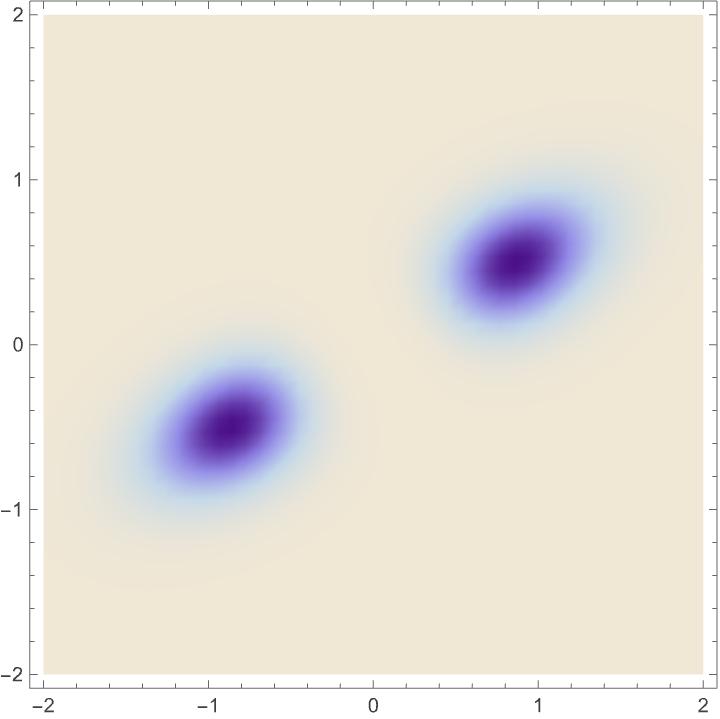}
      \end{tabular}
      \caption{2-D spatial cross-sections of the of the form
        $|\hat{T}_{\varphi^m}(\omega_1, \omega_2;\; \sigma_{\varphi}, \sigma_{\orth})|
        = | (i \, (\omega_1 \cos \varphi + \omega_2 \sin \varphi))^m \,
        T(\omega_1, \omega_2;\; \sigma_{\varphi}, \sigma_{\orth})|$,
        with the latter entity given by
        $T(\omega_1, \omega_2;\; \sigma_{\varphi}, \sigma_{\orth})=
        \exp \left(-\frac{1}{4} (\sigma_{\varphi}-\sigma_{\orth}) (\sigma_{\varphi}+\sigma_{\orth}) (2
   \omega_1 \, \omega_2 \sin (2 \phi )+\cos (2 \phi ) (\omega_1-\omega_2)
   (\omega_1+\omega_2))-\frac{1}{4} \left(\sigma_{\varphi}^2+\sigma_{\orth}^2\right)
   \left(\omega_1^2+\omega_2^2\right)\right)$,
        of the absolute values of the Fourier transform
        of the 2+1-D spatio-temporal receptive field model
        (\ref{eq-spat-temp-RF-model-der-norm-caus}),
        with $\sigma_{\varphi}$ and $\sigma_{\orth}$ corresponding to
        the amount of smoothing, as specified by the spatial covariance matrix
        $\Sigma$, along the preferred {\em vs.\/}\ the orthogonal directions of
        the receptive field, here for $\varphi = \pi/6$,
        $\sigma_{\varphi}=2$  and $\sigma_{\orth}=4$, and
        with darker tones corresponding to higher values.
        (Horizontal axes: angular frequency variable $\omega_1 \in [-2, 2]$
        along the horizontal direction.
        Vertical axes: angular frequency variable $\omega_2 \in [-2, 2]$
        along the vertical direction.)}
      
    \end{subfigure}

    \medskip
    
    \begin{subfigure}{0.99\textwidth}
      \begin{tabular}{cccc}
        {\footnotesize $\hat{T}_{\varphi}(\omega_{\varphi}, \omega_t;\; \sigma_{\varphi}, \sigma_{t}, v)$}
        & {\footnotesize $\hat{T}_{\varphi\varphi}(\omega_{\varphi}, \omega_t;\; \sigma_{\varphi}, \sigma_{t}, v)$}
        & {\footnotesize $\hat{T}_{\varphi\varphi\varphi}(\omega_{\varphi}, \omega_t;\; \sigma_{\varphi}, \sigma_{t}, v)$}
        & {\footnotesize $\hat{T}_{\varphi\varphi\varphi\varphi}(\omega_{\varphi}, \omega_t;\; \sigma_{\varphi}, \sigma_{t}, v)$} \\
        \includegraphics[width=0.22\textwidth]{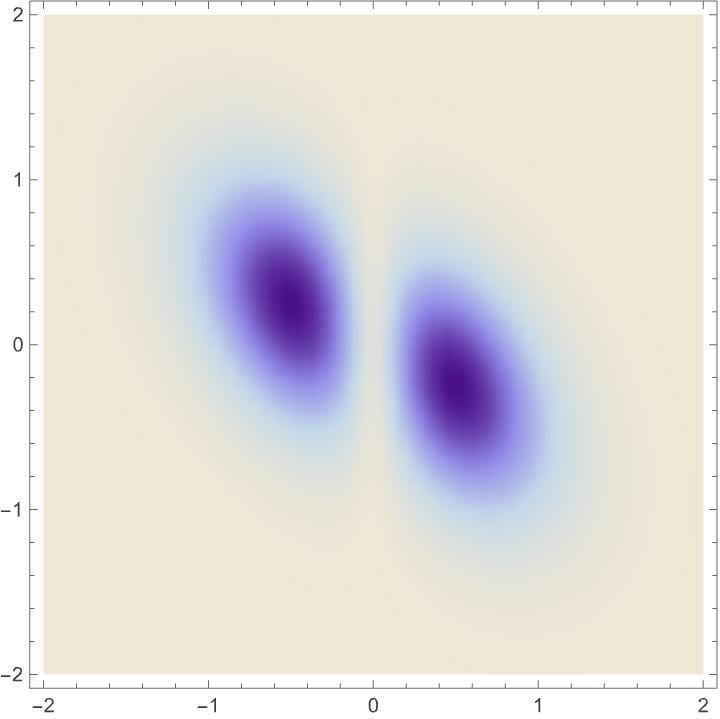}
        & \includegraphics[width=0.22\textwidth]{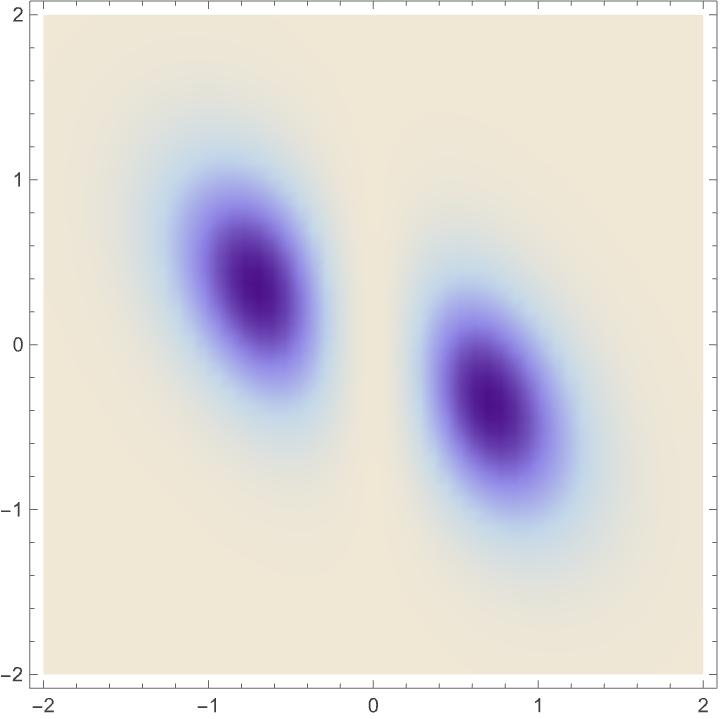}
        & \includegraphics[width=0.22\textwidth]{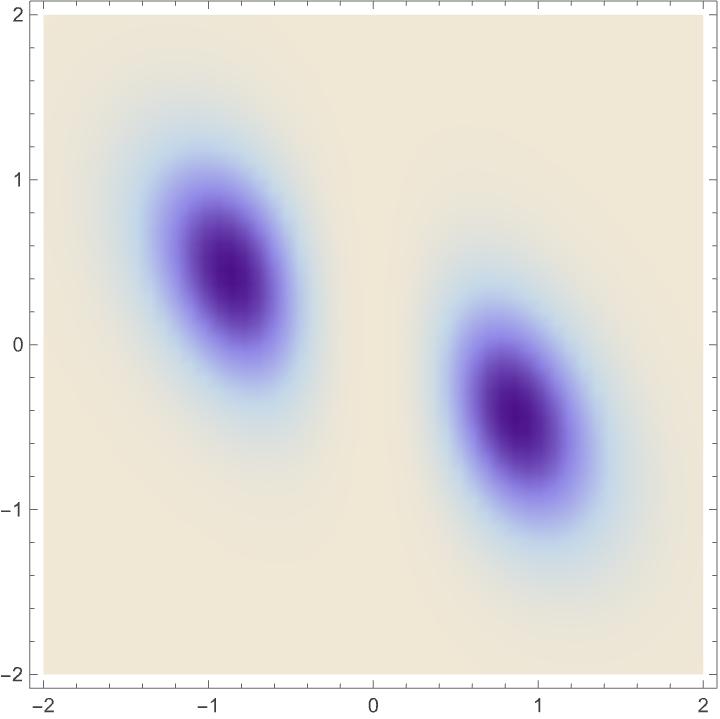}
        & \includegraphics[width=0.22\textwidth]{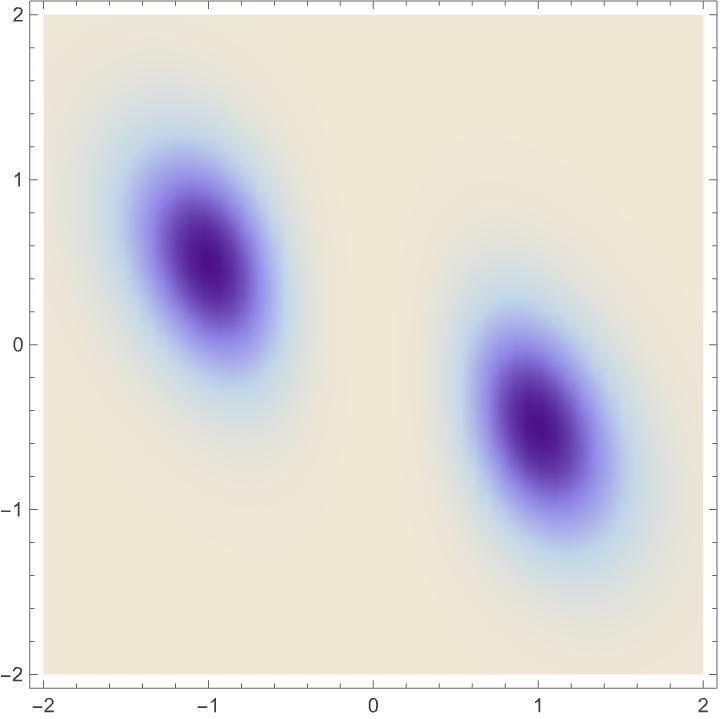}
      \end{tabular}
      \caption{1+1-D spatio-temporal cross-sections of the form
        $|\hat{T}_{\varphi^m}(\omega_{\varphi}, \omega_t;\; \sigma_{\varphi}, \sigma_{t}, v)|
        = | (i \, \omega_{\varphi})^m \exp(-\frac{1}{2} \sigma_{\varphi}^2 \omega_{\varphi}^2
        -\frac{1}{8} \sigma_t^2 (\omega_{\varphi}+2 \omega_t)^2)|$
        of the absolute values of the Fourier transforms
        of the 2+1-D spatio-temporal receptive field model
        (\ref{eq-spat-temp-RF-model-der-norm-caus})
        along the preferred direction over the spatial domain,
        here for $\varphi = \pi/6$, $\sigma_{\varphi}=2$,
        $\sigma_t=2$ and $v=1/2$, and
        with darker tones corresponding to higher values.
        (Horizontal axes: spatial angular frequency variable along the preferred
        direction of the receptive field $\omega_{\varphi} \in [-2, 2]$.
        Vertical axes: temporal angular frequency variable $\omega_t \in [-5, 5]$.)}
    \end{subfigure}
  \end{center}
  \caption{Visualizations of the Fourier transforms of the
    velocity-adapted affine Gaussian derivative filters used
    both as (i)~idealized models of simple cells
    of the form (\ref{eq-spat-temp-RF-model-der-norm-caus})
    and (ii)~for computing the directional quasi
    quadrature measures according to
    (\ref{eq-quasi-quad-dir-vel-adapt-spat-temp-12}) and
   (\ref{eq-quasi-quad-dir-vel-adapt-spat-temp-34}) in the idealized
   models of complex cells.
   Specifically, this figure shows the Fourier transforms of the
   either {\bf (a)} purely 2-D spatial cross-sections or the
   {\bf (b)} joint 1+1-D cross-sections of the spatio-temporal
   receptive fields shown in Figure~\ref{fig-filters-compl-cell-models}.}
   \label{fig-ft-filters-compl-cell-models}
\end{figure*}

\section{Fourier transforms of the idealized models for simple cells}
\label{app-fourier-transf} 

For readers who may find a complementary frequency analysis valuable,
Figure~\ref{fig-ft-filters-compl-cell-models} shows the
Fourier transforms of the cross-sections of the
spatio-temporal receptive fields shown in
Figure~\ref{fig-filters-compl-cell-models}.

\bibliographystyle{abbrvnat}

{\footnotesize
\bibliography{defs,tlmac}
}

\end{document}